\documentclass[12pt,preprint]{aastex}

\slugcomment{accepted by AJ April 23, 2006}
\shorttitle{UDF BPZ Catalog}
\shortauthors{Coe et al.}
\newcommand{\galfit}{{\tt galfit}}
\newcommand{\sex}{{\tt SExtractor}}

\begin{document}

\title{Galaxies in the Hubble Ultra Deep Field:
I. Detection, Multiband Photometry, Photometric Redshifts, and Morphology}

\author{Dan Coe\altaffilmark{1,2}, Narciso Ben\'itez\altaffilmark{1,2},
Sebasti\'an F. S\'anchez\altaffilmark{3}, Myungkook Jee\altaffilmark{1}, 
Rychard Bouwens\altaffilmark{4}, Holland Ford\altaffilmark{1}}

\email{
coe@iaa.es, 
benitez@iaa.es, 
sanchez@caha.es, 
mkjee@pha.jhu.edu, 
bouwens@ucolick.org,
ford@pha.jhu.edu
}

\altaffiltext{1}
{Johns Hopkins University, 
Dept. of Physics \& Astronomy, 
3400 N. Charles St., Baltimore, MD 21218, USA}

\altaffiltext{2}{Instituto de Astrof\'isica de Andaluc\'ia (CSIC), 
Camino Bajo de Hu\'etor 50, Granada 18008, Spain}

\altaffiltext{3}{Calar Alto Observatory, Almer\'ia E-040004, Spain}

\altaffiltext{4}{University of California, 
Astronomy Dept., Santa Cruz, CA 95064}

\begin{abstract}
We present aperture-matched PSF-corrected $BVi\arcmin z\arcmin JH$ photometry
and Bayesian photometric redshifts ({\tt BPZ}) for objects detected in the Hubble
Ultra Deep Field (UDF), 8,042 of which are detected at the 10-$\sigma$ level (e.g.,
$i\arcmin<29.01$ or $z\arcmin<28.43$).
Most of our objects are defined identically to those in the public STScI catalogs,
enabling straightforward object-by-object comparison.
We have combined detections from $i\arcmin$, $z\arcmin$, $J$+$H$, 
and $B$+$V$+$i\arcmin$+$z\arcmin$ images into a single comprehensive segmentation map.
Using a new program called {\tt SExSeg} we are able to force this segmentation map
into {\tt SExtractor} for photometric analysis.
The resulting photometry is corrected for the wider NIC3 PSFs using our {\tt ColorPro} software.
We also correct for the ACS $z\arcmin$-band PSF halo.
Offsets are applied to our NIC3 magnitudes, 
which are found to be too faint relative to the ACS fluxes.
Based on {\tt BPZ} SED fits to objects of known spectroscopic redshift,
we derived corrections of $-0.30\pm0.03$ mag in $J$ and $-0.18\pm0.04$ mag in $H$.
Our offsets appear to be supported by a recent recalibration of the UDF NIC3 images
combined with non-linearity measured in NICMOS itself.

The UDF reveals a large population of faint blue galaxies
(presumably young starbursts), 
bluer than those observed in the original Hubble Deep Fields (HDF).
To accommodate these galaxies, we have added two new starburst templates to the
SED library used in previous {\tt BPZ} papers.
The resulting photometric redshifts are accurate to within $0.04(1+z_{spec})$ out to $z < 6$.
Our {\tt BPZ} results include a full redshift probability distribution for each galaxy.
By adding these distributions, we obtain the redshift probability histogram 
for galaxies in the UDF.
Median redshifts are also provided for different magnitude limited samples.
Finally, we measure galaxy morphology, including S\'ersic index and asymmetry.
Simulations allow us to quantify the reliability of our morphological results.
Our full catalog along with our software packages {\tt SExSeg} and {\tt ColorPro}
are available at \texttt{http://adcam.pha.jhu.edu/\textasciitilde{}coe/UDF/}.

\end{abstract}


\section{Introduction}

The Hubble Ultra Deep Field (UDF) provides us with our deepest view
to date of the visible universe. It is located within one of the best
studied areas of the sky: the Chandra Deep Field South (CDF-S). 
With a total of 544 orbits, it is one of the largest time allocations 
with HST, and indeed the filter coverage, depth, and exquisite quality 
of the UDF ACS and NICMOS images provide an unprecedented data set 
for galaxy evolution studies.

A comprehensive picture of galaxy formation and evolution must match 
the observed population statistics of integrated galaxy properties.
These include the galaxy luminosity function, size distribution, and star formation rates
all as functions of both redshift and environment.
We must also be able to explain observed internal galactic structure,
including bulge-to-disk ratio, asymmetry, and nuclear properties.

Large-area HST/ACS multiband surveys 
such as GEMS \citep{GEMS}, GOODS \citep{GOODS}, and COSMOS (Scoville et al., in prep.)~have 
contributed significantly to our understanding of galaxy evolution.
These studies demonstrate the utility of high resolution multiband imaging.
Multiband photometry allows robust determinations of photometric redshifts and even 
star formation rates, while high resolution imaging enables morphological 
classifications out to distant redshifts.
The unparalleled depth and spatial resolution of the UDF dataset allow astronomers 
to extend studies like these to higher redshift.

To date, 76 spectroscopic redshifts have been
obtained for galaxies within the UDF (see \S\ref{subsub:spec-z}),
and more will surely be forthcoming. But, as was the case with the
original Hubble Deep Fields (HDF-N \citealt{HDF-N}; HDF-S \citealt{HDF-S}),
most of the objects detected in this field 
will elude spectroscopy for years to come.
(We detect over 8,000 galaxies at 10-$\sigma$ in the UDF.)

\defcitealias{FLY99}{FLY99}

The original Hubble Deep Field (HDF-N) gave impetus to photometric redshifts, 
transforming the method from {}``A Poor Person's
Z Machine'' \citep{Koo85} to the cosmological workhorse it is today.
Spectroscopic redshifts are simply unattainable for about 95\% of
the objects in the HDF-N; these objects are too faint ($I\gtrsim25$),
beyond the spectroscopic limits of today's telescopes.
\citet{Steidel92} had already demonstrated the powerful {}``dropout
technique'' for identifying high redshift galaxies based on rest
frame Lyman-$\alpha$ absorption. And with the public availability
of extremely high quality multi-band WFPC2 photometry (and subsequent
near-IR observations from the ground), astronomers quickly refined 
the photometric redshift technique (from \citet{Gwyn96}
to \citet[hereafter \citetalias{FLY99}]{FLY99} and \citet{BPZ00}).
Today, photometric redshifts are an essential tool for 
measuring galactic distances when spectroscopic redshifts are unavailable.
In fact, high quality photometric redshifts based on multi-band photometry
may be more robust than spectroscopic redshifts of low confidence
\citep{FernandezSoto01}.

High quality photometry is the key to obtaining robust photometric
redshifts. The UDF images are somewhat of a challenge in that respect,
as the NICMOS images have wider PSF widths than the ACS images. If
not handled properly, the measured NICMOS fluxes will be
understated, by as much as 1 magnitude or more for small, faint objects. 
Our {\tt ColorPro}
software package enables us to obtain consistent aperture-matched
and PSF-corrected photometry across all filters.
The ACS $z\arcmin$-band also sports a PSF halo
which typically loses 0.1 magnitudes or more for faint objects.
When properly accounted for, this extra $z\arcmin$-band flux
may provide a slight boost to measurements of star formation rate density 
at $z\sim 6$ (Paper II).

After obtaining robust $B V i\arcmin z\arcmin J H$ photometry,
we use {\tt BPZ} \citep{Benitez04} to obtain
Bayesian photometric redshifts of the UDF galaxies. Spectral energy
classifications are also obtained (e.g., elliptical, spiral, starburst).
The Bayesian method not only yields more reliable photo-z's than traditional
$\chi^{2}$ methods but also provides a measure of that reliability
for each photo-z. In fact, {\tt BPZ} returns an entire
probability distribution $P(z)$ for each galaxy, which can then be
summarized in terms of a most likely redshift and a confidence level
and confidence interval for that redshift. The new version of {\tt BPZ}
takes the summary of $P(z)$ a step further by providing up to three high 
probability redshifts (the three highest peaks of $P(z)$) along
with confidence levels and intervals for each.
By adding the full redshift distributions $P(z)$, we obtain the
redshift probability histogram for galaxies in the UDF.
A markedly different (and less accurate) histogram emerges 
if one simply bins the single value best fit redshifts.

The main purpose of this paper is to present our method and catalog to the astronomical community.
In \S\ref{sec:Observations} we describe the UDF observations. 
\S\ref{sec:Catalogs} describes our method for obtaining the photometric catalog. 
Our morphological measurements are described in \S\ref{sub:morph}.
\S\ref{sec:BPZ} presents our Bayesian photometric redshifts. 
And finally, we give a summary in \S\ref{sec:Summary}.
Our catalog and software are available at {\tt http://adcam.pha.jhu.edu/\~{}coe/UDF/}.
\notetoeditor{Website is password protected: name=udf, pass=shmoodf}
In Paper II (Coe et al., in prep.) we examine the role of different galaxy types
in the star formation history of the universe, as observed within the UDF.

\section{Observations}
\label{sec:Observations}

The UDF
(RA=$03^{h}32^{m}39\fs0$, Dec=$-27\arcdeg47\arcmin29\farcs1$ (J2000))
was observed by the Wide Field Camera (WFC) of
Hubble's Advanced Camera for Surveys (ACS, \citealt{ACS})
for a total of 400 orbits: 56 orbits
each in the $B$ \& $V$-bands (F435W \& F606W) and 144 orbits each
in $i\arcmin$ \& $z\arcmin$ (F775W \& F850LP) (P.I. Steven Beckwith%
\footnote{Director's Discretionary Cycle 12 Programs 9978 \& 10086: 9/24/03
- 1/16/04%
}). These images cover $12.80\textrm{ arcmin}^{2}$, over twice the
area of each of the previous Hubble Deep Fields 
(HDF-N \citealt{HDF-N}; HDF-S \citealt{HDF-S}).
We prune our catalog to the central $11.97\textrm{ arcmin}^{2}$
of the ACS images, which has at least half the average depth of the
whole image. The $B$, $V$, \& $i\arcmin$ UDF images are also $\sim1.0$,
0.9, \& 1.4 mags deeper than the respective HDF images. A filter similar
to $z\arcmin$ was not available to image the HDF, and its presence
allows us to probe the UDF for $i\arcmin$-band dropout galaxies at
$5.7\la z\la7$. \emph{}

\begin{figure}
\plotone{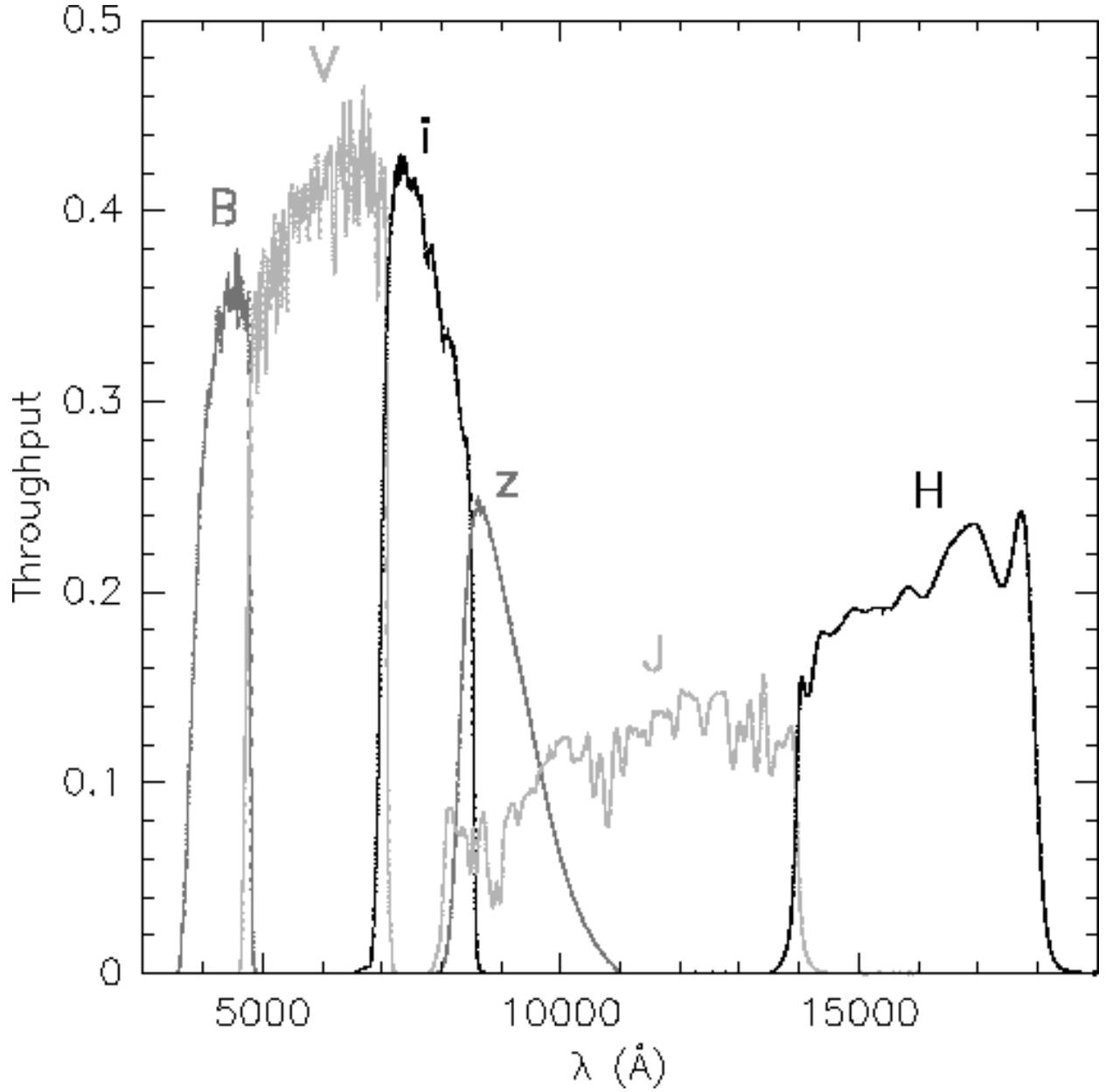}
\caption{\label{cap:transmission}Transmission curves for the ACS $BVi\arcmin z\arcmin$
\& NIC3 $JH$ filters. The $i\arcmin$ \& $z\arcmin$ filters are
identical to those used on SDSS. The $J$ filter extends much further
blueward than traditional ground-based $J$ filters.}
\end{figure}

For still higher redshift study, NICMOS's camera C3 {}``NIC3'' was
trained on this same patch of sky for an additional 144 orbits (P.I.
Rodger Thompson%
\footnote{Cycle 12 Treasury Program 9803: 8/31/03 - 11/27/03}). 
While only covering $5.76\textrm{ arcmin}^{2}$, or about half
the ACS FOV, the NIC3 observations, split equally between the $J$
\& $H$-bands (F110W \& F160W), have the potential to reveal $z\arcmin$-dropouts
with redshifts $>7$. Transmission curves of the filters are shown
in Fig.~\ref{cap:transmission}. Note that the filter we refer to
as $J$ (or $J_{110}$) is actually much bluer than traditional ground-based
$J$-band filters, fully overlapping the $z\arcmin_{850}$ filter and extending
to $\lambda\sim8000\textrm{\AA}$.

See Tables \ref{cap:cameras} and \ref{cap:filters} for a summary
of the observations.
The extinction corrections in Table \ref{cap:filters}
are derived from the dust maps of \citet{dustmaps},
for which we obtained $E(B-V)=0.0079$.

\clearpage

\begin{deluxetable}{cccc}
\tablecaption{\label{cap:cameras}UDF Imaging: Cameras}
\tablewidth{0pt}
\tablehead{
\colhead{Camera/Detector}  & 
\colhead{Resolution}&
\colhead{Drizzled} &
\colhead{Area}
}
\startdata
ACS/WFC&
$.05\arcsec$/pix&
$.03\arcsec$/pix&
12.80 sq $\arcmin$\\
NICMOS/C3 ({}``NIC3'')&
$.20\arcsec$/pix&
$.09\arcsec$/pix&
5.76 sq $\arcmin$
\enddata
\end{deluxetable}
\begin{deluxetable}{ccccccc}
\tablecaption{\label{cap:filters}UDF Imaging: Filters}
\tablewidth{0pt} 
\tablehead{
\colhead{Camera}&
\colhead{Filter}&
\colhead{Orbits}&
\colhead{Zeropoint\tablenotemark{a}}&
\colhead{Galactic\tablenotemark{b}}&
\colhead{Offset\tablenotemark{c}}&
\colhead{Depth\tablenotemark{d}}\\
\colhead{}&
\colhead{}&
\colhead{}&
\colhead{(AB)}&
\colhead{Extinction}&
\colhead{}&
\colhead{(AB)}
}
\startdata
ACS&
$B$ (F435W)&
56&
25.673&
0.0326&
\ldots&
28.71\\
ACS&
$V$ (F606W)&
56&
26.486&
0.0232&
\ldots&
29.13\\
ACS&
$i\arcmin$ (F775W)&
144&
25.654&
0.0160&
\ldots&
29.01\\
ACS&
$z\arcmin$ (F850LP)&
144&
24.862&
0.0117&
\ldots&
28.43\\
NIC3&
$J$ (F110W)&
72&
23.4034&
0.0071&
0.30&
28.30\\
NIC3&
$H$ (F160W)&
72&
23.2146&
0.0046&
0.18&
28.22%
\enddata
\tablenotetext{a}{Provided in B04's {\tt wfc\_README.txt} and T04's NICMOS image headers.}
\tablenotetext{b}{Subtracted from the zeropoints.}
\tablenotetext{c}{Empirically derived in \S\ref{sub:NIC3magoffsets}; subtracted from the NIC3 magnitudes.}
\tablenotetext{d}{10-$\sigma$ limiting AB magnitude within a 0.2sq$\arcsec$
(0.5$\arcsec$ diameter) aperture, after subtracting extinction and offsets.}
\end{deluxetable}


\clearpage

The ACS images were reduced at STScI by \citet[hereafter B04, 
where 2004 refers to the release date]{UDFACS}. 
The original images of $0.05\arcsec$ resolution were combined and drizzled \citet{drizzle}
to an even finer resolution of $0.03\arcsec$/pixel.
Pixel integrity was maintained by setting \texttt{pixfrac = 0}. Meanwhile,
the reduction of the NIC3 images was performed by \citet{UDFNIC3}.
The original $0.20\arcsec$/pixel images have been drizzled
to $0.09\arcsec$/pixel resolution. \texttt{pixfrac} was set to \texttt{0.6},
which (as Thompson et al.\ point out) introduces correlation between 
neighboring pixels, and therefore
artificially reduces the measured noise in the final NIC3 images.
We use the method of \citet{drizzleRMS} to restore the NIC3 noise
maps to their true levels (see \S \ref{sub:dm}).
The reduced images and noise maps are available to the public at
\texttt{http://www.stsci.edu/hst/udf}.

Throughout this paper we use Thompson et al.'s version 1 NIC3 image reductions
and catalog (hereafter T04).  We have compared version 1 to two other reductions.
\citet{UDFNIC3} present version 2 featuring improved masking of bad pixels 
and slightly better alignment to the ACS images.
To keep pace, we visually inspect objects detected in the version 1 NIC3 images 
and remove any obviously spurious sources.
We also correct the slight version 1 alignment offset (\S \ref{sub:photometry}).
Otherwise, there are no magnitude offsets or other significant differences between version 1 and version 2.
Meanwhile, Louis Bergeron has performed an independent reduction of the UDF NIC3 images (priv.\ comm.).
This version yields objects between 0.04 and 0.08 magnitudes brighter in the $J$-band
(based on analyses performed both by us and by Bahram Mobasher, priv.\ comm.).
This issue appears to have been settled by a recent recalibration of the zeropoints
of the Thompson et al.\ UDF images \citep{Thompsoncal}.
In \S\ref{sub:NIC3recal} we discuss this recalibration 
as well as a count-rate dependent non-linearity that affects the calibration of all NICMOS images.

\section{Catalogs}
\label{sec:Catalogs}

Along with the reduced images, B04 and T04 also
released photometric catalogs at \texttt{http://www.stsci.edu/hst/udf}.
The two catalogs were generated independently, one being based on
the ACS images and the other being based on the NIC3 images. Thus,
object detections and aperture definitions in each filter are in general
inconsistent, and accurate ACS-NIC3 colors cannot be obtained from
these catalogs (except perhaps for the brightest objects). 

We have built our work upon the object detections performed by the
two previous teams, in an effort to avoid an unnecessary proliferation
of different catalogs with small differences among themselves. 
For most objects, our isophotal aperture definitions are identical to 
those used in the B04 catalog (given their``segmentation maps''
(\S\ref{sub:detection})).
This allows direct comparison of our results on an object-by-object basis. 
To these objects we have added those detected in the T04 NIC3 segmentation map.
And finally, we perform our own ACS and NIC3 detections, 
adding any {}``new'' objects to complete our segmentation map.

Using a new program we have developed called {\tt SExSeg},
we are able to force all of these object definitions into {\tt SExtractor}
(version 2.2.2; \citealt{SExref}) for photometric analysis (\S\ref{sub:SExSeg}).
The resulting ACS \& NIC3 photometry has been obtained within consistent isophotal
apertures in every filter. 
Isophotal apertures have been shown to produce the most robust colors,
performing slightly better than circular apertures and much better than
{\tt SExtractor}'s {\tt MAG\_AUTO} for faint objects \citep{Benitez04}.

Our NIC3 photometry is also corrected to match the ACS PSF, 
yielding robust ACS-NIC3 colors (\S\ref{sub:photometry}).
All photometry is performed on images in the highest resolution frame
(the NIC3 images are remapped to the ACS frame).
And photometry is performed on undegraded images whenever possible. 
Rather than degrade every image to the worst PSF, we only degrade our detection
image enough to match the PSF of each individual filter.

Based on our {\tt BPZ} fits to objects with known spectroscopic redshifts,
we find disagreement between the ACS and NIC3 calibrations (\S\ref{sub:NIC3magoffsets}).
To correct for this, we apply simple offsets of 
$-0.30\pm0.03$ and $-0.18\pm0.04$ mag to the NIC3 $J$ and $H$-bands, respectively.
The latest recalibration efforts (of the NIC3 images and of NICMOS itself)
appear to support our derived offsets (\S\ref{sub:NIC3recal}).

Our detection and photometric catalogs are presented in Tables
\ref{cap:catdet} \& \ref{cap:catphot}, respectively.
These are also available as a single catalog which also includes the {\tt BPZ} results.
This catalog may be downloaded from \texttt{http://adcam.pha.jhu.edu/\~{}coe/UDF}.
Our {\tt ColorPro} photometric software and {\tt SExSeg} package 
are also available via this website.

Finally, our measurements of galaxy morphology are described in \S\ref{sub:morph},
and our morphological catalog is presented in Table \ref{cap:catmorph}.
This catalog contains only those objects detected in B04's $i\arcmin$-band catalog.

\clearpage

\begin{deluxetable}{lcrrrrrrcrc}
\tabletypesize{\small}
\tablewidth{0pt}
\tablecaption{\label{cap:catdet}Catalog: Detection}
\tablehead{
\colhead{ID\tablenotemark{a}}&
\colhead{altered\tablenotemark{a}}&
\colhead{$\Delta i\arcmin_{ST}$\tablenotemark{b}}&
\colhead{area}&
\multicolumn{2}{c}{RA \& DEC (J2000)}&
\colhead{$x$\tablenotemark{c}}&
\colhead{$y$\tablenotemark{c}}&
\colhead{wfcexp\tablenotemark{d}}&
\colhead{sig\tablenotemark{e}}&
\colhead{$\rm{stel}_{i\arcmin}$\tablenotemark{f}}\\
\colhead{}&
\colhead{}&
\colhead{(mag)}&
\colhead{(pix)}&
\multicolumn{2}{c}{(degrees)}&
\colhead{(pix)}&
\colhead{(pix)}&
\colhead{}&
\colhead{}&
\colhead{}
}
\startdata
1&
 0&
  0.0003&
  5693&
 53.16551208&
-27.82847977&
 4932.80&
  802.88&
2.01&
  551.4&
0.03\\
2*&
 1&
 -0.3040&
   103&
 53.16449738&
-27.82928467&
 5040.27&
  706.25&
1.84&
   13.4&
0.00\\
3*&
 1&
 -0.8914&
    76&
 53.16319275&
-27.82922173&
 5178.82&
  713.79&
2.05&
   10.9&
0.00\\
4&
 0&
 -0.0164&
    77&
 53.16295624&
-27.82913971&
 5203.87&
  723.62&
2.06&
   10.1&
0.17\\
5&
 0&
  0.0010&
   269&
 53.16403580&
-27.82889175&
 5089.33&
  753.53&
1.86&
   55.5&
0.03
\enddata
\tablecomments{Table \ref{cap:catdet} is published in its entirety in the electronic version of
the Astronomical Journal.
A portion is shown here for guidance regarding its form and content.}
\tablenotetext{a}{ID numbers below 41000 correspond to B04 \& T04 detections; 
asterisks ({*}) indicate that object definitions have been altered (\S\ref{sub:detection}).}
\tablenotetext{b}{Rough guide to the degree of alteration: 
difference between our $i\arcmin$-band magnitude and that from the B04 catalog.}
\tablenotetext{c}{Coordinates in the B04 ACS images ($0.03\arcsec / \rm{pix}$).}
\tablenotetext{d}{Exposure time in the ACS detection image $d$ 
normalized to the average depth of the whole image.
For our analyses, we prune $\tt{wfcexp}>0.5$.}
\tablenotetext{e}{Maximum detection significance from our 5 detections.}
\tablenotetext{f}{{\tt SExtractor stellarity} measured in the $i\arcmin$-band image.}
\end{deluxetable}
\begin{deluxetable}{lrrrrrl}
\tablewidth{0pt}
\tablecaption{\label{cap:catphot}Catalog: Photometry}
\tablehead{
\colhead{ID}&
\colhead{$B_{435}$}&
\colhead{$V_{606}$}&
\colhead{$i\arcmin_{775}$}&
\colhead{$z\arcmin_{850}$}&
\colhead{$J_{110}$}&
\colhead{$H_{160}$}
}
\startdata
1&
$24.10\pm0.01$&
$23.32\pm0.00$&
$22.80\pm0.00$&
$22.68\pm0.00$&
$-99.00\pm0.00$&
$-99.00\pm0.00$\\
2*&
$29.70\pm0.20$&
$29.26\pm0.10$&
$29.43\pm0.12$&
$29.12\pm0.17$&
$-99.00\pm0.00$&
$-99.00\pm0.00$\\
3*&
$29.60\pm0.17$&
$29.79\pm0.14$&
$30.14\pm0.20$&
$29.73\pm0.25$&
$-99.00\pm0.00$&
$-99.00\pm0.00$\\
4&
$99.00\pm31.55$&
$29.56\pm0.12$&
$29.33\pm0.10$&
$29.34\pm0.18$&
$-99.00\pm0.00$&
$-99.00\pm0.00$\\
5&
$28.04\pm0.07$&
$27.35\pm0.03$&
$26.93\pm0.02$&
$26.96\pm0.04$&
$-99.00\pm0.00$&
$-99.00\pm0.00$
\enddata
\tablecomments{%
Table \ref{cap:catphot} is published in its entirety in the electronic version of
the Astronomical Journal.
A portion is shown here for guidance regarding its form and content.
Magnitudes are ``total'' AB magnitudes with isophotal colors: 
NIC3 magnitudes are corrected to the PSF of the ACS images (\S\ref{sub:photometry}). 
We have also applied offsets of ($J$: $-0.30\pm0.03$, $H$: $-0.18\pm0.04$) 
to the NIC3 magnitudes (\S\ref{sub:NIC3magoffsets}).
Non-detections (listed, for example, as $99.00\pm31.55$) 
quote the 1-$\sigma$ detection limit of the aperture used on the given object.
A value of $-99.00$ is entered for unobserved magnitudes: 
outside the NIC3 FOV or containing saturated or other bad pixels.
}
\end{deluxetable}
\begin{deluxetable}{crrrrrrcrc}
\tabletypesize{\scriptsize}
\tablecaption{\label{cap:catmorph}Catalog: Morphology in the UDF $i\arcmin$-band Image}
\tablewidth{0pt} 
\tablehead{
\colhead{ID}&
\colhead{$\chi^{2}/\nu$}&
\colhead{$i\arcmin_{775}$}&
\colhead{$R_{e}$}&
\colhead{$a/b$}&
\colhead{$\theta$}&
\colhead{$n$}&
\colhead{dist}&
\colhead{Asym.}&
\colhead{Number}\\
\colhead{}&
\colhead{}&
\colhead{(mag)}&
\colhead{(pixels)}&
\colhead{}&
\colhead{(degrees)}&
\colhead{(S\'ersic)}&
\colhead{(pixels)}&
\colhead{Index}&
\colhead{Companions}
}
\startdata
1&
 1.835&
 $22.72\pm0.01$&
 $41.43\pm0.18$&
 $0.13\pm0.00$&
 $94.98\pm0.03$&
 $1.28\pm0.01$&
 0.07&
 0.120&
 0\\
 2&
 1.098&
 $29.34\pm0.10$&
 $2.28\pm0.50$&
 $0.71\pm0.21$&
 $39.13\pm32.81$&
 $0.5\pm0.75$&
 0.11&
 0.109&
 0\\
 3&
 1.143&
 $30.22\pm0.37$&
 $1.89\pm1.51$&
 $1.99\pm1.94$&
 $28.53\pm37.99$&
 $0.80\pm1.98$&
 1.07&
 0.081&
 0\\
 4&
 1.198&
 $29.37\pm0.33$&
 $0.53\pm0.73$&
 $0.12\pm1.13$&
 $2.43\pm70.24$&
 $3.98\pm11.78$&
 1.10&
 0.234&
 0\\
 5&
 1.220&
 $26.88\pm0.02$&
 $3.17\pm0.07$&
 $0.55\pm0.02$&
 $7.64\pm2.11$&
 $0.67\pm0.07$&
 0.30&
 0.129&
 0
\enddata

\tablecomments{%
Table \ref{cap:catmorph} is published in its entirety in the electronic version of
the Astronomical Journal.
A portion is shown here for guidance regarding its form and content.
Only galaxies in the B04 catalog are analyzed.
ID numbers correspond to that catalog.
The magnitude $i\arcmin_{775}$,
effective radius $R_{e}$, semiaxis ratio $a/b$, position angle $\theta$,
S\'ersic index $n$, and badness of fit $\chi^{2}/\nu$ are
all derived from \galfit. 
The distance between \galfit's best fit centroid and that from B04 is
given here as ``dist''.
This distance is restricted to fewer than 2 pixels; 
$\rm{dist}>1000$ indicates a misfit.
The asymmetry index and number of companions
are measured as described in \S\ref{sub:morph}.
Additional columns in the electronic version are RA \& Dec 
(based on the B04 catalog)
and \galfit's best fit centroid ($x, y$).
}

\end{deluxetable}

\clearpage

\subsection{Synthesized $BVi\arcmin z\arcmin JH$ Detection}
\label{sub:detection}

\begin{figure*}
\plotone{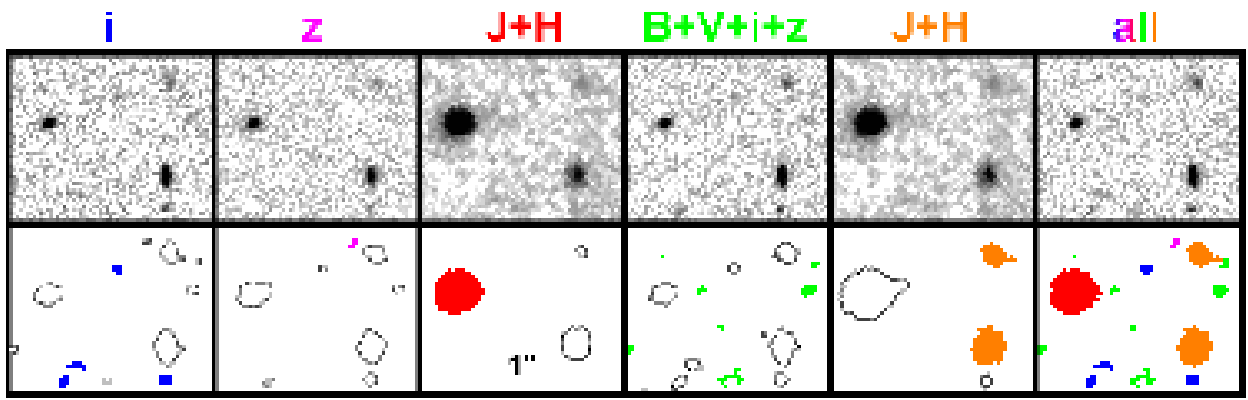}
\caption{\label{cap:aperturesall}Comprehensive detection of faint objects
demonstrated in a small region of the UDF. We begin with the $i\arcmin$-band
image (top left) and B04's corresponding segmentation
map (bottom left) which defines their detections in that image. We then
add {}``new'' segments from four other detections. (The first
three detections were performed by the B04 and T04 teams, 
and the final two (ACS $B$+$V$+$i\arcmin$+$z\arcmin$
and NIC3 $J$+$H$) are our own.) Some of these {}``new'' objects
are completely new, while others are simply re-definitions of objects
previously detected to allow for larger apertures (see \S\ref{sub:detection}).
The colored (filled) segments are the new segments in each detection that ``survive''
to the final comprehensive segmentation map (bottom right). This final
segmentation map defines the photometric apertures that will be applied
to all images.}
\end{figure*}

Our catalog combines the results of five independent detections: two
performed by B04 on the ACS image ($i\arcmin$, $z\arcmin$)%
\footnote{We neglect B04's {}``supplemental'' $i\arcmin$-band
detection as neither the {\tt SExtractor} parameters nor a segmentation
map was readily available for this detection. However, we do serendipitously
{}``re-discover'' 5 of those 100 objects with our other detections.
We reassign B04's IDs (in the 20000 range) to these objects.
B04's 95 other ``supplemental'' objects are not found in our catalog; 
they remain blended with other segments.%
}, the T04 NIC3 detection ($J$+$H$) and two performed
by us ($B$+$V$+$i\arcmin$+$z\arcmin$, $J$+$H$) (Table \ref{cap:det}
and Fig.~\ref{cap:aperturesall}). Segmentation maps for the B04 and T04 
detections were obtained from {\tt http://www.stsci.edu/hst/udf}.
Using their object definitions allows us to compare our photometry,
photometric redshifts, etc. on an object-by-object basis,
knowing that we have used identical apertures.%
\footnote{Using {\tt SExtractor} alone, we were able to emulate B04's
main $i\arcmin$-band catalog fairly well, but not exactly.  Any
attempt to reproduce another's catalog quickly becomes a lesson in
{\tt SExtractor}'s sensitivity to input parameters.  Beckwith et al.\ plan
to publish their full set of input parameters in an upcoming paper.
But we skirt the issue by applying their segmentation maps directly.%
} Future groups may also wish to use these object definitions to facilitate
comparison.

Our $B$+$V$+$i\arcmin$+$z\arcmin$ ACS detection image {}``$d$''
was created by dividing each image by the RMS of a {}``blank'' region
and then adding the four images. This allows the deepest possible
detection in the ACS images \emph{}for objects detected in all the
filters. Similarly, we create a NIC3 $J$+$H$ detection image (like
the one used by T04). We run {\tt SExtractor}
on these two images using the same parameters used by the UDF teams
(to the best of our knowledge%
\footnote{For {\tt SExtractor} detection of an object, the UDF teams require 9 contiguous pixels
0.61-$\sigma$ above the background. The deblending parameters are
{\tt DEBLEND\_NTHRESH} = 32 and {\tt DEBLEND\_MINCONT} = 0.03.
And (at least for the NIC3 images), no global background is subtracted,
but a local background is subtracted from each object, using an annulus
of width {\tt BACKPHOTO\_THICK} = 24.%
}, including the use of the ACS and NIC3 detection weight maps) producing
our final two segmentation maps. Our NIC3 detection is slightly more
aggressive than that performed by T04, yielding extra detections
and larger isophotal apertures.

For each detection, {\tt SExtractor} produces (upon request)
a \emph{segmentation map}.
A segmentation map defines the pixels belonging to each object. It
is an integer FITS image on the same scale as the detection image.
Each pixel contains the ID number of the object it belongs to. If
a pixel doesn't belong to an object, then it is set to zero. The segmentation
map thus defines the location and extent of objects in the detection
image (see Fig.~\ref{cap:aperturesall}).

\clearpage

\begin{deluxetable}{ccrrrr}
\tablecaption{\label{cap:det}Comprehensive Object Detection}
\tablewidth{0pt} 
\tablehead{
\multicolumn{2}{c}{Detection}&
\colhead{Starting}&
\colhead{Objects}&
\colhead{Objects in}&
\colhead{$\geq$10-$\sigma$ in}\\
\colhead{Filter}&
\colhead{Author}&
\colhead{ID}&
\colhead{Detected}&
\colhead{Final Catalog\tablenotemark{d}}&
\colhead{Final Catalog}
}
\startdata
$i\arcmin$&
B04&
1\tablenotemark{a}&
10045&
9989&
6968\\
$z\arcmin$&
B04&
30001\tablenotemark{a}&
7016&
451&
42\\
$J$+$H$&
T04&
40001\tablenotemark{b}&
926&
6&
6\\
$J$+$H$& this paper &
41001\tablenotemark{c}&
1414&
71&
28\\
$B$+$V$+$i\arcmin$+$z\arcmin$& this paper &
50001\tablenotemark{c}&
17692&
8184&
993\\\tableline
TOTAL&
\nodata&
\nodata&
37093&
18706\tablenotemark{e}&
8042\tablenotemark{e}
\enddata

\tablenotetext{a}{ID numbers below 40000 correspond to the B04 catalog.
Segments that have been altered are flagged in our catalog and their
ID numbers marked with an asterisk ({*}) in this paper. However, most
of the B04 objects (6,955 of their $i\arcmin$-band detections)
do retain their original definitions (segments).}

\tablenotetext{b}{We have added 40000 to the T04 ID numbers.}

\tablenotetext{c}{The order of our final two detections is swapped in the catalog
(cf. Fig.~\ref{cap:aperturesall}).}

\tablenotetext{d}{Number of segments that survive more or less intact to our final catalog.}

\tablenotetext{e}{%
The astute reader will have noticed that there are 5 extra objects
in the total numbers. These correspond to objects in B04's
supplemental $i\arcmin$-band catalog that were serendipitously {}``discovered''
and defined by our other detections. These objects retain their ID
numbers (in the 20000 range) from B04's catalog.}
\end{deluxetable}

\clearpage

After remapping the NIC3 segmentation maps to the ACS frame,
the five segmentation maps are combined using an automated procedure. 
ST's {}``main'' $i\arcmin$-band segmentation map 
serves as the starting point, and the other segmentation maps
are compared to it: new segments are added and some old segments are
enlarged (Fig.~\ref{cap:aperturesall}). To be more precise, a given
segment is added if at least some fraction (we used $\onethird$)
of its pixels are {}``new'' (don't already belong to an object).
So not only are entirely new segments added, but we also add some
segments that overlap with existing segments. {}``Disputed'' pixels
are always reassigned to the new segment. We are able to add any segment that
overlaps just slightly with an existing segment. We also add any segment
that is over 50\% larger than its predecessor. The old object is discarded
whenever $\twothirds$ of its pixels have been consumed by the new
object.

\begin{figure}
\epsscale{0.7}
\plotone{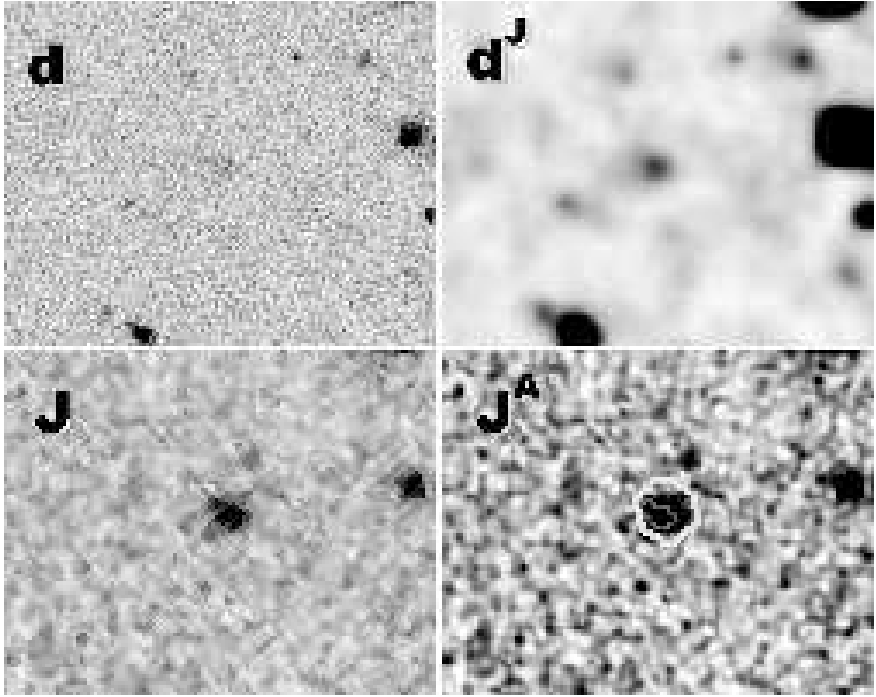}
\epsscale{1.0}
\caption{\label{cap:apertures}
Four image stamps of the same region, centered
on object \#1820{*}. This object was a faint detection in $d$, but
it is much brighter in the NIC3 images. Top-left: the ACS detection
image ($B$+$V$+$i\arcmin$+$z\arcmin$). Top-right: the same image
degraded to match the PSF of the NIC3 $J$-band image (bottom-left).
Bottom-right: the $J$-band image re-mapped to the ACS frame and pixel
scale; isophotal apertures are overlaid: the pink (inner) aperture
was defined in $d$ while the yellow (outer) aperture was defined in
the NIC3 detection image $J+H$ (and then re-mapped to the ACS frame).
This object is significantly brighter in $J+H$ than in $d$. Thus
its $J+H$ isophotal aperture is significantly larger than its isophotal
aperture in $d$. The $d$ aperture is much smaller than the size
of the object, requiring an unnecessarily large PSF correction. Our
automated procedure replaces it with the larger $J+H$ aperture,
taking advantage of the full signal for a more secure measurement of $d-J$. 
The asterisk ({*}) after the ID number indicates that the B04 segment 
for \#1820 has been altered, or replaced as in this case
(\S\ref{sub:detection}).
(Despite the large color decrement ($z\arcmin-J=1.55$),
object \#1820{*} is probably not at high redshift. Its photometry
is well fit by the SED of an elliptical galaxy at $z=1.98\pm0.35$.) 
It should be emphasized that this figure illustrates a rare occurrence.
For most objects, the isophotal aperture is larger in $d$ than in
$J$ because the ACS images are deeper than the NIC3 images.}
\end{figure}

Replacing apertures with larger versions aids in obtaining robust
photometry of dropout galaxies. If an object detected in the $i\arcmin$-band
image is brighter in $J$+$H$ and has a ($>50$\%) larger isophotal
area in that image, then its larger $J$+$H$ segment will replace
the {}``original'' smaller $i\arcmin$ segment. The larger segment
takes advantage of the full $J$+$H$ signal. (Capturing the full
signal is one of the reasons isophotal apertures outperform circular
apertures, as mentioned in the introduction to this section \S\ref{sec:Catalogs}.
The smaller segment would not do the dropout galaxy justice, capturing
only a fraction of its light in $J$ \& $H$ and requiring a larger
(and more uncertain) PSF correction (see Fig.~\ref{cap:apertures}).)
Perhaps an even better strategy would be to enlarge apertures \emph{every}\ time, 
regardless of how much larger the {}``new'' segment is. 
Thus, a ``maximal isophotal aperture'' would be used for every object.
We may explore this strategy in future work, but one of our goals for
this paper was to maintain the integrity of objects defined in the
catalogs released by STScI.%
\footnote{{\tt SExSeg} also gives us the ability to {}``correct''
{\tt SExtractor}'s segmentation. We can actually redraw
segments (to deblend objects, eliminate star spikes, etc.) and force
{\tt SExtractor} to analyze objects in the new corrected
segments. We did not take advantage of this ability in this paper.%
}

The only drawback to enlarging objects in this way
is that deblended objects are occasionally recombined.
For example, if a $J$+$H$ aperture is $\geq \onethird$ new
it will be added to the segmentation map,
regardless of the current segmentation in its footprint.
Usually just one object (if any) will be supplanted.
But occasionally multiple segments will be consumed
(and thus united) by the new segment.
(In the latest version of our software we do provide the option to forgo aperture enlargements
in the event that multiple objects would be re-blended into one.)
In the case of our catalog,
56 B04 $i\arcmin$-band detections and 1 $z\arcmin$-band detection 
are thus consumed by neighboring objects.
Of course perfect deblending was never the goal of this paper.
Instead we are satisfied to base our catalog on the
B04 and T04 detections,
maintaining the majority of those definitions,
while enlarging apertures and adding objects where deemed appropriate.

The final segmentation map is comprehensive, being formed by segments
from the five independent detections (see Table \ref{cap:det}).
It defines the (isophotal) apertures that will be used for our photometric
analysis of the ACS and NIC3 images, which we describe in the next
subsection (\S\ref{sub:SExSeg}). ID numbers in the segmentation map correspond to those
from the B04 catalog, except in the cases of {}``new''
objects (undetected by B04). These ID numbers are carried
through to our catalog. Each B04 object is also flagged
as to whether any alterations were made to the segment (whether any
pixels were lost or the segment was replaced by a larger version).
This flag takes the form of an asterisk ({*}) appended to ID numbers 
in the text of this paper.

\subsubsection{New Galaxies}

We pause from describing our technique to consider what we have gained from our comprehensive object detection.  Our automated procedure began with B04's $i\arcmin$-band segmentation map and added new objects from each of four other detections (\S\ref{sub:detection}, Table \ref{cap:det}).  Here we describe these new objects and the value they add to our investigation.

B04's $z\arcmin$-band segmentation map adds 42 new objects detected at the 10-$\sigma$ level (Table \ref{cap:det}, last column).  Upon visual inspection, most of these do appear to be legitimate $i\arcmin$-band dropouts.  And {\tt BPZ} verifies that they probably lie beyond $z\gtrsim 6$ (\S\ref{sub:high-z}).  4 of these objects appear to be spurious, while another 2 appear to be legitimate new objects now ``de-blended'' from larger B04 $i\arcmin$-band segments.

T04's NIC3 $J$+$H$ segmentation map yields 6 ``new objects'' at 10-$\sigma$, including \#40819, the famous massive old $z\sim 6.5$ candidate galaxy, also known as HUDF-JD2 \citep{Mobasher05}.  It is for objects such as this that incorporation of T04's segmentation map is essential.  Another potentially interesting object \#40925 fills in a very red patch amongst at least three other small galaxies.  But \#40925 and its neighbors all appear to be at a redshift (or redshifts) of 2 or so.  The other 4 ``new objects'' in this detection appear to be spurious: either spurious detections (from the glare of neighbors) or spurious re-segmentations.  By ``spurious re-segmentation'', we mean that the object was previously detected, and now it is being re-detected slightly offset from the original.  The new detection covers enough ``new'' pixels to be added to our final segmentation map, but leaves enough ($> \onethird$) of the ``old'' segment uncovered that it survives as well (although missing a good chunk).  These spurious re-segmentations could have perhaps been avoided with a tweaking of the \onethird\ parameter, or with a more sophisticated algorithm for combining segmentation maps.  This proves to be a tricky business, akin to {\tt SExtractor}'s object de-blending.  Our algorithm has room for improvement. But for now we allow for a handful of objects with poor segmentation out of a catalog of thousands.

In our own $J$+$H$ detection, we add 28 ``objects'' at 10-$\sigma$.
Three of these (also featured in Table \ref{cap:highz}) don't correspond to optical detections, and if their NIC3 detections are confirmed could turn out to have very high redshifts indeed: \#41107 ($z_{b}=8.57_{-.83}^{+1.08}$), \#41092 ($z_{b}=7.73_{-.60}^{+1.31}$), and \#41066 ($z_{b}=7.13_{-.54}^{+1.13}$; with a faint $z\arcmin$-band detection).
The rest of our 28 detections appear to be spurious: a few new false detections, 
but mostly ``spurious re-segmentations'', as discussed above.
This occurs when the NIC3 segment is slightly offset from the ACS segment.
The most likely explanation for this is that part of the galaxy appears brighter in the near-IR than the rest, which could be interesting in its own right.
More exciting possibilities are that these are supernovae or other activity (between the time the ACS and NIC3 images were taken), or even chance alignments of galaxies slightly offset from more distant ones at very high redshift.
But we will not be pursuing those possibilities here.

Finally, we discuss our $d$=$B$+$V$+$i\arcmin$+$z\arcmin$ detection, which is supposed to allow the deepest possible detection in the ACS images for objects detected in all the filters.  Most of the 993 10-$\sigma$ objects in this detection are simply outside the field of view studied by B04.\footnote{Those authors trimmed the edges of the ACS field to avoid regions of low signal to noise.  Our catalog contains objects detected all the way out to the edge of the image.  We only trim the edges as part of our analysis, and then we trim less area than B04.  After trimming this detection, we're still left with 708 ``new objects''.}  But the interesting ones are the 127 objects that we find inside B04's search area.  Some are spurious re-segmentations, and there are a few wispy detections that are almost undoubtedly false.  But many of these objects are faint blue galaxies, with $i\arcmin$ and $z\arcmin$-band fluxes too faint to be detected in these bands.  Given the large population of faint blue galaxies visible in the UDF (see \S\ref{sub:faintblue}), it is important to include a detection such as this based (at least in part) on the bluer bands $B$ and $V$.

\subsection{{\tt SExSeg}}
\label{sub:SExSeg}

Armed with our single comprehensive segmentation map (the definition
of objects and their extents), we need the ability to obtain multicolor
photometry given these object definitions. To this end, we have developed
a new program called {\tt SExSeg} (part of the {\tt ColorPro}
package; Coe et al., in prep.), which forces {\tt SExtractor}
to run using a pre-defined segmentation map. We have chosen not to
modify the {\tt SExtractor} code itself, which although
perhaps more straightforward, would involve changing a software which
has become a de facto standard and is well understood by many astronomers.
Instead,  {\tt SExSeg} alters the input detection image
based on the input segmentation map. When {\tt SExtractor}
is run on this new detection image it is forced to acknowledge the
desired segments. {\tt SExtractor} is then run in double-image
mode with this new detection image and the desired photometric analysis
image.

The input segmentation map is altered slightly by inserting gaps between
neighboring objects. This ensures {\tt SExtractor}'s accurate
and stable reproduction of the segmentation. Gaps are always created
by discarding pixels from the larger of the two neighbors. And the
number of pixels lost (if any) by each object is recorded in the catalog.
But it must be emphasized that these slight segment alterations do
not adversely affect our color measurements (Coe et al., in prep.),
as we discuss below.

\begin{figure*}
\begin{center}
\epsscale{0.9}
\plottwo{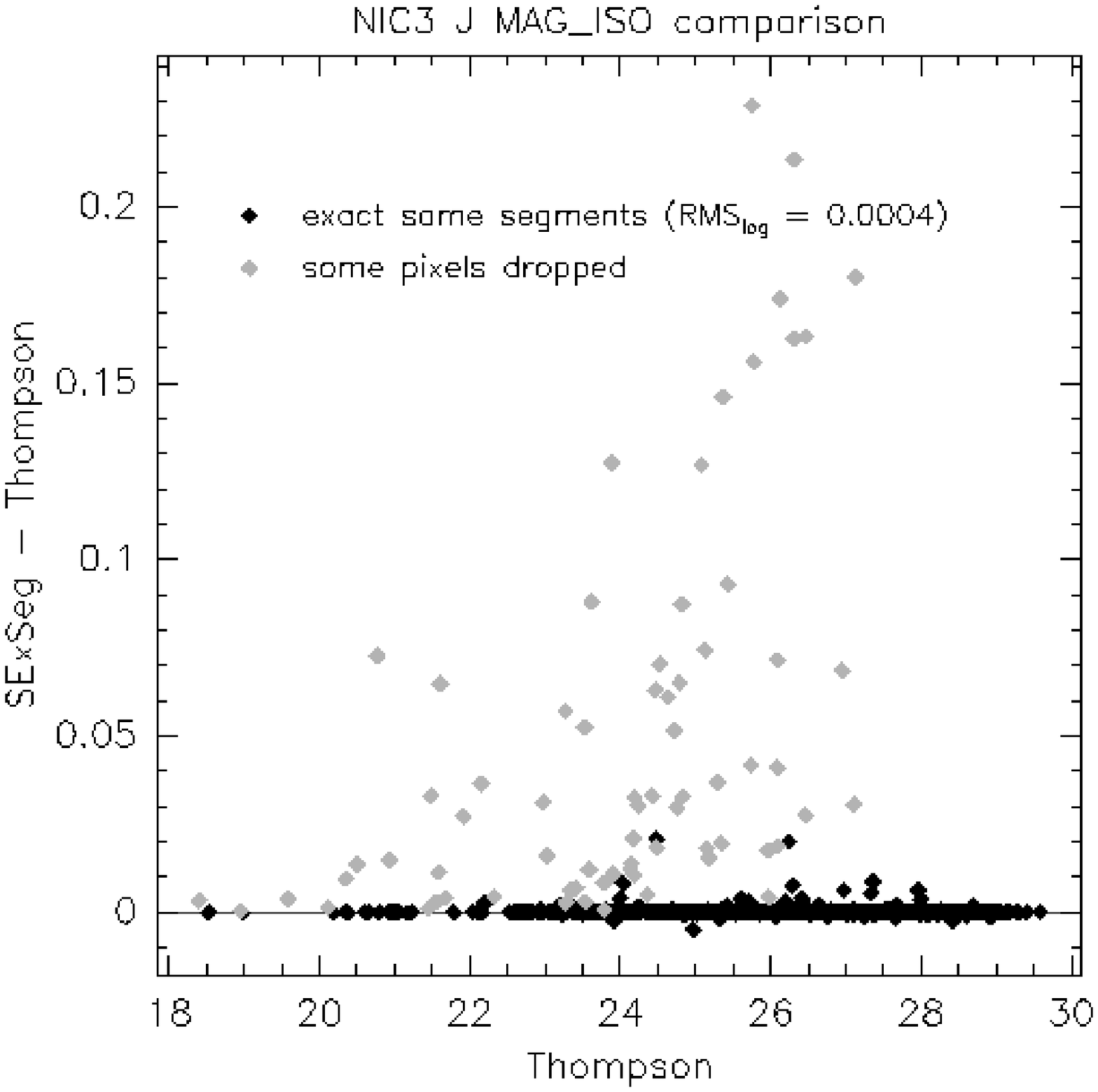}{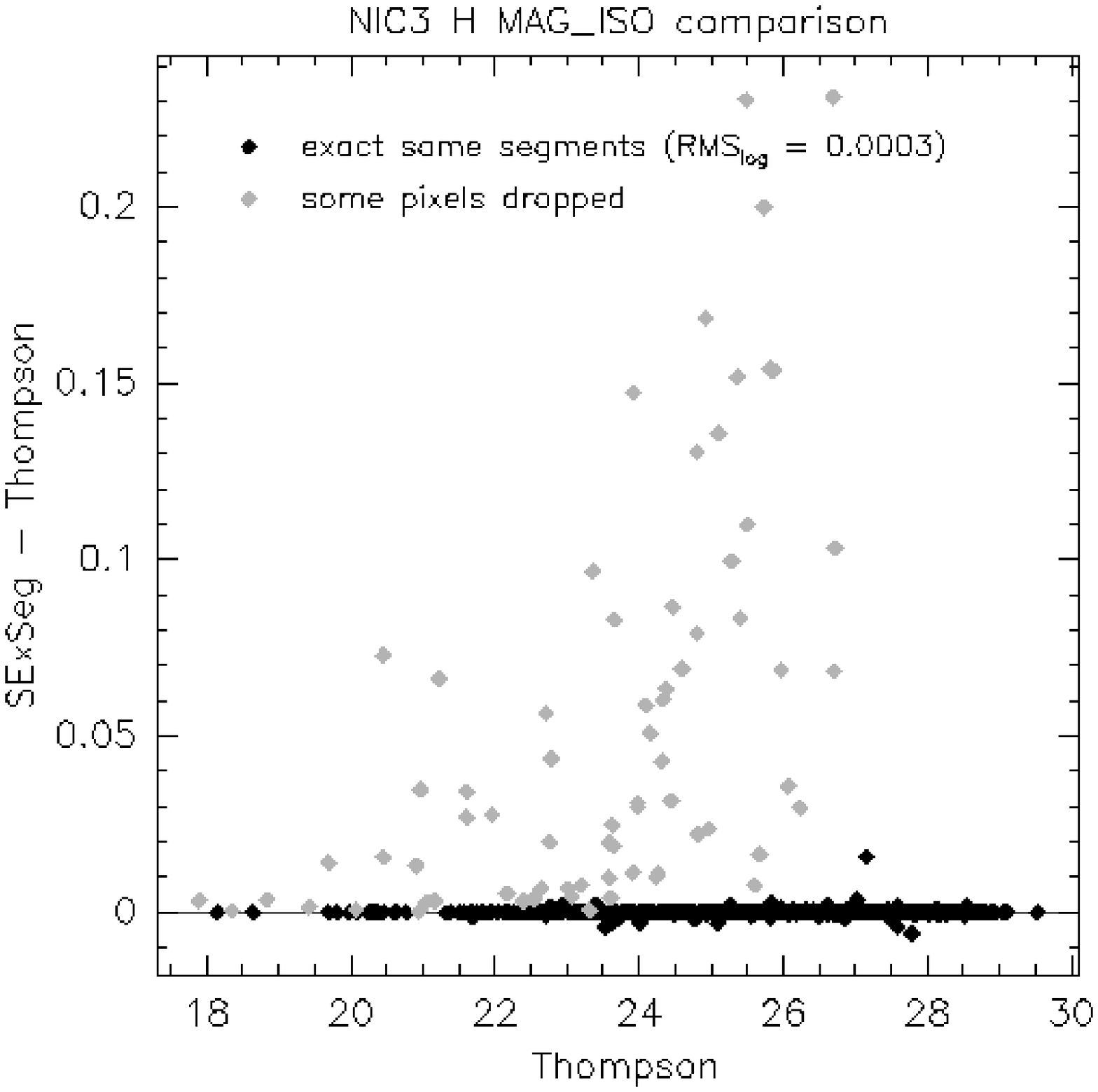}
\plottwo{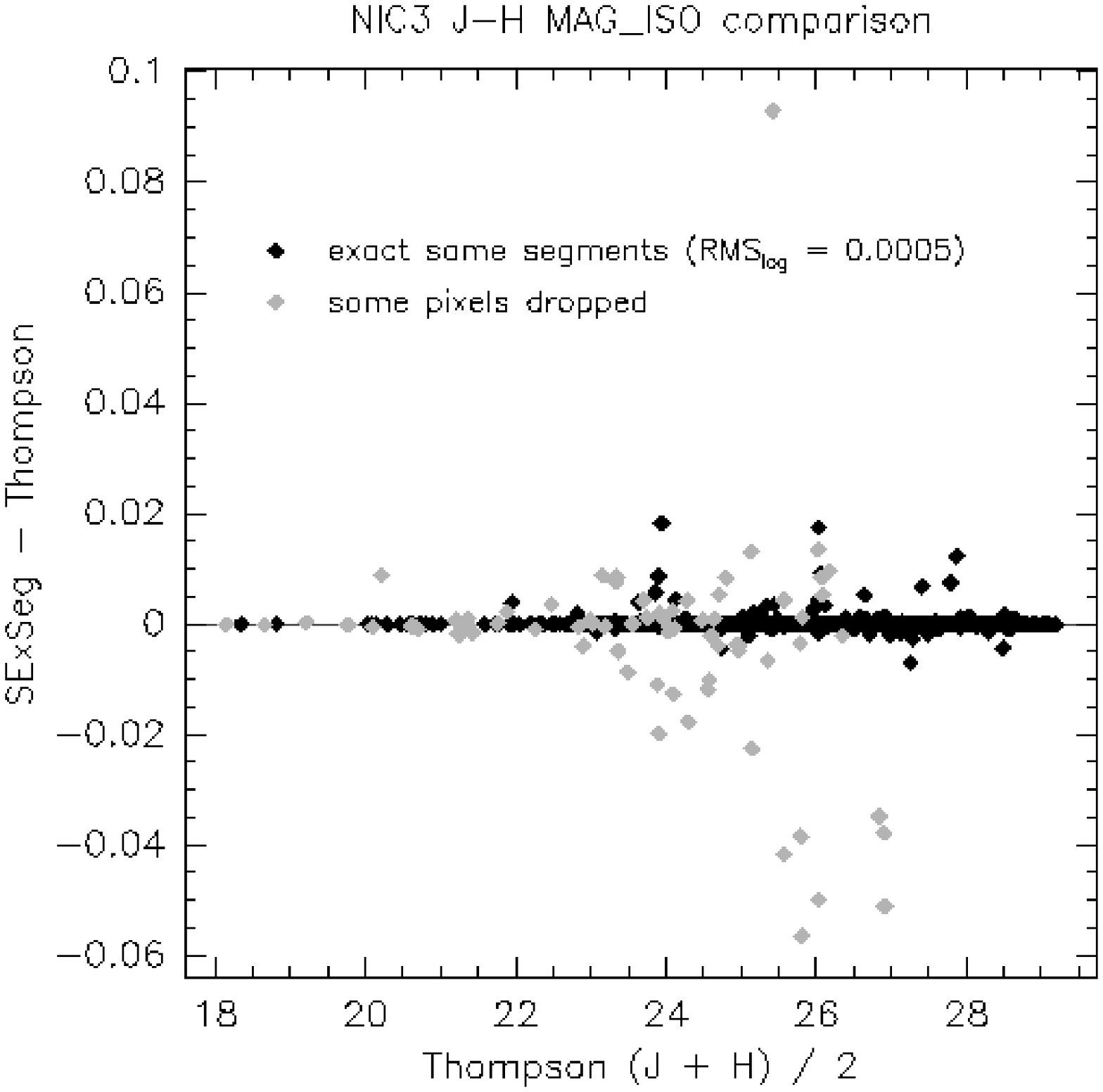}{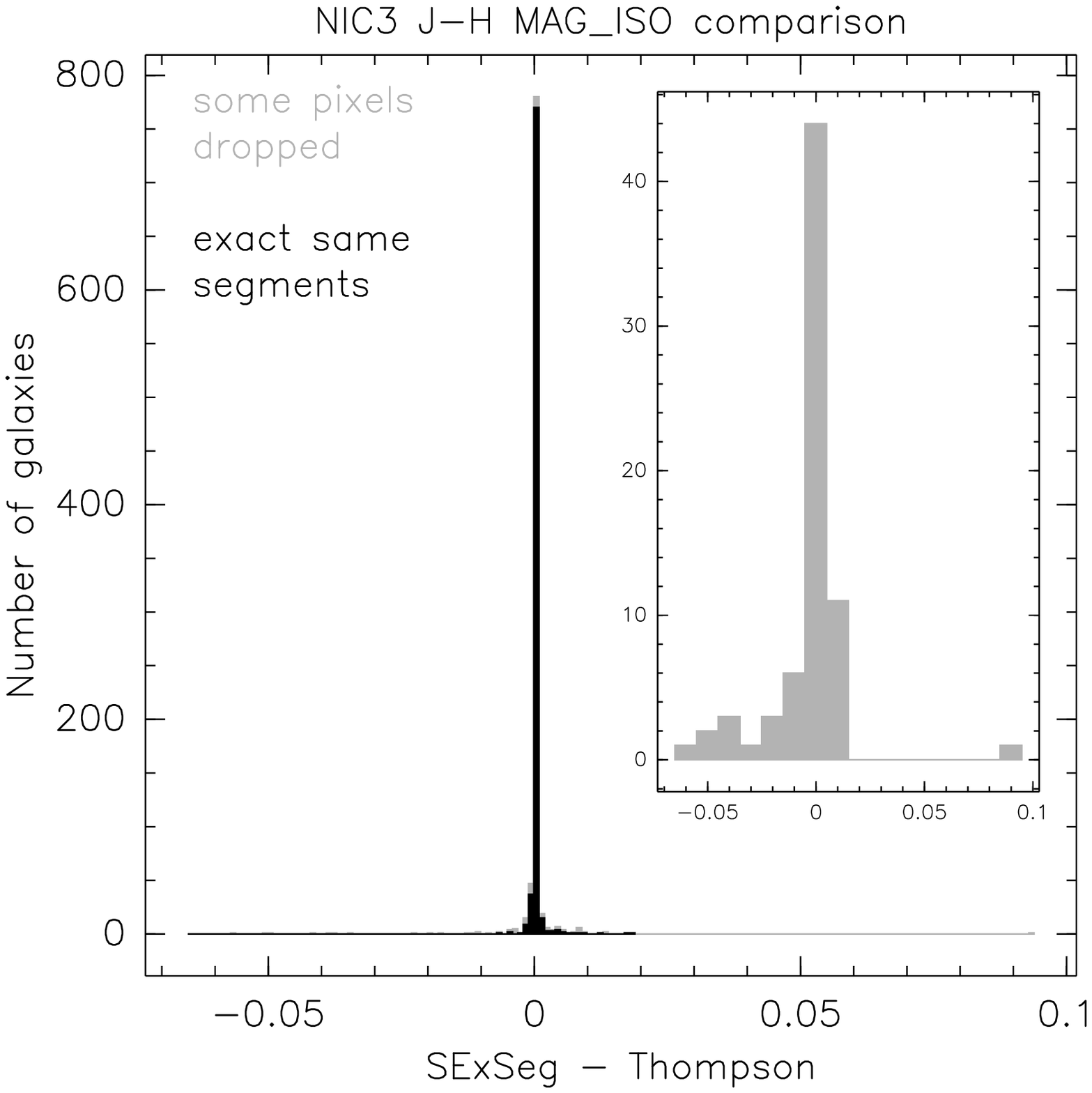}
\epsscale{1.0}
\end{center}
\caption{\label{cap:SExSeg}NIC3 {\tt SExSeg} isophotal magnitudes
and colors are compared to those derived directly from {\tt SExtractor}
by T04. {\tt SExSeg} inserts gaps to
separate neighboring objects; these altered segments are plotted in
green. The lost pixels normally result in lost flux (higher magnitudes).
However the main purpose of {\tt SExSeg} is to measure
accurate colors, and when apertures are slightly altered, they are
still used consistently across filters. The resulting colors may be
slightly different, but they are no less accurate given the effects
of photometric noise (as verified by simulations, Coe et al., in prep.)
Meanwhile, unaltered segments (black) usually yield identical magnitudes
and colors (with occasional slight variations: logarithmic RMS values
are on the order of $10^{-4}$). The histogram on the bottom right
emphasizes that most objects have {\tt SExSeg} $J-H$
colors identical to those measured by T04.}
\end{figure*}

To demonstrate {\tt SExSeg}'s accuracy, we ran {\tt SExSeg}
on the original NIC3 images using the segmentation map provided by
T04 ({\tt http://www.stsci.edu/hst/udf}). We compare
the resulting magnitudes to those given in the T04 catalog.
For the majority of the objects, our magnitudes match T04's magnitudes
exactly (Fig.~\ref{cap:SExSeg}). The only significant variations
in magnitude arise from objects whose segments have been altered (where
gaps were inserted between neighboring objects). These objects do
get flagged in the catalog, but their colors should not be considered
wrong or {}``off''. The inserted gaps make our isophotal apertures
slightly smaller than those used by T04 for these objects.
But by consistently applying our apertures to all images (here $J$
\& $H$), we ensure accurate color measurements. All of our $J-H$
color measurements match T04's measurements to within
0.1 mags (most match to within 0.01 mags). But where our color measurements
disagree, we cannot say which method obtained the more accurate measurement.
In other words, Thompson et al.\ can't say our method is {}``off''
any more than we can say their method is {}``off''. Simulations
verify that {\tt SExSeg} colors are just as accurate {\tt SExtractor}
colors (given the limits of photometric noise) when the segment has
been altered (Coe et al., in prep.). Of course when the segment has
not been altered (as is the case for the majority of objects in most
images) the {\tt SExSeg} colors are (almost always) identical
to the {\tt SExtractor} colors.

\subsection{Robust Aperture-Matched, PSF-corrected $BVi\arcmin z\arcmin JH$
Photometry}
\label{sub:photometry}

Aperture-matched PSF-corrected photometry is essential to obtaining robust colors
across images with varied PSF (see e.g. \citealt{Benitez99}, \citealt{VanzellaHDF-S}). 
Galaxy images blur as the PSF is degraded. 
The photometry of bright galaxies is not significantly affected,
as we use large ``maximal isophotal apertures'' (\S\ref{sub:detection}).
But for faint objects (with small isophotal apertures),
the scant flux gets spread too thin, 
much of it getting swept under the ``rug'' that is the noise floor.

\begin{figure}
\plotone{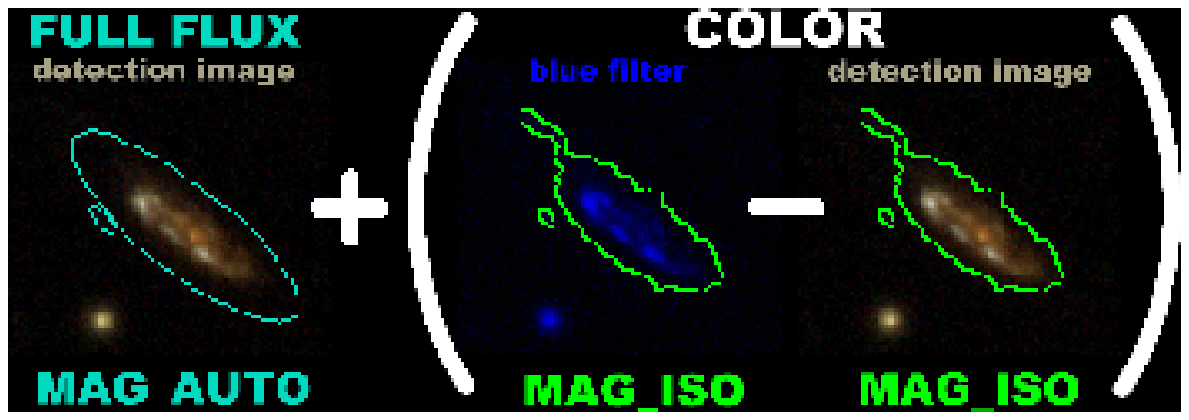}\\
\plotone{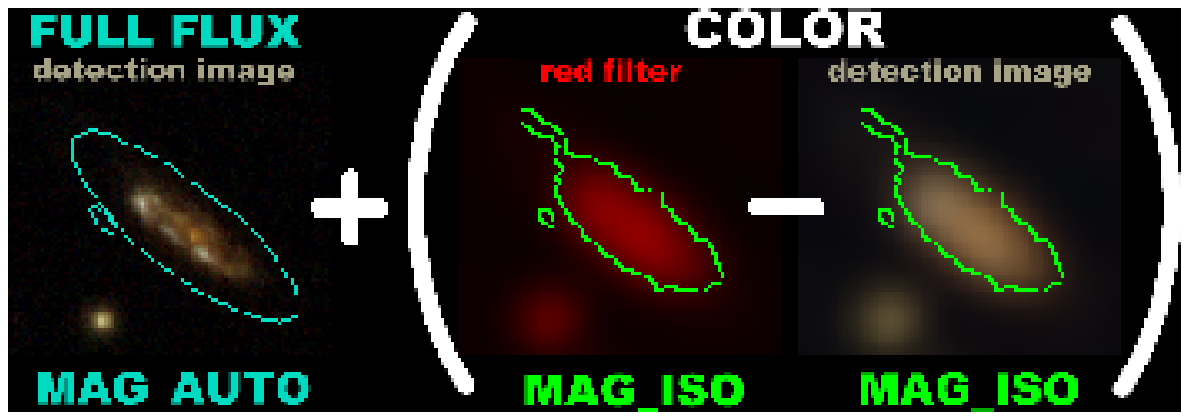}

\caption{\label{cap:colors}PSF-corrected isophotal aperture matched 
photometry. In the top frame, 
the blue magnitude $B=d_{AUTO}+(B_{ISO}-d_{ISO})$, i.e. the
total ({\tt MAG\_AUTO}) flux in the $BVi\arcmin z\arcmin$ detection
image $d$ plus a color term measured within the object's isophotal
aperture. In the bottom frame, we encounter a blurry red image. The
red magnitude $J=d_{AUTO}+(J_{ISO}-d_{ISO}^{J})$, where we have degraded
our detection image to match the PSF of the blurry image.
This PSF-corrected magnitude is the magnitude that would have been measured 
in the $J$-band image if it had the sharper $B$-band PSF.
The resulting $B-J$ color measurement is robust.}
\end{figure}

To estimate the flux loss,
we degrade our best (ACS detection) image of the galaxy to the poor (NIC3) PSF 
and observe how much flux is lost.
We then correct our observed NIC3 flux by the same amount (Fig.~\ref{cap:colors}).

This procedure relies on the assumption 
that the ACS detection image is a good model for the NIC3 images.
But what if a galaxy has a large internal color gradient?
The ACS detection image is a stacked $B$+$V$+$i\arcmin$+$z\arcmin$ image.
The resulting galaxy light profiles are the average of those in the four ACS filters.
Thus they are less sensitive to internal color gradients. 
Also note that this is a non-issue for bright galaxies,
for which the PSF corrections are small, regardless of internal color gradients.

We will now describe our process in more detail, 
as it is implemented in our {\tt ColorPro} software.

\begin{figure*}
\includegraphics[width=0.32\hsize]{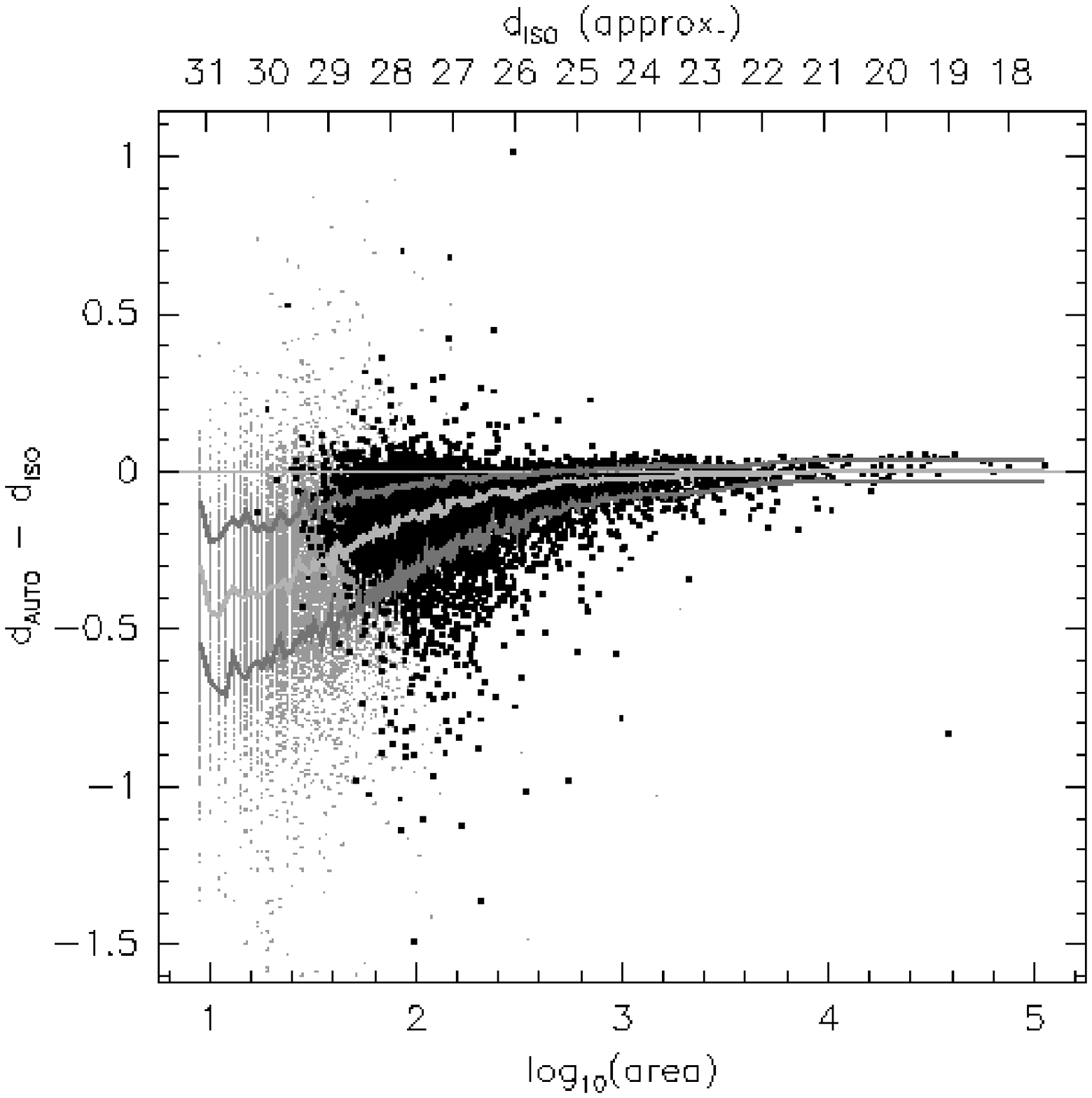} \includegraphics[width=0.32\hsize]{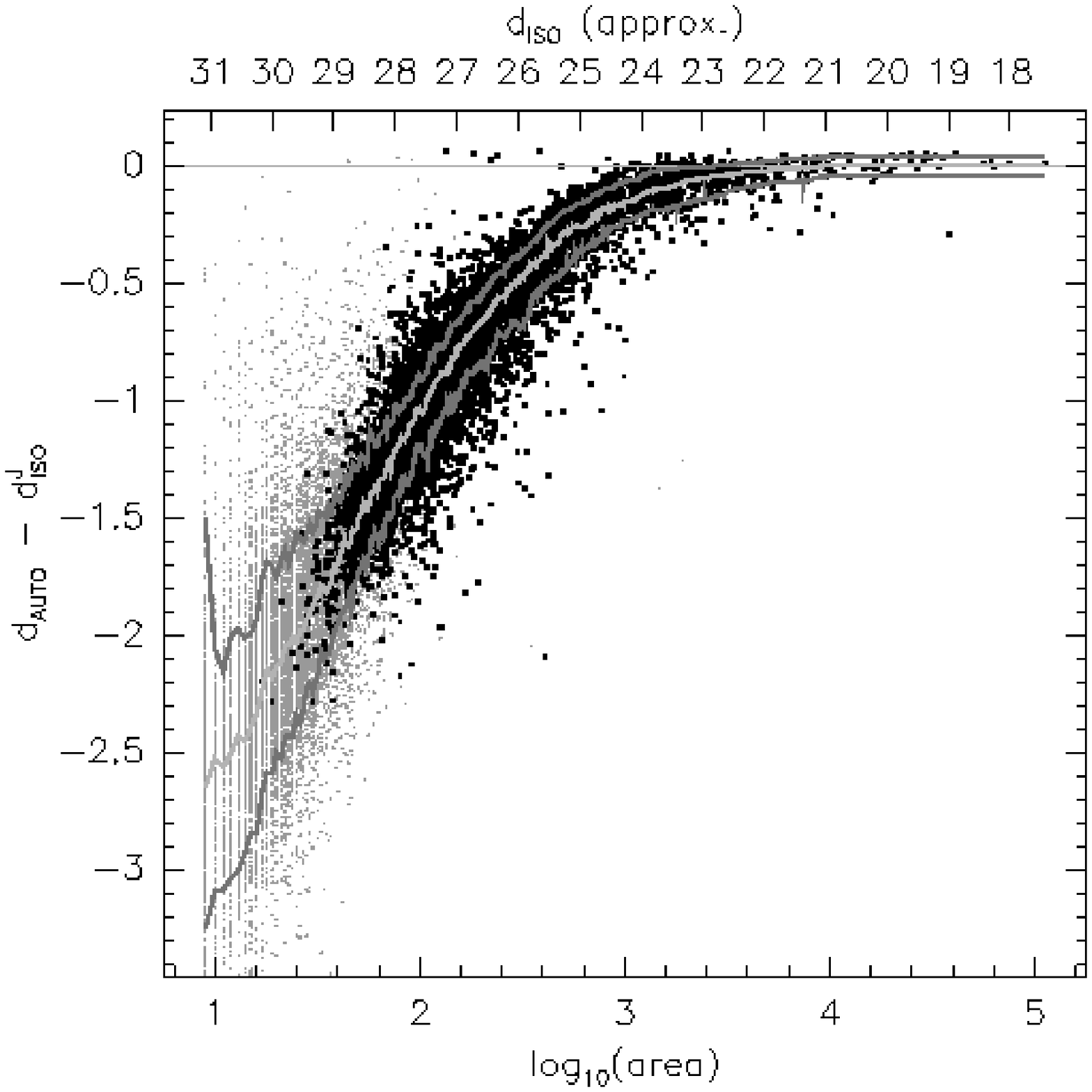} \includegraphics[width=0.32\hsize]{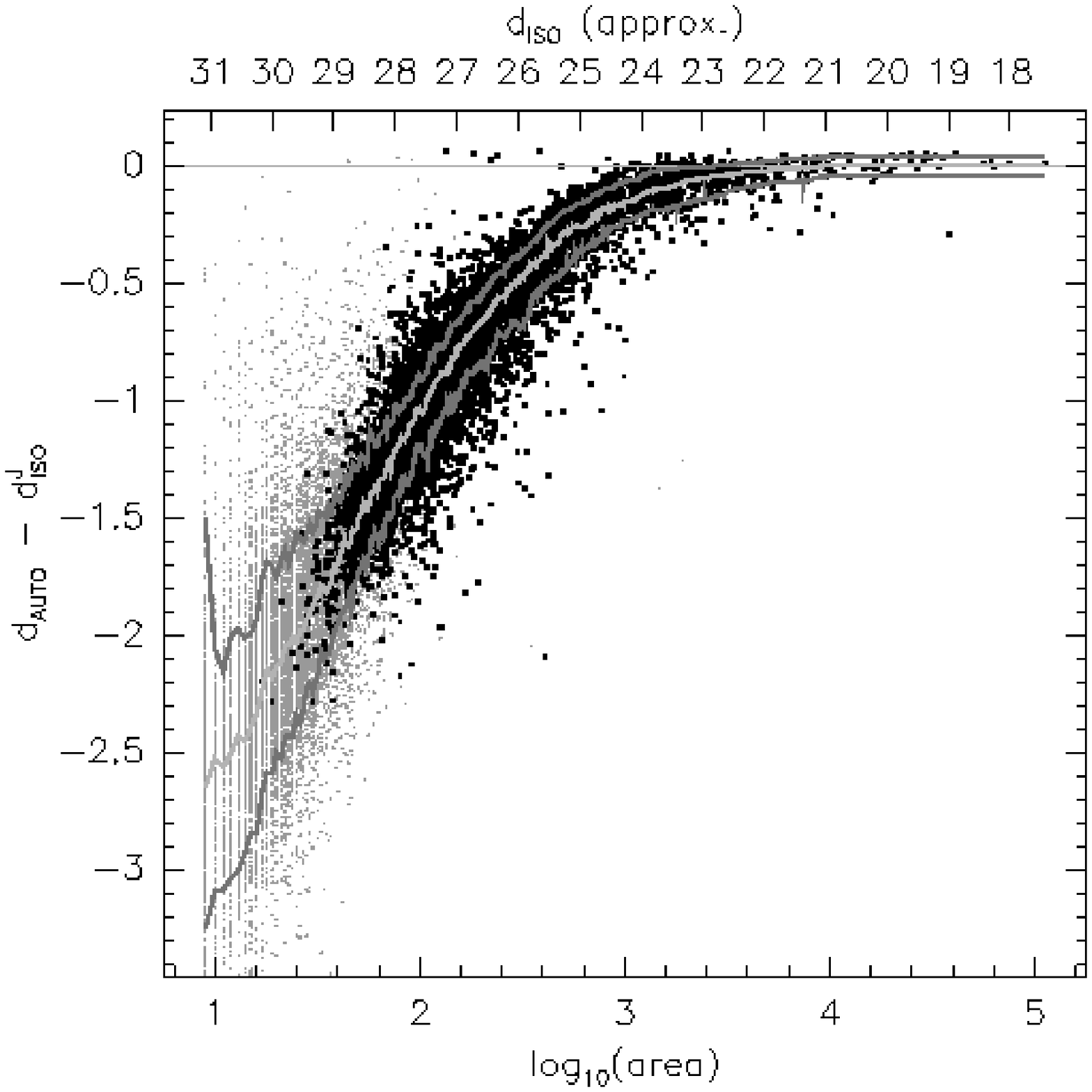}
\caption{\label{cap:PSFcor}Left: Aperture corrections ({\tt MAG\_AUTO}
- {\tt MAG\_ISO}) in the detection image $d$ plotted vs.~isophotal
aperture area in pixels. We take the liberty of labeling the top
axis with approximate values for $d_{ISO}$, as isophotal magnitude
and isophotal area are tightly correlated. The cyan line gives the
median correction of each data point's 250 closest neighbors along
the x-axis. (Near the extrema in area, the number of neighbors is
relaxed to as low as 100.) The magenta lines give the scatter (1-$\sigma$)
of these neighbors. \textit{All} galaxies are included in this plot,
but only those detected at 10-$\sigma$ are plotted in black. Lesser
detections are plotted in grey, if at all (the y-axis does not extend
to accommodate all of them). These $<10\textrm{-}\sigma$ detections
do not significantly affect the average corrections, except to add
more data points at low area. Center: aperture corrections in $d^{J}$
($d$ degraded to the PSF of $J$). Right: aperture corrections in
$d^{H}$, to the same scale as the center plot.}
\end{figure*}

The NIC3 $J$ image is mapped to the higher resolution ACS frame using
{\tt IRAF}'s {\tt wregister}%
\footnote{The {\tt fits} images released by ST contain accurate WCS information
in their headers and thus aligned almost perfectly after {\tt wregister}
re-mapping. Perfect alignment was achieved by shifting the NIC3 WCS
headers by a half pixel in both x \& y.%
}, taking care to preserve each object's flux by setting {\tt fluxconserve=yes}
and {\tt interp=spline3}. The resulting image is referred to as
$J^{A}$ (see Fig.~\ref{cap:apertures}). Next, we degrade the ACS
detection image $d$ (the $B$+$V$+$i\arcmin$+$z\arcmin$ image)
to the PSF of $J^{A}$, the result being $d^{J}$.%
\footnote{This degradation must be performed carefully to avoid significant
errors (of a magnitude or more) for faint objects. We discuss our
robust procedure in the appendix.%
} For a given object, an identical aperture is used in $J^{A}$, $d$,
and $d^{J}$, namely the isophotal aperture defined by the segmentation
map via {\tt SExSeg} (\S\ref{sub:SExSeg}).
Thus we measure magnitudes $J_{ISO}^{A}$, $d_{ISO}$, and $d_{ISO}^{J}$.
The PSF correction is $d_{ISO}-d_{ISO}^{J}$, i.e. the difference
in magnitudes resulting from the object being observed with the PSF
of the NIC3 $J$-band as opposed to the PSF of ACS. This correction
is applied to the $J$ magnitude yielding $J=J_{ISO}^{A}+(d_{ISO}-d_{ISO}^{J})$.
\emph{This PSF-corrected magnitude is the magnitude that would have been
measured in the NIC3 image if it had the sharper ACS PSF.} Thus this
magnitude can be compared with magnitudes measured in the ACS filters,
yielding robust colors $B-J$, $V-J$, $i\arcmin-J$, and $z\arcmin-J$.%
\footnote{The $z\arcmin$-band also requires a small PSF correction which
we discuss in \S \ref{sub:zapcor}.%
} This process is repeated for the $H$-band image, which has a slightly
worse PSF than $J$. It is important to note that the PSF corrections
are different for every object. (This would be the case even if the same
aperture size was used for every object.) And faint objects can
have large PSF corrections of 2 magnitudes or more (see Fig.~\ref{cap:PSFcor}).

This procedure ensures consistent isophotal colors across all filters.
But it is well known that isophotal magnitudes lose some flux; {\tt SExtractor}'s
{\tt MAG\_AUTO} is a better measure of a galaxy's total flux \citep{SExref}.
So we obtain our final {}``total'' magnitudes
by applying a correction of $d_{AUTO}-d_{ISO}$ to each isophotal
magnitude defined above. Rearranging terms, we have:
\begin{eqnarray*}
B&=&(B_{ISO}-d_{ISO})+d_{AUTO}\\
V&=&(V_{ISO}-d_{ISO})+d_{AUTO}\\
i\arcmin&=&(i\arcmin_{ISO}-d_{ISO})+d_{AUTO}\\
z\arcmin&=&(z\arcmin_{ISO}-d_{ISO})+d_{AUTO}+z\arcmin_{apcor}\\
J&=&(J_{ISO}^{A}-d_{ISO}^{J})+d_{AUTO}\\
H&=&(H_{ISO}^{A}-d_{ISO}^{H})+d_{AUTO}
\end{eqnarray*}

Note that a given color across ACS filters is simply the isophotal
color, e.g. $B-V=B_{ISO}-V_{ISO}$ (except for the $z\arcmin$-band,
which requires its own PSF correction $z\arcmin_{apcor}$ (\S
\ref{sub:zapcor})). But a color between ACS \& NIC3 filters contains
the PSF correction term described above, e.g. $B-J=B_{ISO}-J_{ISO}^{A}+(d_{ISO}-d_{ISO}^{J})$.

The above magnitude equations may look more familiar when reformulated
as aperture corrections, for example:

$B=B_{ISO}-(d_{ISO}-d_{AUTO})$

where we restore the flux lost as a result of using an isophotal aperture
(assuming {\tt MAG\_AUTO} is our best measure of the total flux).
But we prefer the previous set of equations, as they emphasize that
every color is measured relative to the detection image $d$ in a
consistent aperture, and that for each galaxy, $d_{AUTO}$ is just
a constant added to each color. 

Some objects lack measurements for $d_{AUTO}$, $d_{ISO}$, $d_{ISO}^{J}$,
and/or $d_{ISO}^{H}$, either due to a total non-detection ($<1$-$\sigma$)
or perhaps saturation or other bad pixels. In these cases, we apply
the average magnitude corrections successfully applied to other objects
with those aperture areas (Fig.~\ref{cap:PSFcor}).

\subsubsection{$z\arcmin$-band PSF Corrections}
\label{sub:zapcor}

ACS $z\arcmin$-band images sport a slightly wider PSF than images in the bluer bands.
\citet{Sirianni03} have meticulously quantified the resulting PSF corrections
as a function of both wavelength and aperture size.
We use their results rather than relying on the degradation technique described above.

\begin{figure*}
\plottwo{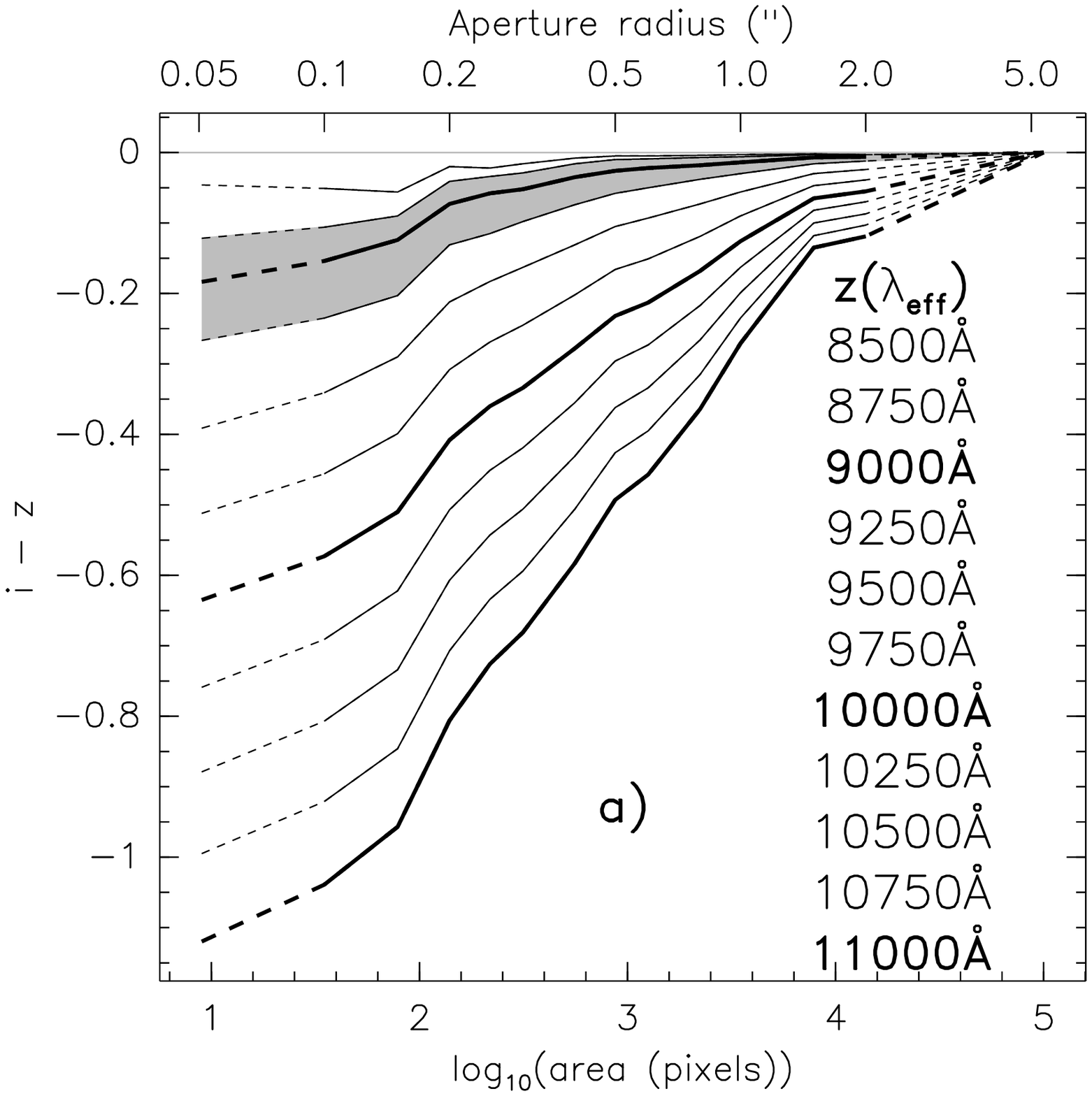}{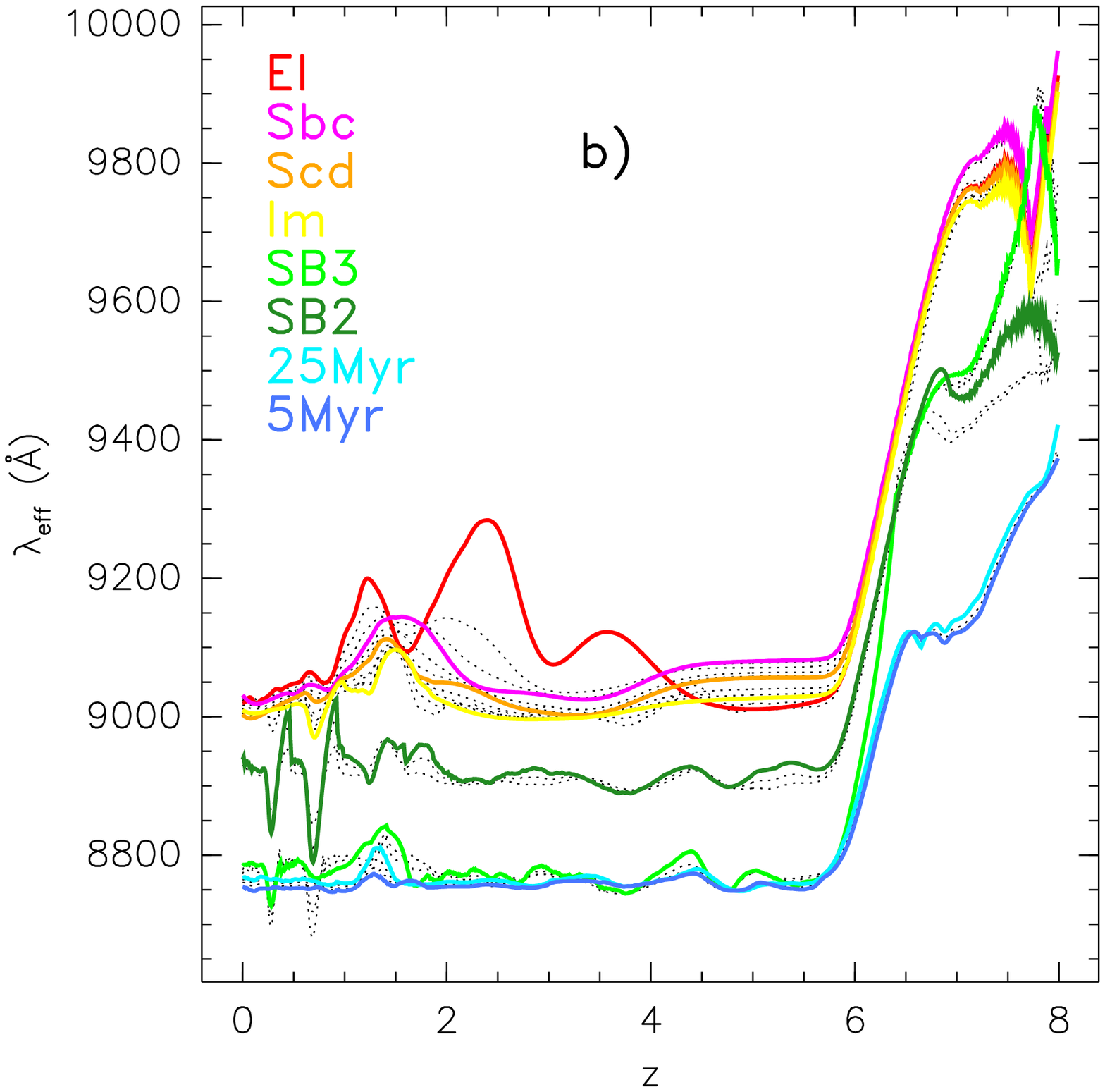}
\caption{\label{cap:zapcor}Left: Aperture corrections applied to the $z\arcmin$-band
photometry. The solid lines are taken from Table 8 of \citet{Sirianni03}.
For a given object, the aperture correction depends on
both the aperture radius (labeled across the top axis, with the corresponding
area in the ACS images labeled across the bottom) and the effective
wavelength $\lambda_{eff}$ of that object in the $z\arcmin$-band
(\S \ref{sub:zapcor}). Redder objects require larger aperture corrections.
The dashed lines are extrapolations to smaller and larger radii. (To
avoid negative aperture corrections, we simply assign zero aperture
correction to $r=5.0\arcsec$.) The thicker lines merely indicate
$\lambda_{eff}$ multiples of $1000\AA$, while the shaded region
is where most galaxies fall, as we see in our next plot.
Right: Effective wavelength $\lambda_{eff}$ as a function of SED
type (\S\ref{sub:faintblue}) and redshift. The colors represent
SED type, as in Fig.~\ref{cap:SEDs}. Intermediate SED types are plotted
as dotted lines. At $z\sim5.7$, objects begin to drop out of the
$z\arcmin$-band, yielding significantly higher $\lambda_{eff}$.
We assign no aperture correction to $z>8$ galaxies, as these have
all but dropped out of the $z\arcmin$-band, yielding meaningless
$\lambda_{eff}$ and $i\arcmin-z\arcmin$.}
\end{figure*}

All ACS CCD detectors scatter light longward of $\sim7500\textrm{\AA}$ into a halo. 
The degree of scatter increases with wavelength. 
For a given galaxy observed in a given filter, 
we define the effective wavelength 
$\lambda_{eff}=\int d\lambda\,\lambda^{2}\, F_{\lambda}(\lambda)\, R(\lambda)\,/\,\int d\lambda\,\lambda\, F_{\lambda}(\lambda)\, R(\lambda)$,
where $F_{\lambda}(\lambda)$ is the object's observed flux per unit
wavelength, and $R(\lambda)$ is the response curve of the given filter
(Fig.~\ref{cap:transmission}). Table 8 of \citet{Sirianni03}
provides aperture corrections (to infinite aperture size) as a function
of aperture radius and effective wavelength $\lambda_{eff}$. These
corrections are roughly independent of $\lambda_{eff}$ for observations
in the $B$, $V$, and $i\arcmin$ filters, but are much greater in
the $z\arcmin$-band. We subtract the $z\arcmin$-band corrections
from the $i\arcmin$-band corrections (using a nominal value of $\lambda_{eff}=7750\textrm{\AA}$
for the $i\arcmin$-band), yielding the aperture corrections $z\arcmin_{apcor}$
that will bring our $z\arcmin$-band magnitudes back in line with
the other ACS filters. We plot these corrections in Fig.~\ref{cap:zapcor}a
for the expected range of $z\arcmin$-band $\lambda_{eff}$
(Fig.~\ref{cap:zapcor}b). Note that the aperture corrections are much smaller
than those for the NIC3 filters.

Since we do not know a galaxy's $\lambda_{eff}$ until we assign an
SED and redshift, we use the middle 9000\AA ~curve as an initial
guess, including an appropriate uncertainty: taking the top and bottom
curves as our 95\% (2-$\sigma$) confidence interval. Using this photometry,
we run {\tt BPZ}. Then, given each galaxy's SED and redshift,
we re-calculate $\lambda_{eff}$ and thus $i\arcmin-z\arcmin$ for
each galaxy.%
\footnote{This time, the uncertainties for $z\arcmin_{apcor}$ are the result
of a Monte Carlo simulation:
we reassign galaxy redshifts and SEDs given their {\tt BPZ}
probability distributions $P(z,t)$. Each realization yields values
for $\lambda_{eff}$ and thus $i\arcmin-z\arcmin$. The 1-$\sigma$
scatter of these $i\arcmin-z\arcmin$ values (for each galaxy) give
us our aperture correction uncertainty, which is added (in quadrature)
to the $z\arcmin$ magnitude uncertainty. These simulations were not
carried out for galaxies detected at $<10$-$\sigma$. For these galaxies,
we use the mean aperture correction uncertainties of 0.0086 for $z<5.7$
galaxies and 0.04 for $z>5.7$ galaxies.%
} Finally, with our updated photometry, we re-run {\tt BPZ}.

\subsubsection{Magnitude Uncertainties and Significance}
\label{sub:dm}

{\tt SExtractor} calculates magnitude uncertainties using
the weight maps released with the ACS images and the noise (RMS) maps
released with the NIC3 images. The NIC3 noise maps were corrected
for drizzling following \citet{drizzleRMS}.%
\footnote{The NIC3 flux uncertainties are divided by $\sqrt{F_{A}}$ from Equation
(A13) of \citet{drizzleRMS}. For $l>p$, $\sqrt{F_{A}}=1-p/3l$.
For $p>l$, $\sqrt{F_{A}}=(l/p)\cdot(1-l/3p)$. For the NIC3 images,
$p$={\tt pixfrac=0.6}. For the object's linear size, we use $l=\sqrt{area}$,
where $area$ is the aperture size measured in input pixels (pre-drizzling:
$0.20\arcsec$/pixel).%
} No such correction was necessary for the ACS images which were drizzled
with {\tt pixfrac=0}.

The NIC3 magnitude uncertainties must also account for the uncertainty
of the PSF corrections. This uncertainty is difficult to measure directly,
so we estimate it as the (1-$\sigma$) scatter of PSF corrections
for a given aperture size (see Fig.~\ref{cap:PSFcor}). We then
add this uncertainty in quadrature to the magnitude uncertainty reported
by {\tt SExtractor}. Also added in quadrature are uncertainties
($J:0.025$, $H:0.042$) from our NIC3 magnitude offsets (\S\ref{sub:NIC3magoffsets}).

As we are using isophotal apertures, we generally report isophotal
magnitude uncertainties. However, some isophotal apertures are actually
smaller than the PSF of the image (that is, a circle with a diameter
of twice the FWHM of the PSF). Thus we also measure magnitude uncertainties
within a circular aperture of each image's PSF size. We use {\tt FLUXERR\_APER}
in place of {\tt FLUXERR\_ISO} whenever the isophotal aperture
is smaller than the PSF. These area thresholds are 28 and 355 pixels ($0.03\arcsec$/pix),
respectively for the ACS and NIC3 images.

\begin{figure}
\plotone{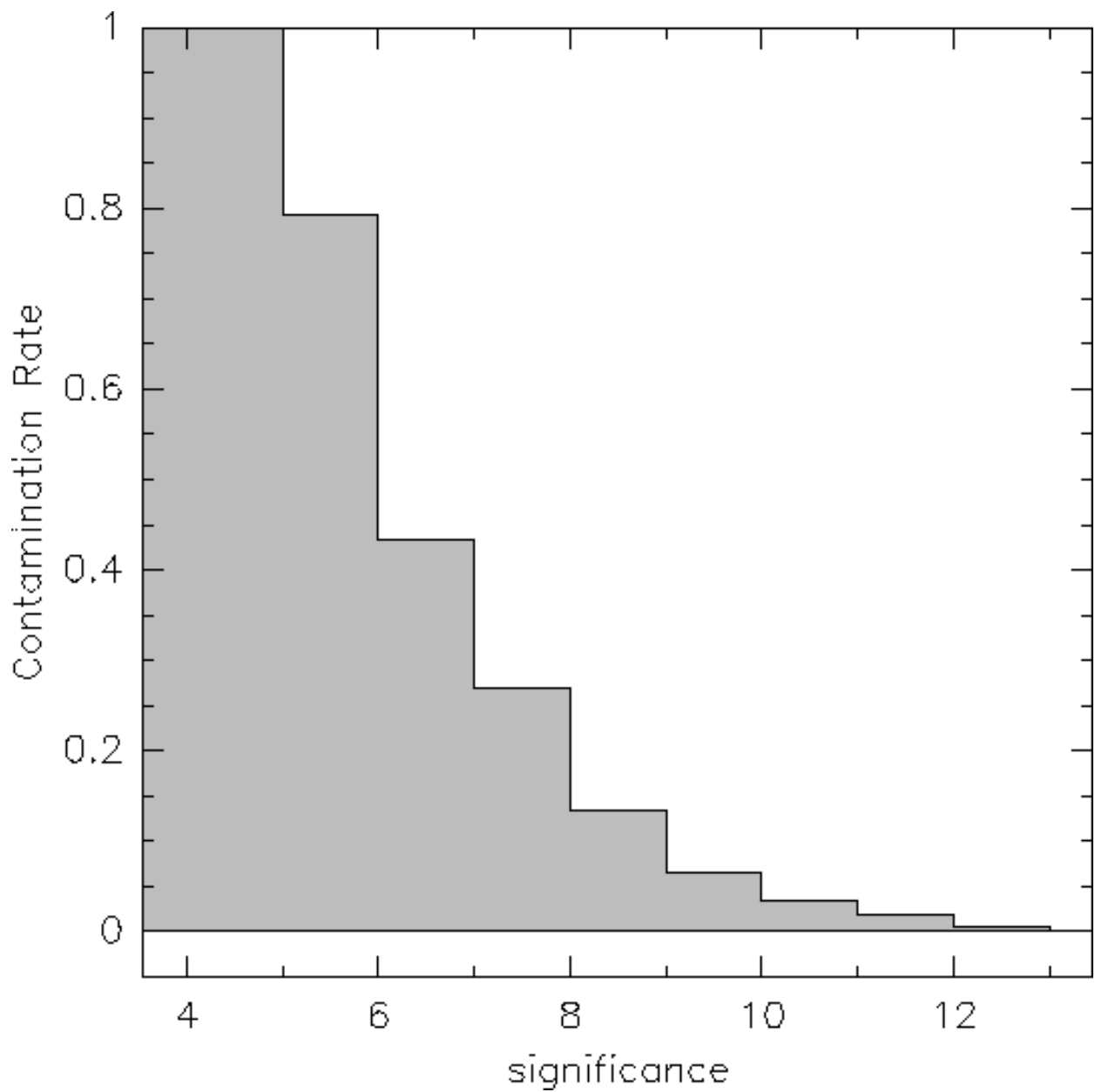}
\caption{\label{cap:sig-1} Spurious detection fraction in $d$ as a function
of significance. For much of our analysis that follows, we prune our
catalog at 10-$\sigma$. 599 objects have been detected between 10-
and 11-$\sigma$ vs.~21 objects found in the negative image of $d$,
yielding a 3.5\% rate of contamination in that significance bin. Those
interested in fainter sources may probe our catalog to as low as 6-$\sigma$.
The majority (57\%) of sources detected at 6- to 7-$\sigma$ will
still be real.}
\end{figure}

We measure the significance of each detection in each filter as {\tt FLUX\_ISO}
/ {\tt FLUXERR} ({\tt FLUXERR\_ISO} or {\tt FLUXERR\_APER},
depending on the aperture size). Most of our published results in
\S\ref{sec:BPZ} and Paper II employ a conservatively pruned
catalog: any object without a 10-$\sigma$ detection in any filter
or detection image is discarded. Analysis of the inverted ACS detection
image ($d$ multiplied by -1) yields 36 objects detected at the 10-$\sigma$
level or higher. These are negative noise peaks, and we can expect
to find a similar number of positive noise peaks (spurious objects)
in our detection catalog. This is an insignificant level of contamination:
36 / 7,565 = 0.5\%. Even among the faintest of our pruned detections, between 10-
and 11-$\sigma$, we only expect 3.5\% to be spurious (599
objects vs.~21 found in the negative image, see also Fig.~\ref{cap:sig-1}).
Those interested may comb our full catalog for fainter sources. For
example, the majority (57\%) of sources detected at 6- to 7-$\sigma$ will
still be real.

A non-detection in any filter ($<1$-$\sigma$; {\tt FLUX\_ISO} $>$ {\tt FLUXERR})
is assigned a flux of zero and a flux uncertainty (upper
limit) equal to the 1-$\sigma$ detection limit. In table \ref{cap:filters},
we quote 10-$\sigma$ detection limits within a 0.5sq$\arcsec$ aperture.
The 1-$\sigma$ limits are 2.5 magnitudes fainter. But our isophotal
apertures vary greatly in size, and each aperture has a different
detection limit. Fortunately, {\tt SExtractor} custom-calculates
a detection limit for each non-detection. This is given simply as
{\tt FLUXERR} ({\tt \_ISO} or {\tt \_APER}). 

The upper flux limits assigned to NIC3 non-detections must incorporate
PSF corrections. For example, {\tt FLUXERR} may yield an upper
limit corresponding to $J_{ISO}>29$ for a given aperture. But suppose
this aperture has a $J$-band PSF correction of $d_{ISO}-d_{ISO}^{J}=-1$.
Then an object just barely detectable in this aperture would see its
magnitude corrected from $J_{ISO}=29$ to $J=28$. So a non-detection
should be treated as $J>28$ when fitting SEDs to this object.

Finally, objects unobserved in a given filter (outside the NIC3 FOV
or containing saturated or other bad pixels) are assigned infinite
uncertainties.

\subsection{UDF NIC3 Recalibration}
\label{sub:NIC3recal}

Based on {\tt BPZ} SED fits to objects of known spectroscopic redshift,
we derived corrections of $-0.30\pm0.03$ mag in $J$ and $-0.18\pm0.04$ mag in $H$
(\S\ref{sub:NIC3magoffsets}).
Our derived corrections appear to be supported by two recent recalibrations:
the first pertaining solely to the Thompson et al.\ UDF image reductions (both versions 1 \& 2)
and the second affecting all NICMOS images.
We discuss these recalibrations here.

\citet{Thompsoncal} have recalibrated the zeropoints of their UDF images,
resulting in objects brighter by $\sim 0.08$ and $\sim 0.09$ mag in $J$ and $H$, respectively.
These offsets were due to a $\sim 10\%$ miscalibration of the filter sensitivity curves
in their original analysis.
Their catalogs (both versions 1 \& 2) should be corrected for this recalibration
(and that due to non-linearity, as we discuss below).
However, as the Thompson et al.\ images were not reduced by the standard STScI NICMOS pipeline,
these offsets do not apply to any other (non-UDF) STScI NICMOS image reductions or catalogs.
In fact, this correction brings the measured UDF NIC3 fluxes into better agreement
with those measured internally and independently at STScI (Louis Bergeron, priv.\ comm.).

Meanwhile, STScI has been investigating issues of NICMOS non-linearity dependent on count rate \citep{NICMOScal}.
(This is not to be confused with the non-linearity inherent in all IR detectors which is dependent on total counts.
This effect is well understood and corrected for in the NICMOS pipeline.)
Apparently, brighter objects (with higher count rates) register slightly higher total fluxes than expected in NICMOS images,
while fainter objects register slightly lower fluxes than expected.
This effect was first discovered by \citet{Bohlin05}, followed up \citep{Bohlin06}, 
and recently confirmed by robust lamp on/off tests \citep{lamptests}.
The results from this latter report show that 
for each dex (2.5 mag) decrease in incident flux, NIC3-observed $J$-band magnitudes drop $\sim 0.048$ more than expected.
$H$-band magnitudes suffer a similar but weaker non-linearity of $\sim 0.016$ mag / dex.
This presumably applies to all NICMOS images.

The UDF NIC3 images were calibrated relative to standard stars of $\sim 12^{\rm th}$ mag
which is $\sim 4$ dex (10 mag) brighter than the sky-background of the UDF.
Thus sky-dominated objects in the UDF are expected to suffer offsets of
$\sim 0.19$ mag in $J$ and $\sim 0.06$ mag in $H$ due to this count-rate dependent non-linearity.
(By sky-dominated objects, we mean those objects with count rates less than that of the sky background.
The total count rate of these objects (galaxy + sky) is therefore roughly equal to that of the sky itself.)
For brighter UDF objects the offsets should be slightly less, decreasing by $\sim 0.048$ and $\sim 0.016$ mag / dex,
respectively for $J$ and $H$.
Objects with $J \sim 22$ or $H \sim 22$ have roughly the same count rates as the sky in that filter,
yielding total count rates $\sim 2 \times$ that of the sky.
Thus the offset for a $J \sim 22$ object decreases slightly to $0.19 - 0.01 = 0.18$
(where $0.01 \sim 0.048 \times \log_{10}(2)$).
And an object 1 dex fainter than that at $J \sim 19.5$ would have an offset of roughly $0.18 - 0.048 = 0.13$.
But $J \sim 19.5$ objects are very rare in the UDF.
Only 5 objects are brighter than $J < 19.5$ with none brighter than $J < 18$.
In fact there are only 38 objects brighter than $J < 22$.
Thus to correct for this non-linearity, a constant offset of 0.19 mag in $J$ should prove an excellent approximation,
especially for those 2,800+ other objects detectable in $J$ but fainter than $J > 22$.
Similarly, a constant 0.06 mag offset should adequately correct the $H$ band magnitudes.

Proper corrections for non-linearity require corrections on a pixel-by-pixel basis,
which will be implemented into a future version of the STScI NICMOS pipeline.
As of April 2006, a beta version of software capable of performing this correction on NICMOS images
was made available to the public.\footnote{http://www.stsci.edu/hst/nicmos/performance/anomalies/nonlinearity.html}
When run on the UDF, this software yields magnitude offsets similar to those quoted above,
although small uncertainties still remain, pending further calibration tests \citep{NICMOScor}.

When the magnitude offsets due to non-linearity are added to those due to the filter recalibrations described above,
we find total offsets of $\sim 0.27$ and $\sim 0.15$ mag in $J$ and $H$, respectively.
Thus, given the UDF NIC3 images with their original zeropoints,
a $J = 24$, $H = 24$ object would be observed to have $J \sim 24.27$ and $H \sim 24.15$.
Note that these offsets are very similar to those we quoted above,
as derived empirically in \S\ref{sub:NIC3magoffsets} from SED fitting using {\tt BPZ}
(based on the assumption that the ACS photometry was accurate).
Thus we are encouraged to proceed with our analysis given our derived offsets:
$-0.30\pm0.03$ in $J$ and $-0.18\pm0.04$ in $H$.
(The uncertainties are added in quadrature to each object's NIC3 magnitude uncertainties.)

\subsection{Morphology}
\label{sub:morph}

To increase the utility of our catalog, 
we have included measures of several morphological parameters
that are useful in automatic galaxy classification.
These include S\'ersic \citeyearpar{Sersic} index $n$, asymmetry, 
and number of nearby neighbors.

For isolated and undisturbed galaxies, 
the S\'ersic index $n$ alone
is a fairly reliable indicator of morphological
type \citep[e.g.,][]{Andredakis95}.
We adopt $n=2.5$ as the dividing line between disk- ($n<2.5$) and 
spheroidal-dominated ($n>2.5$) galaxies
({}``late'' and {}``early'' type, respectively), 
consistent with the analysis conducted by the Sloan Digital Sky Survey 
\protect(SDSS; see \citealt{Shen03}), and more recently 
the Galaxy Evolution by Morphology and SEDs (GEMS) Survey \citep{GEMS}. 
Simulations (\S\ref{sub:morphsims}) indicate that 80-95\% of galaxies 
in our catalog with $\sigma n/n<1$ (confident measures of $n$) 
have a correct morphological classification
(late vs.~early type, assuming that $n=2.5$ is a perfect discriminator). 
And this cut only discards $\sim8$\% of the catalog.

Less well behaved galaxies, including mergers and irregulars,
generally do not have well defined S\'ersic indices.
Fortunately these galaxies can generally be weeded out (or selected for)
by measuring their large asymmetries \citep[e.g.,][]{Conselice03}.
Meanwhile, neighbors in projection can also affect the model fitting
(stymieing even the most careful attempts to mask the neighbors out).
Thus, in our catalog we also provide counts of nearby neighbors,
which may be used to select well isolated galaxies, 
or alternatively, to help find interacting galaxies.
Any reliable morphological classification should take all three parameters into account:
S\'ersic index, asymmetry, and nearby neighbors.

\begin{figure}
\plotone{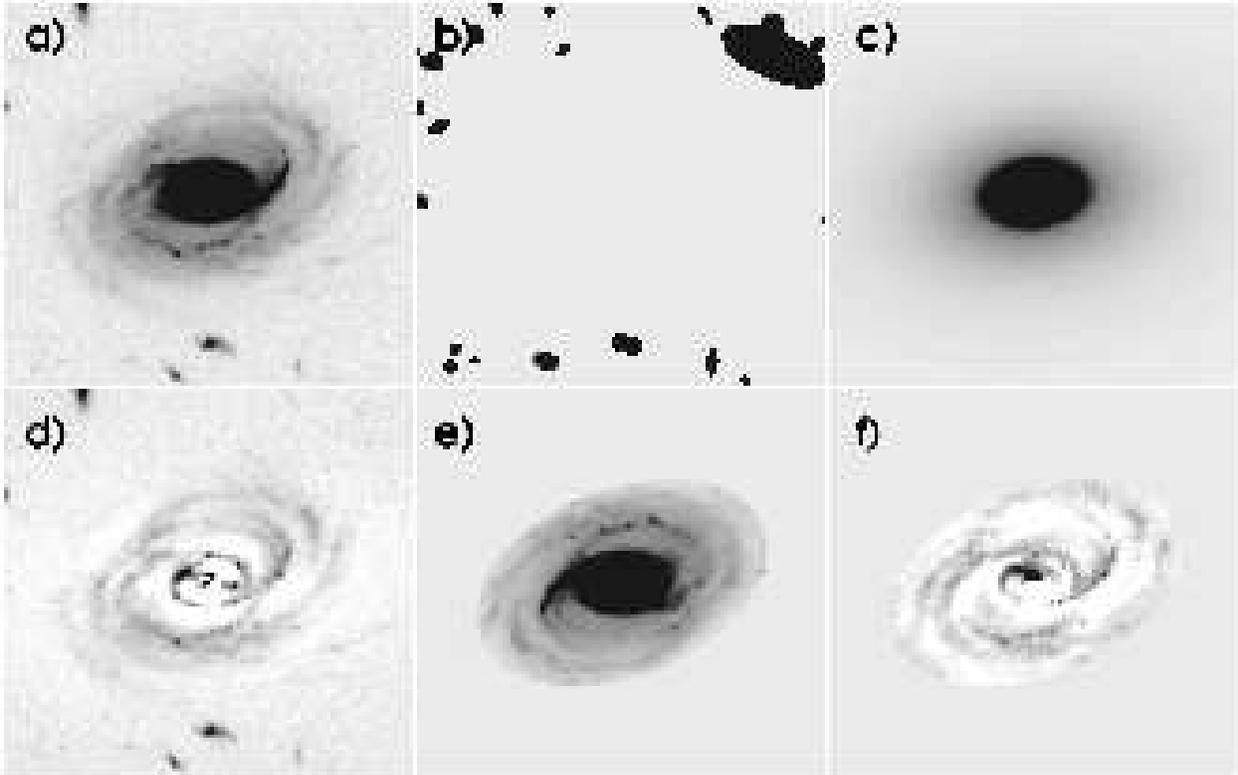}
\caption{\label{cap:recursive}
Recursive procedure used to obtain morphological measurements.
$(a)$: $i\arcmin$-band image.
$(b)$: Ellipses used to mask out neighbors from the model fitting.
$(c)$: Resulting single component S\'ersic model from \galfit.
$(d)$: Model subtracted from $i\arcmin$-band image.
This galaxy will ``remain'' subtracted for the subsequent modelling of all fainter galaxies.
$(e)$: Galaxy rotated by $180{}^{o}$, and framed within an ellipse of $b = 4 \times r_{50}$.
$(f)$: Difference of $(a)$ and $(e)$, used to measure galaxy asymmetry.  This spiral galaxy shows a fair amount of asymmetry, but not enough to be flagged as ``Irregular'' or a merger
(see Fig.~\ref{cap:morphtypes}).
}
\end{figure}

All of our morphological measurements are obtained from the $i\arcmin$-band
image (the deepest ACS image).
We analyze every object in B04's $i\arcmin$-band catalog%
\footnote{The relationship between our catalog and the ST catalog is well defined,
with most objects being defined identically (\S\ref{sub:detection}).%
}, beginning with the brightest galaxy and working our way down to the faintest.
Along the way we subtract each galaxy model from the $i\arcmin$-band image 
(see Fig.~\ref{cap:recursive}).

Thus we begin by creating a postage stamp, $5 \times r_{50}$ on a side, for the brightest galaxy,
where $r_{50}$ is the galaxy's half-light radius, as given by \sex.
Within that postage stamp, neighboring galaxies are masked out using ellipses,
each ellipse given a minor axis length $b = 2 \times r_{50}$ for that galaxy.
(Note that we do not use our segmentation map (\S\ref{sub:detection})
to measure morphological parameters.
We have not studied the effects of segmentation on such measurements,
and thus we opt for a more traditional approach.)
Using \galfit\ \citep{galfit}, the brightest galaxy is fit to a single component S\'ersic model
$\Sigma(r) \propto \exp(-\kappa_n[(R/R_{e})^{1/n}-1])$,
where $\kappa = \kappa(n)$ is a normalization constant
and $R_e$ is the effective radius.
The fit is constrained to $0.2<n<8$ and 
$0.3<R_{e}<500$ pixels, and the centroid is
confined to within 2 pixels of the position derived by \sex.
As initial guesses for the \galfit\ parameters, 
we use the \sex\ output parameters given in B04's $i\arcmin$-band catalog.
Lacking estimates for the S\'ersic index from \sex, we start all fits with $n=1.5$. 

Having been calculated for the brightest galaxy, 
the S\'ersic model is subtracted from the $i\arcmin$-band image.
This subtraction benefits the subsequent modelling of all fainter nearby galaxies.
We proceed to model the second-brightest galaxy,
and continue in order of decreasing brightness,
modelling and subtracting every galaxy in B04's $i\arcmin$-band catalog.
Of the 9,339 objects with ${\tt stellarity}<0.9$,
\galfit\ derives meaningful output for 8,805, or about 94\% of the objects.
Table \ref{cap:catmorph} summarizes the resulting
fit parameters and their uncertainties: magnitude $i\arcmin$, effective
radius $R_{e}$, ellipticity $a/b$, position angle $\theta$, and
S\'ersic index $n$. We also give the {}``badness'' of each fit $\chi^{2}/\nu$.

\begin{figure}
\epsscale{0.85}
\plotone{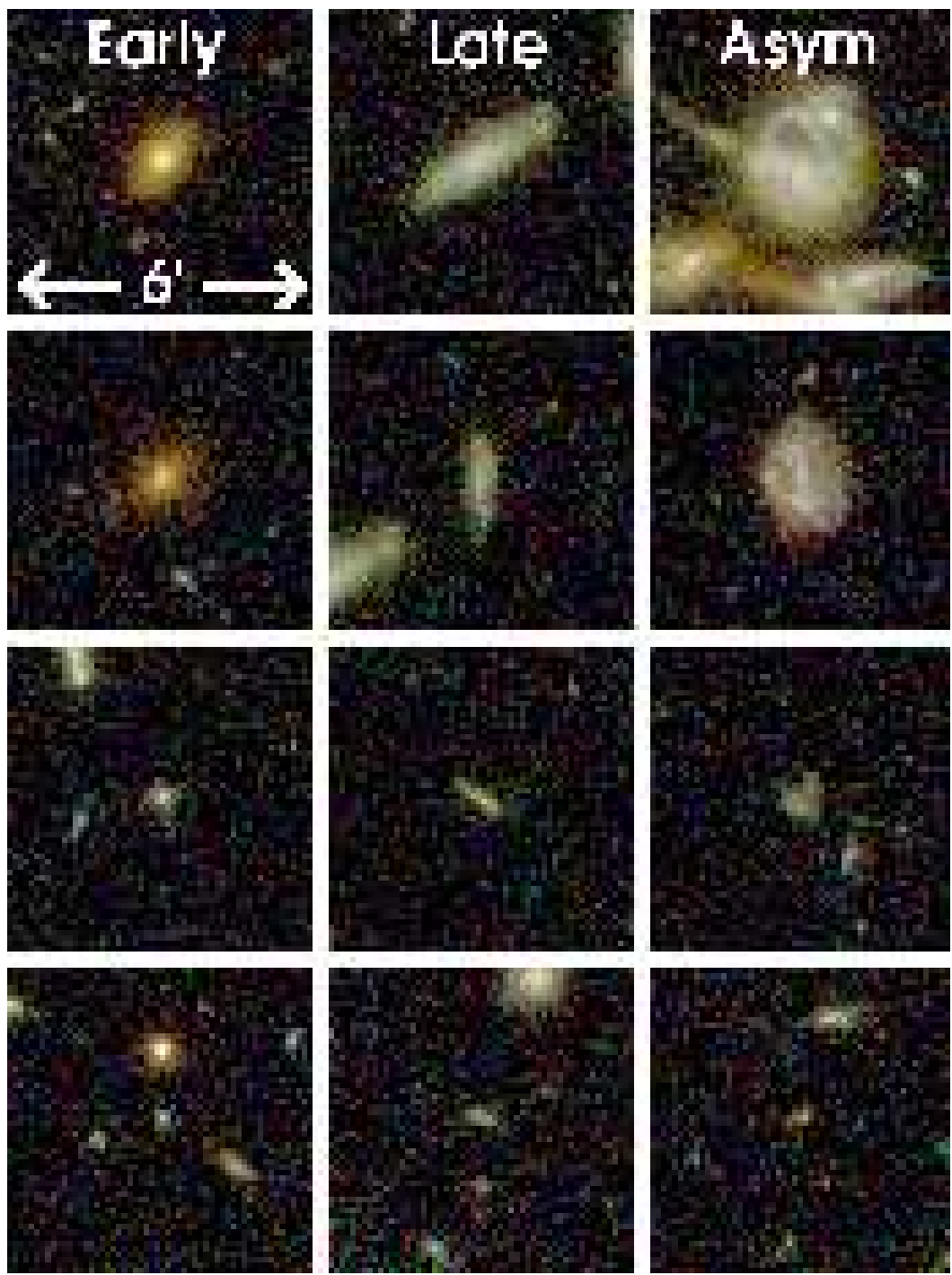}
\epsscale{1.0}
\caption{\label{cap:morphtypes}
Examples of early type, late type, and highly asymmetrical galaxies.
All postage stamps are 6\arcsec$\times$6\arcsec, 
taken from our $B V i\arcmin z\arcmin$ 4-color image.
The first two columns show isolated and symmetrical galaxies 
with reliable measures of S\'ersic index ($\sigma_n/n<$1).
Galaxies in the first column are morphologically classified as early ($n>2.5$),
while those in the second are classified as late ($n<2.5$).
The third column shows galaxies with clear asymmetries ($A>0.25$).
Galaxies in this column should not be classified by S\'ersic index alone.
Galaxy magnitudes range here range from roughly $i\arcmin \sim 22.5$ to $i\arcmin \sim 26.5$.
}
\end{figure}

Examples of early type ($n>2.5$), late type ($n<2.5$), 
and highly asymmetrical galaxies are given in Fig.~\ref{cap:morphtypes}.
For the latter, S\'ersic fits often prove unreliable, as mentioned above.
Thus we measure asymmetry:

\[
A=\frac{\Sigma|I_{i,j}-I_{i,j}^{rot}|}{2\Sigma|I_{i,j}|}\]

where $I_{i,j}$ are the pixel values 
and $I_{i,j}^{rot}$ is the image rotated by $180{}^{o}$ 
\citep{Schade95,Abraham96,Conselice00}.
These measurements are obtained within an ellipse of $b = 4 \times r_{50}$
drawn around the galaxy
(with neighbors masked out and brighter galaxies subtracted as above, Fig.~\ref{cap:recursive}e).
This index proves to be a good estimate of asymmetry for galaxy images with good
signal-to-noise \citep{Conselice00}. 
Our method does not minimize the asymmetry,
and in that respect it is slightly different from the method of \citet{Conselice00}.

For each galaxy, we also give the number of nearby neighbors, or companions.
Two galaxies are identified as companions if their centroids lie within twice the
sum of their effective radii and their $i\arcmin$-band photometry matches to within 0.5 mag.

The morphological parameters in our catalog may be used, for example,
to address questions of ``Nature vs.~Nurture'', 
including the well-studied morphology-density relation \citep[e.g.,][]{Kauffmann04}.
Are galaxy morphologies dictated mainly by their formation epoch,
or are they shaped more by their environment (e.g., cluster vs.~field)?
We may also investigate the contributions of different galaxy types
to star formation rates \citep[e.g.,][Paper II]{Wolf05}.

\section{Bayesian Photometric Redshifts ({\tt BPZ})}
\label{sec:BPZ}

We obtained photometric redshifts of the objects in our catalog using
an updated version of the Bayesian photometric redshift software {\tt BPZ}
\citep{BPZ00}. In addition to re-calibrated SED (spectral energy distribution)
templates introduced in \citealt{Benitez04}, this new version also produces
an enhanced summary of the redshift probability distribution $P(z)$
for each galaxy, reporting up to three peaks where warranted, along
with their widths and relative probabilities. And in this paper, we
advocate the addition of two new templates to the SED library (\S\ref{sub:faintblue}).

We have experienced some numerical instabilities in {\tt BPZ}
for the extreme redshift and magnitude ranges present in the UDF. 
Future versions of {\tt BPZ} will correct this problem,
which lies in the normalization factor of the likelihood function
$p(C|z,T)\propto F_{TT}(z)^{-1/2}\exp(-\onehalf\,\chi^{2}(z,\, T,\, a_{m}))$
(Eq. 12 of \citealt{BPZ00}; $C$ represents the observed colors and $z,\, T,\, a_{m}$
are the model redshift, template, and amplitude, respectively). But
for now, we simply remove the normalization factor, effectively reverting
to the {}``frequentist'' (ML) expression $p(C|z,T)\propto\exp(-\onehalf\,\chi^{2}(z,\, T,\, a_{m}))$.
Of course every other aspect of the Bayesian method is retained, including
the use of priors (which we have modified to accommodate our new templates).

Our {\tt BPZ} catalog is available in Table \ref{cap:catBPZ}.
Redshift probability distributions $P(z)$ are available via
{\tt http://adcam.pha.jhu.edu/\~{}coe/UDF/}.

\clearpage

\begin{deluxetable}{lcccccccccccccc}
\tabletypesize{\scriptsize}
\rotate
\tablewidth{0pt}
\tablecaption{\label{cap:catBPZ}Catalog: BPZ}
\tablehead{
\colhead{ID}&
\colhead{$z_b$\tablenotemark{a}}&
\colhead{$t_b$\tablenotemark{b}}&
\colhead{ODDS\tablenotemark{c}}&
\colhead{$\chi^{2~}$\tablenotemark{d}}&
\colhead{$\chi^2_{mod}$\tablenotemark{e}}&
\colhead{$z_b1$\tablenotemark{f}}&
\colhead{$t_b1$\tablenotemark{b}}&
\colhead{ODDS1\tablenotemark{g}}&
\colhead{$z_b2$\tablenotemark{f}}&
\colhead{$t_b2$\tablenotemark{b}}&
\colhead{ODDS2\tablenotemark{g}}&
\colhead{$z_b3$\tablenotemark{f}}&
\colhead{$t_b3$\tablenotemark{b}}&
\colhead{ODDS3\tablenotemark{g}}
}
\startdata
1&
$ 0.48\pm0.17$&
3.67&
1.000&
  2.429&
  0.087&
\ldots&
\ldots&
\ldots&
\ldots&
\ldots&
\ldots&
\ldots&
\ldots&
\ldots\\
2*&
$ 2.71^{+0.49}_{-2.45}$&
6.00&
0.500&
  0.118&
  0.669&
$ 2.71^{+0.82}_{-0.55}$&
6.00&
 0.584&
$ 0.35^{+0.47}_{-0.24}$&
 6.67&
 0.095&
$ 1.81^{+0.35}_{-0.89}$&
 4.00&
 0.315\\
3*&
$ 1.29^{+1.31}_{-1.03}$&
7.33&
0.359&
  0.147&
  0.689&
$ 1.29^{+1.78}_{-0.49}$&
7.33&
 0.873&
$ 0.55^{+0.25}_{-0.54}$&
 7.67&
 0.127&
\ldots&
\ldots&
\ldots\\
4&
$ 3.80^{+0.56}_{-0.87}$&
7.00&
0.945&
  0.079&
  0.297&
$ 3.80^{+0.52}_{-1.10}$&
7.00&
 0.984&
$ 0.32^{+0.20}_{-0.16}$&
 3.67&
 0.016&
\ldots&
\ldots&
\ldots\\
5&
$ 0.46^{+2.78}_{-0.30}$&
6.00&
0.592&
  0.586&
  0.431&
$ 0.46^{+0.12}_{-0.08}$&
6.00&
 0.590&
$ 0.22^{+0.08}_{-0.12}$&
 5.00&
 0.180&
$ 3.18^{+0.15}_{-0.21}$&
 5.00&
 0.185
\enddata
\tablecomments{%
Table \ref{cap:catBPZ} is published in its entirety in the electronic version of
the Astronomical Journal.
A portion is shown here for guidance regarding its form and content.
The new version of {\tt BPZ} summarizes each galaxy's redshift probability distribution $P(z)$
by giving the three highest peaks, where warranted.
Here, galaxy \#1 is well fit to a single redshift $z_b=0.48\pm0.17$ with {\tt ODDS}=1.0 and $\chi^2_{mod}=0.087$.
Galaxy \#2{*} instead may be anywhere between $ 2.71^{+0.49}_{-2.45}$ (95\% confidence limits).
The three most likely redshifts for galaxy \#2{*} are given 
along with the redshift ranges for each peak and the fractions of $P(z)$ within those ranges.
Due to space limitations, the last two columns of the table are not shown: $z_{ML}$ \& $t_{ML}$, 
the maximum-likelihood redshift and SED fit.
}
\tablenotetext{a}{Most likely redshift and 95\% confidence interval.}
\tablenotetext{b}{SED fit: 1=El, 8=25Myr (Fig. \ref{cap:SEDs}).}
\tablenotetext{c}{$P(z)$ contained within $0.12(1+z_b)$.}
\tablenotetext{d}{Poorness of {\tt BPZ} fit: observed vs. model fluxes.}
\tablenotetext{e}{Modified $\chi^2$: model fluxes given error bars.}
\tablenotetext{f}{Top three most likely redshifts and ranges.}
\tablenotetext{g}{$P(z)$ contained within the redshift range of each peak.}
\end{deluxetable}

\clearpage

\subsection{Faint Blue Galaxy SEDs}
\label{sub:faintblue}

The SED template library of \citet{BPZ00} includes six templates
for photometric redshifts, namely the \citet{CWW80}
templates (used, for example, by \citetalias{FLY99} 
in their analysis of the HDF-N), plus two starburst templates
from \citet{Kinney96}. These starburst templates were added to accommodate
a population of ``faint blue'' galaxies revealed in the HDF-N.
The addition of these templates significantly improved the accuracy
of the photometric redshifts measured in the HDF-N \citep{BPZ00}.

\begin{figure}
\plotone{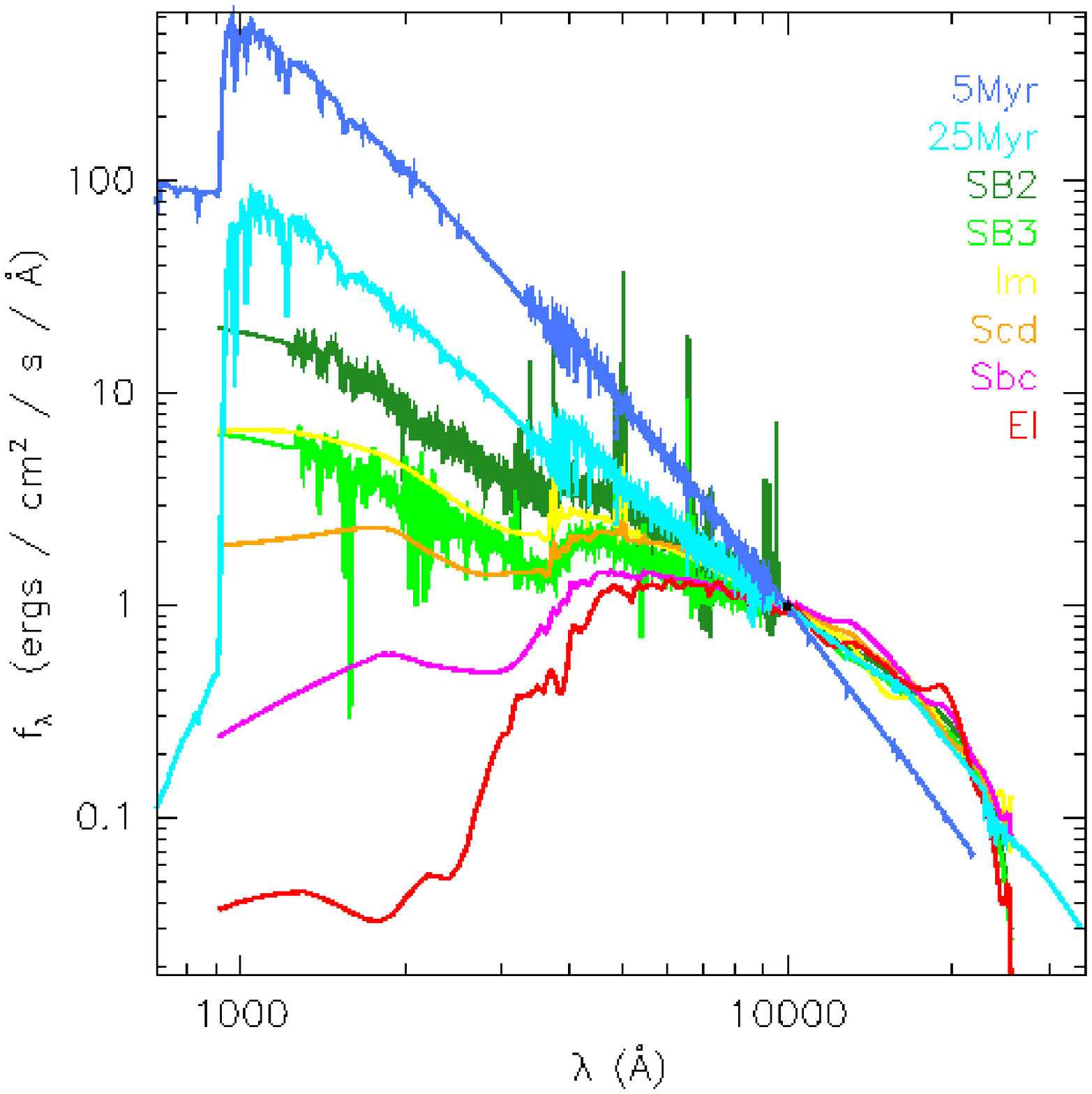}
\caption{\label{cap:SEDs}SED template set used with {\tt BPZ}
in this paper. All SEDs are normalized to $F_{\lambda}=1$ at $\lambda=10,000\textrm{\textrm{\AA}}$.
The bottom 6 are from \citet{Benitez04}. They are modified versions of
the ``CWW+SB'' templates: El, Sbc, Scd, \& Im from \citet{CWW80}
and SB3 \& SB2 starburst galaxies from \citet{Kinney96}.
The steep ({}``blue'') 25Myr \& 5Myr {}``SSP'' SEDs \citep{BC03}
have been added to accommodate the large population of faint blue
galaxies observed in the UDF. 
Between each set of adjacent templates, we interpolate 
an additional two (not shown).}
\end{figure}

The vast majority of galaxies in the HDF-N catalog \citepalias{FLY99}
can be roughly fit to one of these six templates
(hereafter CWW+SB, Fig.~\ref{cap:SEDs}). However, there are systematic
differences between the observed and predicted colors of galaxies
not only in the HDF-N catalog, but also in other spectroscopic catalogs.
This issue was addressed in \citet{Benitez04}. The shapes of the
CWW+SB templates were re-calibrated to more accurately reflect observed
galaxy colors.

\begin{figure}
\plotone{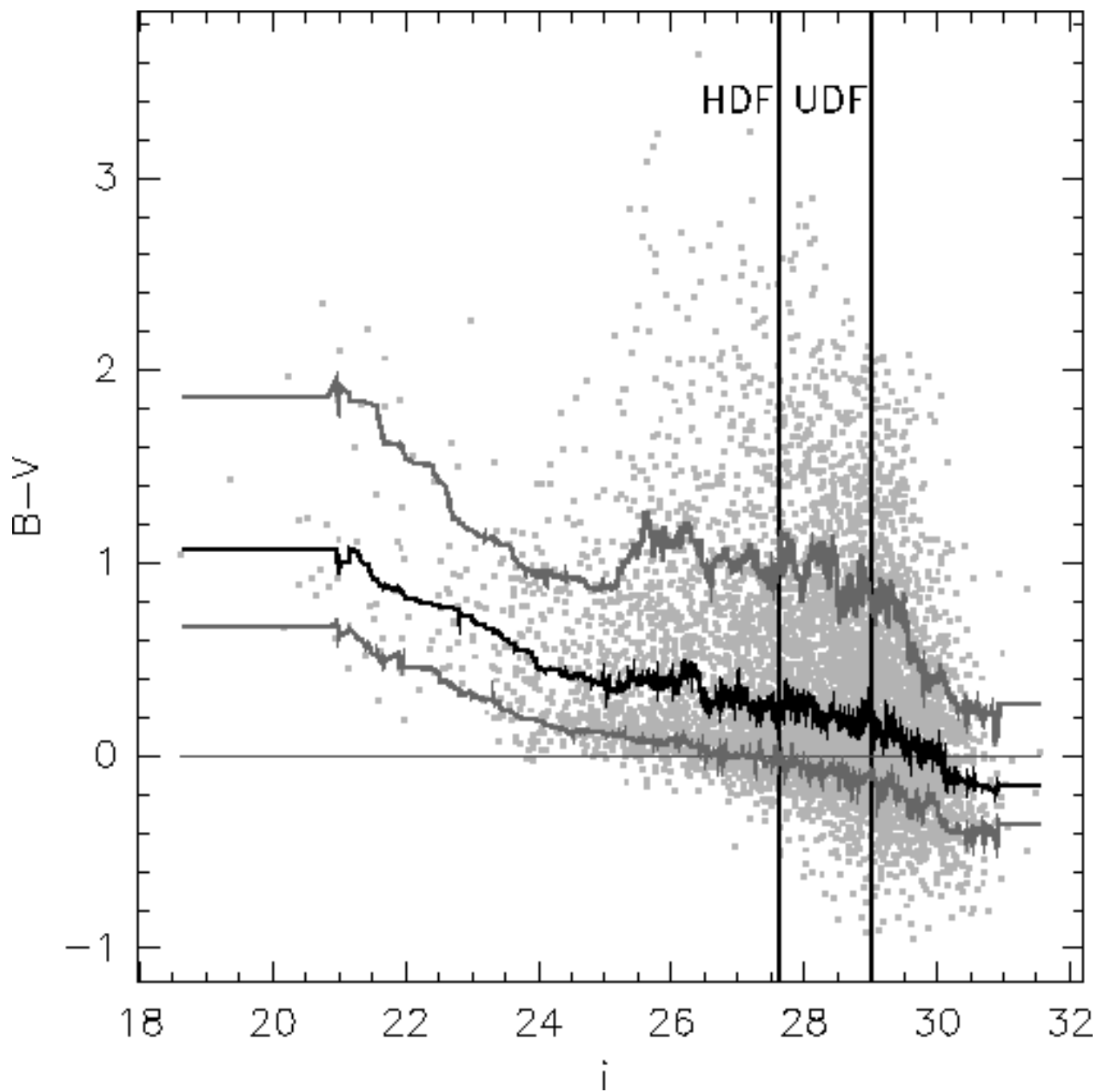}
\caption{\label{cap:BVi}$B-V$ vs.~$i\arcmin$ for galaxies
(${\tt stellarity}<0.8$) in our 10-$\sigma$ catalog. 
The red line is a moving average
(median) of 200 galaxies (or as few as 25 at the edges), while the
magenta lines contain 68\% (1-$\sigma$) of the galaxies. The vertical
lines indicate the 10-$\sigma$ detection limits for the HDF and UDF
in the $i\arcmin$-band (0.5sq$\arcsec$ aperture). As we probe to
fainter magnitudes, we encounter bluer galaxies. }
\end{figure}

\begin{figure*}
\plottwo{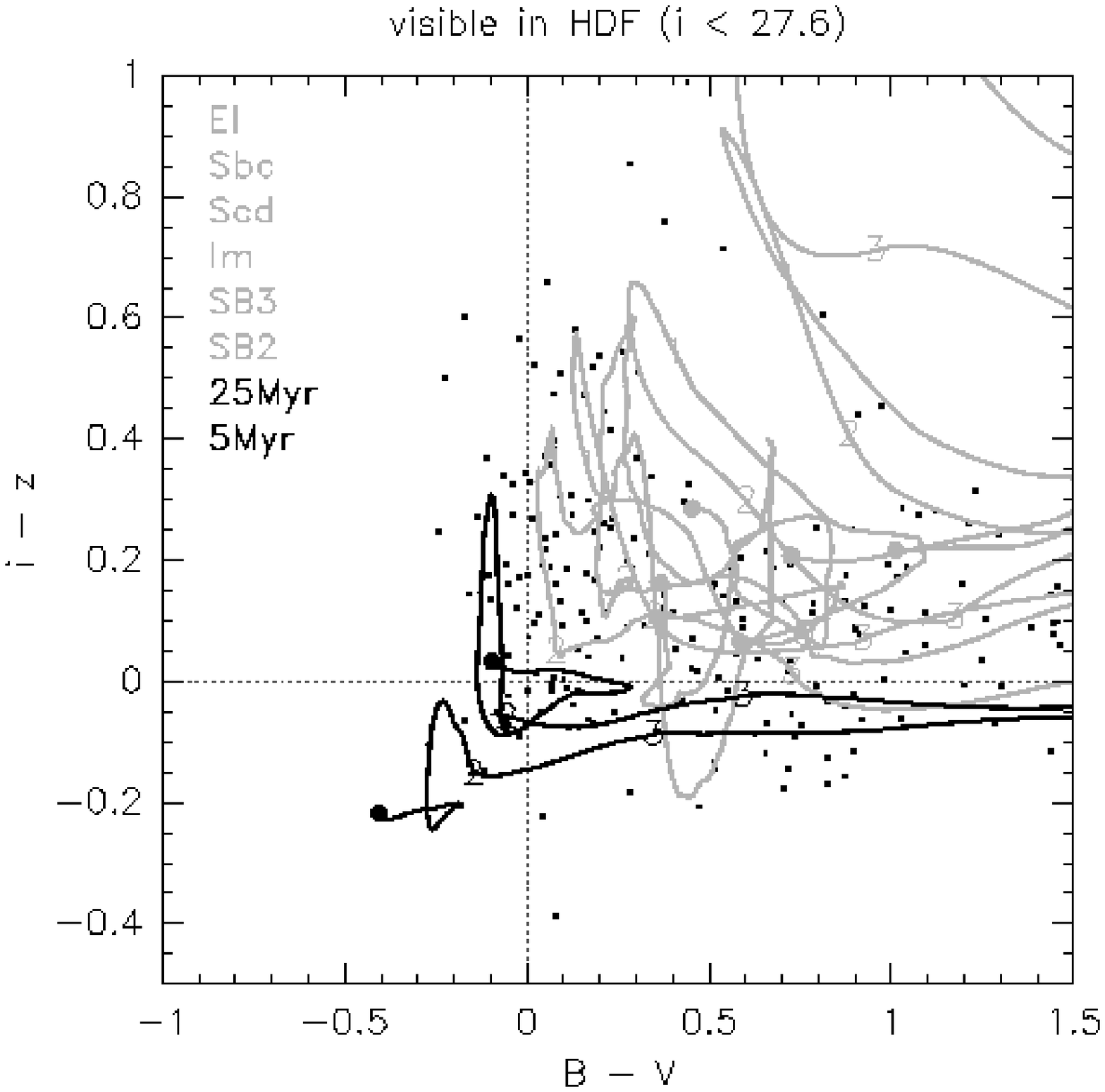}{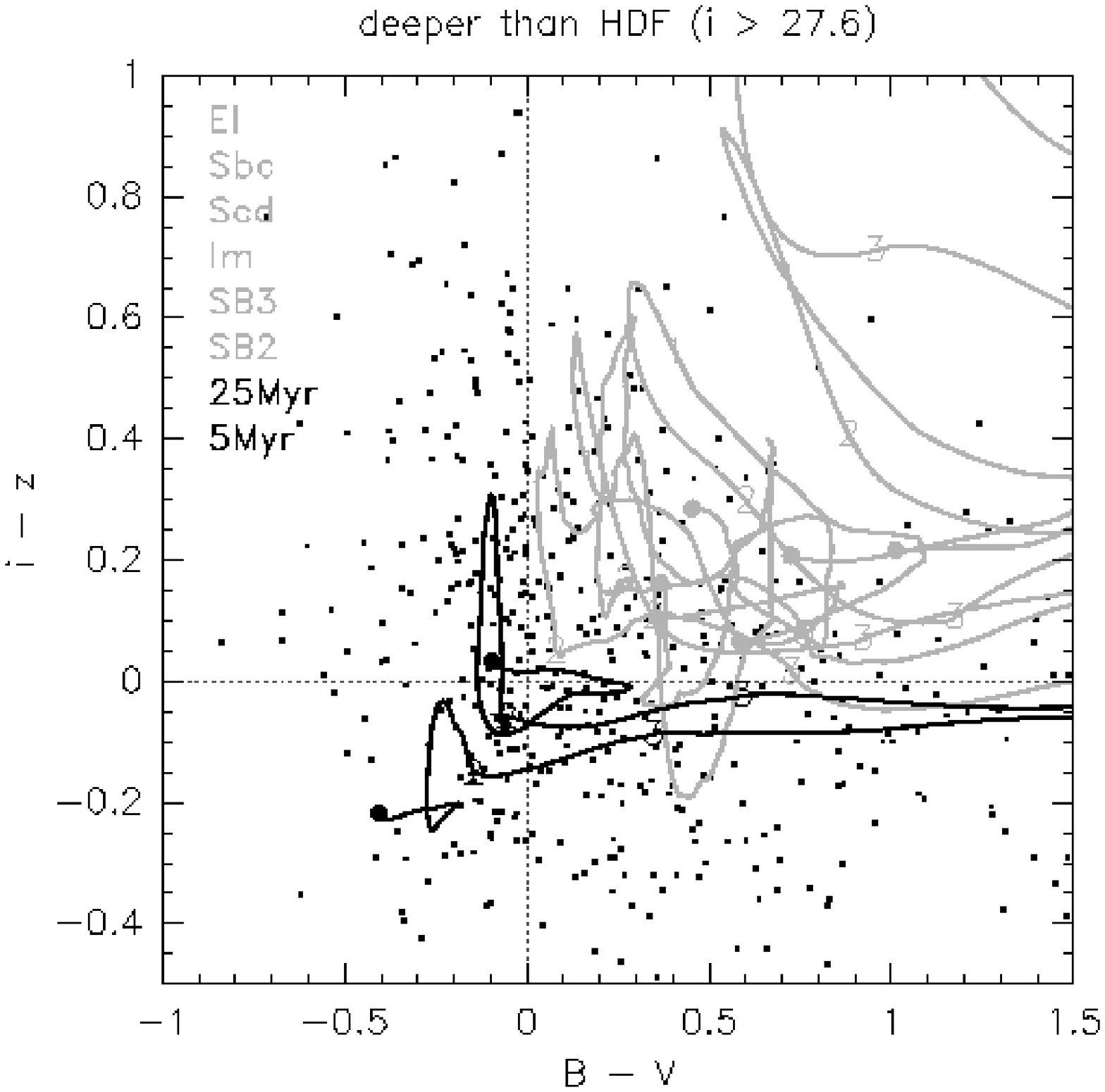}
\caption{\label{cap:color-color}Color-color tracks for our SED templates
plotted against observed colors of galaxies brighter (left) and fainter
(right) than the HDF $i\arcmin$-band detection limit $i\arcmin=27.6$.
For clarity, only 1/10 of the galaxies are plotted (small black squares).
Each template's color-color track begins with a colored circle at
$z=0$, and numbers along the track indicate other redshifts (most
of these numbers are lost in the clutter of the plots). The young
starburst \citetalias{BC03} templates (25Myr \& 5Myr) are required to fit the colors
of the faint blue galaxy population revealed in the UDF.
(These templates also slightly improve photo-z determinations in the HDF.)}
\end{figure*}

But with the increased depth of the UDF, we have discovered a large
population of galaxies even {}``bluer'' than those observed in the
HDF-N (Fig.~\ref{cap:BVi}), and bluer than any of the (re-calibrated)
CWW+SB templates (see Fig.~\ref{cap:color-color}). We are compelled
to add SED templates to fit these galaxies.

\defcitealias{BC03}{BC03}

This time we turn to {\tt GALAXEV}, the synthetic template
set produced and released by 
\citet[hereafter \citetalias{BC03}]{BC03}.
The simple stellar population {}``SSP'' models of \citetalias{BC03} span ages
from 5Myr to 12Gyr and have metallicities of $Z=0.08$, 0.2, and 0.5
(i.e., $Z=0.4Z_{\odot}$, $Z_{\odot}$, and $2.5Z_{\odot}$). 

We experiment with the \citetalias{BC03} templates using the extensive spectroscopic
redshift library of 1,800+ galaxies in the GOODS-N field 
\citep{Cowie04,Wirth04}. High quality ACS $BVi\arcmin z\arcmin$
photometry for these galaxies is available via \\ 
{\tt http://www.stsci.edu/science/goods/}
\citep{GOODS}. We first note that 3\% of these galaxies
are in fact bluer than the CWW+SB templates.  When we run {\tt BPZ}
on the GOODS-N photometry with the CWW+SB templates, our photometric
redshifts match the spectroscopic redshifts with an RMS of $\Delta z=0.06(1+z_{spec})$.

We then add \citetalias{BC03} templates to our CWW+SB template set one at a time
to see if the accuracy and reliability improve. We also note how {}``popular''
a given template is, i.e. how many galaxies {}``choose'' the template
as their best fit over the other six CWW+SB templates. The prior assigned
to the template is exactly the same as that applied to our two SB
templates. We set {\tt INTERP=2}, so that two templates are interpolated
between each set of adjacent templates.

The {}``best'' template is the 5Myr old $Z=0.08=0.4Z_{\odot}$ SSP
template. The addition of this template improves the accuracy of the
photo-z's to an RMS of $\Delta z=0.04(1+z_{spec})$. The {}``second
best'' template is the 25Myr old $Z=0.08$ SSP template. Both of
these templates are bluer than the CWW+SB template set. Adding more
templates does not improve the results, in fact it slightly worsens
them \citetext{see \citealt{BPZ00} for a discussion about the risks and meager
benefits of including larger number of templates in the spectral library}.
Therefore, we decide we will incorporate both of these templates into
our {\tt BPZ} analysis of the UDF, with the reasonable
expectation that they will describe a significant fraction of the
very blue galaxy population. Our final set of 8 templates is shown
in Fig.~\ref{cap:SEDs}.

We expect our UDF {\tt BPZ} accuracy to be $\Delta z=0.04(1+z_{spec})$
or better, given the results obtained with the GOODS-N field, which
do not include near-IR photometry. Our GOODS-N {\tt BPZ}
results selected for {\tt ODDS} $\geq0.99$ has a catastrophic
error rate of $<1\%$, and that includes objects misclassified because
they have AGN spectra. Of course it would be unwarranted to extend
this statistic to magnitudes much fainter than the spectroscopic redshift
limit, but it gives a good idea of the robustness of the {\tt BPZ}
results.

Unfortunately, a proper analysis of the nature of these faint blue galaxies 
is beyond the scope of this paper.
Their redshift distribution is given in \S\ref{sub:bpzhist}
and their contribution to the star formation rate density over time
will be discussed in Paper II.

\subsection{Comparison with Spectroscopic Redshifts
and COMBO-17}
\label{sub:BPZ_vs_spec-z}

\subsubsection{Spectroscopic Redshift Catalog}
\label{subsub:spec-z}

Alessandro Rettura at ESO has compiled a list of all of the publicly
available spectroscopic redshifts within the CDF-S.%
\footnote{{\tt http://www.eso.org/science/goods/spectroscopy/\\CDFS\_Mastercat/}} 
76 of these galaxies (and 3 stars) fall within the UDF ACS FOV.
\footnote{Another 3 galaxies are either on or near the edge of the ACS FOV.
These yield magnitudes in only 2 ACS filters, and we discard them
in this paper.}
The GOODS VLT/FORS2 survey \citep{FORS2} obtained 22 of
these redshifts. 20 of these are considered {}``solid'' or {}``likely''
(quality flags {}``A'' or {}``B''). 
The VIMOS VLT Deep Survey (hereafter, VVDS; \citealt{VVDS})
contributes another 41 redshifts,
25 of these being assigned 95\% or 100\% confidence. 7 redshifts come
from \citet{Szokoly04}. 4 of these are deemed {}``reliable''
(quality flags {}``2'' or {}``2+''). These 3 surveys yield 20+25+4=49
{}``confident'' redshifts. To those 49, we add 6 redshifts which
were not assigned confidence levels: 5 obtained by \citet{Croom01}
and a $z=1.30$ type Ia supernova named {}``Aphrodite'' by \citet{Strolger04}.
Our final catalog contains 55 spectroscopic redshifts
shown in Table \ref{cap:zspec} along with our photometry measurements and
{\tt BPZ} results.

\clearpage

\begin{deluxetable}{ccccccccccccc}
\tabletypesize{\scriptsize}
\rotate
\tablewidth{0pt}
\tablecaption{\label{cap:zspec}Galaxies with confident spectroscopic redshifts in the UDF}
\tablehead{
\colhead{{\tiny ID}}&
\multicolumn{2}{c}{{\tiny RA \& DEC (J2000)}}&
\colhead{{\tiny Survey}}&
\colhead{{\tiny $z_{spec}$}}&
\colhead{{\tiny $z_b$}}&
\colhead{{\tiny $\chi^2_{mod}$}}&
\colhead{{\tiny $B_{435}$}}&
\colhead{{\tiny $V_{606}$}}&
\colhead{{\tiny $i\arcmin_{775}$}}&
\colhead{{\tiny $z\arcmin_{850}$}}&
\colhead{{\tiny $J_{110}$}}&
\colhead{{\tiny $H_{160}$}}
}
\startdata
{\tiny 3088*}&
{\tiny 03:32:36.432}&
{\tiny -27:47:50.64}&
{\tiny FORS2}&
{\tiny $0.127$}&
{\tiny $ 0.19^{+0.25}_{-0.14}$}&
{\tiny 0.12}&
{\tiny $23.68\pm0.01$}&
{\tiny $22.85\pm0.00$}&
{\tiny $22.48\pm0.00$}&
{\tiny $22.37\pm0.00$}&
{\tiny $22.25\pm0.05$}&
{\tiny $22.08\pm0.06$}\\
{\tiny 57290}&
{\tiny 03:32:42.576}&
{\tiny -27:45:50.04}&
{\tiny FORS2}&
{\tiny $0.218$}&
{\tiny $ 0.30^{+0.15}_{-0.21}$}&
{\tiny 0.00}&
{\tiny $23.59\pm0.01$}&
{\tiny \ldots}&
{\tiny $22.36\pm0.00$}&
{\tiny $22.19\pm0.00$}&
{\tiny \ldots}&
{\tiny \ldots}\\
{\tiny 1375}&
{\tiny 03:32:33.000}&
{\tiny -27:48:29.52}&
{\tiny FORS2}&
{\tiny $0.664$}&
{\tiny $ 0.58\pm0.19$}&
{\tiny 0.16}&
{\tiny $24.27\pm0.01$}&
{\tiny $23.64\pm0.00$}&
{\tiny $23.01\pm0.00$}&
{\tiny $22.89\pm0.00$}&
{\tiny \ldots}&
{\tiny \ldots}\\
{\tiny 8810}&
{\tiny 03:32:37.248}&
{\tiny -27:46:10.20}&
{\tiny FORS2}&
{\tiny $0.736$}&
{\tiny $ 0.72\pm0.20$}&
{\tiny 2.48}&
{\tiny $23.65\pm0.01$}&
{\tiny $23.10\pm0.00$}&
{\tiny $22.38\pm0.00$}&
{\tiny $22.11\pm0.00$}&
{\tiny $21.81\pm0.05$}&
{\tiny $21.45\pm0.06$}\\
{\tiny 4142*}&
{\tiny 03:32:44.208}&
{\tiny -27:47:33.36}&
{\tiny FORS2}&
{\tiny $0.737$}&
{\tiny $ 0.67\pm0.20$}&
{\tiny 0.16}&
{\tiny $22.93\pm0.00$}&
{\tiny $22.32\pm0.00$}&
{\tiny $21.66\pm0.00$}&
{\tiny $21.45\pm0.00$}&
{\tiny $21.21\pm0.05$}&
{\tiny $20.95\pm0.06$}\\
{\tiny 6206}&
{\tiny 03:32:38.496}&
{\tiny -27:47:02.40}&
{\tiny FORS2}&
{\tiny $0.954$}&
{\tiny $ 0.92\pm0.23$}&
{\tiny 0.02}&
{\tiny $24.70\pm0.02$}&
{\tiny $23.40\pm0.00$}&
{\tiny $21.94\pm0.00$}&
{\tiny $21.20\pm0.00$}&
{\tiny $20.71\pm0.05$}&
{\tiny $20.10\pm0.06$}\\
{\tiny 153*}&
{\tiny 03:32:39.600}&
{\tiny -27:49:09.48}&
{\tiny FORS2}&
{\tiny $0.980$}&
{\tiny $ 0.88\pm0.22$}&
{\tiny 0.03}&
{\tiny $25.06\pm0.04$}&
{\tiny $22.85\pm0.00$}&
{\tiny $21.43\pm0.00$}&
{\tiny $20.57\pm0.00$}&
{\tiny \ldots}&
{\tiny \ldots}\\
{\tiny 8261}&
{\tiny 03:32:35.784}&
{\tiny -27:46:27.48}&
{\tiny FORS2}&
{\tiny $1.094$}&
{\tiny $ 1.02\pm0.24$}&
{\tiny 0.04}&
{\tiny $25.75\pm0.03$}&
{\tiny $24.65\pm0.01$}&
{\tiny $23.53\pm0.00$}&
{\tiny $22.67\pm0.00$}&
{\tiny \ldots}&
{\tiny \ldots}\\
{\tiny 9264*}&
{\tiny 03:32:37.200}&
{\tiny -27:46:08.04}&
{\tiny FORS2}&
{\tiny $1.096$}&
{\tiny $ 1.17\pm0.26$}&
{\tiny 0.00}&
{\tiny $25.26\pm0.04$}&
{\tiny $23.41\pm0.01$}&
{\tiny $21.91\pm0.00$}&
{\tiny $20.84\pm0.00$}&
{\tiny \ldots}&
{\tiny \ldots}\\
{\tiny 8749}&
{\tiny 03:32:34.848}&
{\tiny -27:46:40.44}&
{\tiny FORS2}&
{\tiny $1.099$}&
{\tiny $ 0.82\pm0.21$}&
{\tiny 0.54}&
{\tiny $25.14\pm0.01$}&
{\tiny $24.21\pm0.00$}&
{\tiny $23.33\pm0.00$}&
{\tiny $22.71\pm0.00$}&
{\tiny \ldots}&
{\tiny \ldots}\\
{\tiny 4816*}&
{\tiny 03:32:44.184}&
{\tiny -27:47:29.40}&
{\tiny FORS2}&
{\tiny $1.220$}&
{\tiny $ 1.40\pm0.28$}&
{\tiny 0.52}&
{\tiny $24.93\pm0.02$}&
{\tiny $24.51\pm0.01$}&
{\tiny $23.97\pm0.00$}&
{\tiny $23.43\pm0.01$}&
{\tiny $22.88\pm0.06$}&
{\tiny $22.16\pm0.07$}\\
{\tiny 4396*}&
{\tiny 03:32:35.784}&
{\tiny -27:47:34.80}&
{\tiny FORS2}&
{\tiny $1.223$}&
{\tiny $ 1.26\pm0.27$}&
{\tiny 0.16}&
{\tiny $25.76\pm0.04$}&
{\tiny $25.34\pm0.02$}&
{\tiny $24.36\pm0.01$}&
{\tiny $23.57\pm0.01$}&
{\tiny $22.82\pm0.05$}&
{\tiny $22.11\pm0.07$}\\
{\tiny 1829}&
{\tiny 03:32:40.920}&
{\tiny -27:48:23.76}&
{\tiny FORS2}&
{\tiny $1.244$}&
{\tiny $ 1.29\pm0.27$}&
{\tiny 3.29}&
{\tiny $25.47\pm0.01$}&
{\tiny $25.34\pm0.01$}&
{\tiny $25.07\pm0.01$}&
{\tiny $24.45\pm0.01$}&
{\tiny $24.38\pm0.10$}&
{\tiny $24.22\pm0.12$}\\
{\tiny 1266}&
{\tiny 03:32:34.824}&
{\tiny -27:48:35.64}&
{\tiny FORS2}&
{\tiny $1.245$}&
{\tiny $ 1.40\pm0.28$}&
{\tiny 0.01}&
{\tiny $24.67\pm0.01$}&
{\tiny $24.25\pm0.00$}&
{\tiny $23.65\pm0.00$}&
{\tiny $22.97\pm0.00$}&
{\tiny \ldots}&
{\tiny \ldots}\\
{\tiny 7995}&
{\tiny 03:32:42.264}&
{\tiny -27:46:25.32}&
{\tiny FORS2}&
{\tiny $1.288$}&
{\tiny $ 1.26\pm0.27$}&
{\tiny 0.04}&
{\tiny $23.86\pm0.01$}&
{\tiny $23.58\pm0.00$}&
{\tiny $23.19\pm0.00$}&
{\tiny $22.53\pm0.00$}&
{\tiny \ldots}&
{\tiny \ldots}\\
{\tiny 6188*}&
{\tiny 03:32:42.384}&
{\tiny -27:47:07.80}&
{\tiny FORS2}&
{\tiny $1.314$}&
{\tiny $ 1.15\pm0.25$}&
{\tiny 3.62}&
{\tiny $26.57\pm0.09$}&
{\tiny $25.34\pm0.02$}&
{\tiny $24.14\pm0.01$}&
{\tiny $23.07\pm0.00$}&
{\tiny $22.18\pm0.05$}&
{\tiny $21.35\pm0.06$}\\
{\tiny 7725}&
{\tiny 03:32:35.088}&
{\tiny -27:46:15.60}&
{\tiny FORS2}&
{\tiny $1.316$}&
{\tiny $ 1.31\pm0.27$}&
{\tiny 0.00}&
{\tiny $24.35\pm0.01$}&
{\tiny $24.11\pm0.00$}&
{\tiny $23.73\pm0.00$}&
{\tiny $23.11\pm0.00$}&
{\tiny \ldots}&
{\tiny \ldots}\\
{\tiny 6027*}&
{\tiny 03:32:39.648}&
{\tiny -27:47:09.24}&
{\tiny FORS2}&
{\tiny $1.317$}&
{\tiny $ 1.17\pm0.26$}&
{\tiny 0.31}&
{\tiny $26.01\pm0.06$}&
{\tiny $24.84\pm0.01$}&
{\tiny $23.63\pm0.00$}&
{\tiny $22.68\pm0.00$}&
{\tiny $21.80\pm0.05$}&
{\tiny $21.10\pm0.06$}\\
{\tiny 8461}&
{\tiny 03:32:44.616}&
{\tiny -27:46:32.16}&
{\tiny FORS2}&
{\tiny $1.426$}&
{\tiny $ 1.08^{+0.47}_{-0.24}$}&
{\tiny 0.04}&
{\tiny $24.39\pm0.01$}&
{\tiny $24.10\pm0.00$}&
{\tiny $23.70\pm0.00$}&
{\tiny $23.22\pm0.00$}&
{\tiny \ldots}&
{\tiny \ldots}\\
{\tiny 2225*}&
{\tiny 03:32:40.008}&
{\tiny -27:48:15.12}&
{\tiny FORS2}&
{\tiny $5.820$}&
{\tiny $ 5.76\pm0.80$}&
{\tiny 0.13}&
{\tiny $>30.41$}&
{\tiny $29.34\pm0.25$}&
{\tiny $26.69\pm0.03$}&
{\tiny $25.11\pm0.01$}&
{\tiny $25.09\pm0.09$}&
{\tiny $25.19\pm0.10$}\\
{\tiny 5670}&
{\tiny 03:32:46.536}&
{\tiny -27:47:08.88}&
{\tiny VVDS}&
{\tiny $0.128$}&
{\tiny $ 0.23^{+0.17}_{-0.14}$}&
{\tiny 0.00}&
{\tiny $22.11\pm0.00$}&
{\tiny $21.23\pm0.00$}&
{\tiny $20.84\pm0.00$}&
{\tiny $20.69\pm0.00$}&
{\tiny \ldots}&
{\tiny \ldots}\\
{\tiny 1971*}&
{\tiny 03:32:41.928}&
{\tiny -27:47:57.48}&
{\tiny VVDS}&
{\tiny $0.151$}&
{\tiny $ 0.17\pm0.14$}&
{\tiny 0.22}&
{\tiny $21.12\pm0.00$}&
{\tiny $20.46\pm0.00$}&
{\tiny $20.18\pm0.00$}&
{\tiny $20.09\pm0.00$}&
{\tiny $19.98\pm0.05$}&
{\tiny $19.80\pm0.06$}\\
{\tiny 5620}&
{\tiny 03:32:43.560}&
{\tiny -27:47:16.80}&
{\tiny VVDS}&
{\tiny $0.212$}&
{\tiny $ 0.22\pm0.14$}&
{\tiny 0.86}&
{\tiny $23.87\pm0.00$}&
{\tiny $23.42\pm0.00$}&
{\tiny $23.34\pm0.00$}&
{\tiny $23.41\pm0.00$}&
{\tiny $23.28\pm0.07$}&
{\tiny $23.34\pm0.08$}\\
{\tiny 1000}&
{\tiny 03:32:36.744}&
{\tiny -27:48:43.56}&
{\tiny VVDS}&
{\tiny $0.213$}&
{\tiny $ 3.13^{+0.49}_{-2.95}$}&
{\tiny 1.23}&
{\tiny $23.86\pm0.00$}&
{\tiny $23.39\pm0.00$}&
{\tiny $23.30\pm0.00$}&
{\tiny $23.40\pm0.00$}&
{\tiny \ldots}&
{\tiny \ldots}\\
{\tiny 5606}&
{\tiny 03:32:34.104}&
{\tiny -27:47:12.12}&
{\tiny VVDS}&
{\tiny $0.226$}&
{\tiny $ 0.17\pm0.14$}&
{\tiny 0.06}&
{\tiny $22.11\pm0.00$}&
{\tiny $21.14\pm0.00$}&
{\tiny $20.73\pm0.00$}&
{\tiny $20.59\pm0.00$}&
{\tiny $20.36\pm0.05$}&
{\tiny $20.10\pm0.06$}\\
{\tiny 5190}&
{\tiny 03:32:34.824}&
{\tiny -27:47:21.84}&
{\tiny VVDS}&
{\tiny $0.315$}&
{\tiny $ 1.23\pm0.26$}&
{\tiny 0.30}&
{\tiny $24.16\pm0.01$}&
{\tiny $23.97\pm0.00$}&
{\tiny $23.67\pm0.00$}&
{\tiny $23.13\pm0.00$}&
{\tiny $22.88\pm0.07$}&
{\tiny $22.56\pm0.08$}\\
{\tiny 7847}&
{\tiny 03:32:41.760}&
{\tiny -27:46:19.56}&
{\tiny VVDS}&
{\tiny $0.334$}&
{\tiny $ 0.38\pm0.16$}&
{\tiny 0.02}&
{\tiny $23.60\pm0.01$}&
{\tiny $22.00\pm0.00$}&
{\tiny $21.25\pm0.00$}&
{\tiny $20.90\pm0.00$}&
{\tiny \ldots}&
{\tiny \ldots}\\
{\tiny 3492}&
{\tiny 03:32:45.072}&
{\tiny -27:47:38.40}&
{\tiny VVDS}&
{\tiny $0.345$}&
{\tiny $ 0.29\pm0.15$}&
{\tiny 0.67}&
{\tiny $21.75\pm0.00$}&
{\tiny $20.83\pm0.00$}&
{\tiny $20.58\pm0.00$}&
{\tiny $20.39\pm0.00$}&
{\tiny $20.32\pm0.05$}&
{\tiny $20.13\pm0.06$}\\
{\tiny 4267*}&
{\tiny 03:32:48.336}&
{\tiny -27:47:38.76}&
{\tiny VVDS}&
{\tiny $0.347$}&
{\tiny $ 3.15\pm0.49$}&
{\tiny 0.06}&
{\tiny $25.47\pm0.02$}&
{\tiny $24.63\pm0.01$}&
{\tiny $24.45\pm0.01$}&
{\tiny $24.42\pm0.01$}&
{\tiny \ldots}&
{\tiny \ldots}\\
{\tiny 3268}&
{\tiny 03:32:41.400}&
{\tiny -27:47:47.04}&
{\tiny VVDS}&
{\tiny $0.347$}&
{\tiny $ 0.30\pm0.15$}&
{\tiny 0.36}&
{\tiny $22.96\pm0.00$}&
{\tiny $22.11\pm0.00$}&
{\tiny $21.84\pm0.00$}&
{\tiny $21.66\pm0.00$}&
{\tiny $21.56\pm0.05$}&
{\tiny $21.37\pm0.06$}\\
{\tiny 8585*}&
{\tiny 03:32:35.496}&
{\tiny -27:46:27.12}&
{\tiny VVDS}&
{\tiny $0.377$}&
{\tiny $ 1.00\pm0.24$}&
{\tiny 0.00}&
{\tiny $22.36\pm0.00$}&
{\tiny $22.07\pm0.00$}&
{\tiny $21.58\pm0.00$}&
{\tiny $21.11\pm0.00$}&
{\tiny \ldots}&
{\tiny \ldots}\\
{\tiny 900*}&
{\tiny 03:32:44.448}&
{\tiny -27:48:19.08}&
{\tiny VVDS}&
{\tiny $0.417$}&
{\tiny $ 0.43\pm0.17$}&
{\tiny 0.06}&
{\tiny $22.29\pm0.00$}&
{\tiny $21.07\pm0.00$}&
{\tiny $20.39\pm0.00$}&
{\tiny $20.06\pm0.00$}&
{\tiny \ldots}&
{\tiny \ldots}\\
{\tiny 4929}&
{\tiny 03:32:45.120}&
{\tiny -27:47:24.00}&
{\tiny VVDS}&
{\tiny $0.436$}&
{\tiny $ 0.50\pm0.18$}&
{\tiny 1.43}&
{\tiny $22.73\pm0.00$}&
{\tiny $21.53\pm0.00$}&
{\tiny $20.82\pm0.00$}&
{\tiny $20.43\pm0.00$}&
{\tiny $20.09\pm0.05$}&
{\tiny $19.57\pm0.06$}\\
{\tiny 2107}&
{\tiny 03:32:45.792}&
{\tiny -27:48:12.96}&
{\tiny VVDS}&
{\tiny $0.534$}&
{\tiny $ 0.56\pm0.18$}&
{\tiny 0.02}&
{\tiny $24.38\pm0.01$}&
{\tiny $22.76\pm0.00$}&
{\tiny $21.75\pm0.00$}&
{\tiny $21.36\pm0.00$}&
{\tiny \ldots}&
{\tiny \ldots}\\
{\tiny 6747*}&
{\tiny 03:32:38.784}&
{\tiny -27:46:48.72}&
{\tiny VVDS}&
{\tiny $0.619$}&
{\tiny $ 0.56\pm0.18$}&
{\tiny 0.82}&
{\tiny $25.01\pm0.03$}&
{\tiny $22.95\pm0.00$}&
{\tiny $21.67\pm0.00$}&
{\tiny $21.22\pm0.00$}&
{\tiny $20.75\pm0.05$}&
{\tiny $20.21\pm0.06$}\\
{\tiny 2607}&
{\tiny 03:32:43.248}&
{\tiny -27:47:56.04}&
{\tiny VVDS}&
{\tiny $0.666$}&
{\tiny $ 0.63\pm0.19$}&
{\tiny 1.10}&
{\tiny $23.06\pm0.01$}&
{\tiny $21.94\pm0.00$}&
{\tiny $20.99\pm0.00$}&
{\tiny $20.66\pm0.00$}&
{\tiny $20.28\pm0.05$}&
{\tiny $19.77\pm0.06$}\\
{\tiny 968}&
{\tiny 03:32:37.536}&
{\tiny -27:48:38.88}&
{\tiny VVDS}&
{\tiny $0.666$}&
{\tiny $ 0.58\pm0.19$}&
{\tiny 0.12}&
{\tiny $22.17\pm0.00$}&
{\tiny $21.56\pm0.00$}&
{\tiny $20.96\pm0.00$}&
{\tiny $20.84\pm0.00$}&
{\tiny $20.63\pm0.05$}&
{\tiny $20.38\pm0.06$}\\
{\tiny 662*}&
{\tiny 03:32:41.880}&
{\tiny -27:48:54.00}&
{\tiny VVDS}&
{\tiny $0.666$}&
{\tiny $ 0.58\pm0.19$}&
{\tiny 0.01}&
{\tiny $23.17\pm0.00$}&
{\tiny $22.56\pm0.00$}&
{\tiny $21.96\pm0.00$}&
{\tiny $21.85\pm0.00$}&
{\tiny \ldots}&
{\tiny \ldots}\\
{\tiny 355}&
{\tiny 03:32:38.808}&
{\tiny -27:49:09.48}&
{\tiny VVDS}&
{\tiny $0.666$}&
{\tiny $ 0.60\pm0.19$}&
{\tiny 0.17}&
{\tiny $24.49\pm0.01$}&
{\tiny $23.69\pm0.00$}&
{\tiny $22.92\pm0.00$}&
{\tiny $22.72\pm0.00$}&
{\tiny \ldots}&
{\tiny \ldots}\\
{\tiny 53380}&
{\tiny 03:32:29.952}&
{\tiny -27:47:57.12}&
{\tiny VVDS}&
{\tiny $0.667$}&
{\tiny $ 0.62\pm0.19$}&
{\tiny 0.07}&
{\tiny $25.50\pm0.04$}&
{\tiny $23.87\pm0.01$}&
{\tiny $22.68\pm0.00$}&
{\tiny $22.19\pm0.00$}&
{\tiny \ldots}&
{\tiny \ldots}\\
{\tiny 6933}&
{\tiny 03:32:33.432}&
{\tiny -27:46:50.52}&
{\tiny VVDS}&
{\tiny $0.733$}&
{\tiny $ 0.61\pm0.19$}&
{\tiny 0.01}&
{\tiny $24.18\pm0.01$}&
{\tiny $23.67\pm0.00$}&
{\tiny $23.05\pm0.00$}&
{\tiny $22.90\pm0.00$}&
{\tiny \ldots}&
{\tiny \ldots}\\
{\tiny 2525}&
{\tiny 03:32:43.584}&
{\tiny -27:48:04.68}&
{\tiny VVDS}&
{\tiny $0.736$}&
{\tiny $ 0.67\pm0.20$}&
{\tiny 0.02}&
{\tiny $24.21\pm0.01$}&
{\tiny $23.52\pm0.00$}&
{\tiny $22.72\pm0.00$}&
{\tiny $22.50\pm0.00$}&
{\tiny $22.31\pm0.06$}&
{\tiny $22.02\pm0.07$}\\
{\tiny 3372*}&
{\tiny 03:32:42.288}&
{\tiny -27:47:45.96}&
{\tiny VVDS}&
{\tiny $0.996$}&
{\tiny $ 0.81\pm0.21$}&
{\tiny 0.60}&
{\tiny $22.86\pm0.00$}&
{\tiny $22.32\pm0.00$}&
{\tiny $21.62\pm0.00$}&
{\tiny $21.23\pm0.00$}&
{\tiny $20.91\pm0.05$}&
{\tiny $20.58\pm0.06$}\\
{\tiny 5417}&
{\tiny 03:32:39.888}&
{\tiny -27:47:15.00}&
{\tiny VVDS}&
{\tiny $1.095$}&
{\tiny $ 0.99\pm0.23$}&
{\tiny 0.29}&
{\tiny $23.03\pm0.00$}&
{\tiny $22.57\pm0.00$}&
{\tiny $21.95\pm0.00$}&
{\tiny $21.44\pm0.00$}&
{\tiny $21.09\pm0.05$}&
{\tiny $20.73\pm0.06$}\\
{\tiny 797*}&
{\tiny 03:32:35.976}&
{\tiny -27:48:50.40}&
{\tiny VVDS}&
{\tiny $1.306$}&
{\tiny $ 1.44\pm0.29$}&
{\tiny 0.00}&
{\tiny $22.37\pm0.00$}&
{\tiny $22.18\pm0.00$}&
{\tiny $21.99\pm0.00$}&
{\tiny $21.57\pm0.00$}&
{\tiny \ldots}&
{\tiny \ldots}\\
{\tiny 4445}&
{\tiny 03:32:38.784}&
{\tiny -27:47:32.28}&
{\tiny Szokoly}&
{\tiny $0.456$}&
{\tiny $ 0.07^{+0.13}_{-0.07}$}&
{\tiny 4.82}&
{\tiny $21.95\pm0.00$}&
{\tiny $21.50\pm0.00$}&
{\tiny $21.15\pm0.00$}&
{\tiny $20.91\pm0.00$}&
{\tiny $20.68\pm0.05$}&
{\tiny $20.33\pm0.06$}\\
{\tiny 4394}&
{\tiny 03:32:31.368}&
{\tiny -27:47:25.08}&
{\tiny Szokoly}&
{\tiny $0.665$}&
{\tiny $ 0.60\pm0.19$}&
{\tiny 0.05}&
{\tiny $22.45\pm0.00$}&
{\tiny $21.78\pm0.00$}&
{\tiny $21.15\pm0.00$}&
{\tiny $21.00\pm0.00$}&
{\tiny \ldots}&
{\tiny \ldots}\\
{\tiny 8275}&
{\tiny 03:32:36.504}&
{\tiny -27:46:29.28}&
{\tiny Szokoly}&
{\tiny $0.764$}&
{\tiny $ 0.70\pm0.20$}&
{\tiny 0.05}&
{\tiny $22.65\pm0.00$}&
{\tiny $22.13\pm0.00$}&
{\tiny $21.40\pm0.00$}&
{\tiny $21.20\pm0.00$}&
{\tiny $20.98\pm0.05$}&
{\tiny $20.75\pm0.06$}\\
{\tiny 865*}&
{\tiny 03:32:39.672}&
{\tiny -27:48:50.76}&
{\tiny Szokoly}&
{\tiny $3.064$}&
{\tiny $ 3.67^{+0.55}_{-3.34}$}&
{\tiny 30.80}&
{\tiny $27.16\pm0.10$}&
{\tiny $25.30\pm0.01$}&
{\tiny $24.57\pm0.01$}&
{\tiny $24.39\pm0.01$}&
{\tiny $23.78\pm0.07$}&
{\tiny $22.32\pm0.08$}\\
{\tiny 8015}&
{\tiny 03:32:33.528}&
{\tiny -27:46:23.52}&
{\tiny Croom}&
{\tiny $0.276$}&
{\tiny $ 0.34^{+0.16}_{-0.23}$}&
{\tiny 0.00}&
{\tiny $22.89\pm0.00$}&
{\tiny $21.78\pm0.00$}&
{\tiny $21.28\pm0.00$}&
{\tiny $21.05\pm0.00$}&
{\tiny \ldots}&
{\tiny \ldots}\\
{\tiny 3822}&
{\tiny 03:32:44.856}&
{\tiny -27:47:27.60}&
{\tiny Croom}&
{\tiny $0.437$}&
{\tiny $ 0.14\pm0.13$}&
{\tiny 0.02}&
{\tiny $20.18\pm0.00$}&
{\tiny $19.14\pm0.00$}&
{\tiny $18.62\pm0.00$}&
{\tiny $18.44\pm0.00$}&
{\tiny $18.09\pm0.05$}&
{\tiny $17.71\pm0.06$}\\
{\tiny 2387*}&
{\tiny 03:32:35.760}&
{\tiny -27:47:58.92}&
{\tiny Croom}&
{\tiny $0.665$}&
{\tiny $ 0.63\pm0.19$}&
{\tiny 0.11}&
{\tiny $24.49\pm0.02$}&
{\tiny $22.14\pm0.00$}&
{\tiny $20.76\pm0.00$}&
{\tiny $20.31\pm0.00$}&
{\tiny $19.84\pm0.05$}&
{\tiny $19.27\pm0.06$}\\
{\tiny 4587*}&
{\tiny 03:32:40.656}&
{\tiny -27:47:30.84}&
{\tiny Croom}&
{\tiny $0.667$}&
{\tiny $ 0.68\pm0.20$}&
{\tiny 0.26}&
{\tiny $24.79\pm0.02$}&
{\tiny $22.98\pm0.00$}&
{\tiny $21.69\pm0.00$}&
{\tiny $21.24\pm0.00$}&
{\tiny $20.76\pm0.05$}&
{\tiny $20.20\pm0.06$}\\
{\tiny 3677*}&
{\tiny 03:32:37.296}&
{\tiny -27:47:29.40}&
{\tiny Croom}&
{\tiny $0.669$}&
{\tiny $ 0.57\pm0.19$}&
{\tiny 2.45}&
{\tiny $23.59\pm0.01$}&
{\tiny $21.62\pm0.00$}&
{\tiny $20.23\pm0.00$}&
{\tiny $19.76\pm0.00$}&
{\tiny $19.25\pm0.05$}&
{\tiny $18.64\pm0.06$}\\
{\tiny 7705}&
{\tiny 03:32:37.560}&
{\tiny -27:46:46.56}&
{\tiny Strolger}&
{\tiny $1.300$}&
{\tiny $ 1.33\pm0.27$}&
{\tiny 0.29}&
{\tiny $25.83\pm0.02$}&
{\tiny $25.82\pm0.01$}&
{\tiny $25.70\pm0.01$}&
{\tiny $25.15\pm0.01$}&
{\tiny $24.93\pm0.09$}&
{\tiny $24.77\pm0.10$}
\enddata
\tablecomments{
Redshift surveys are 
FORS2 \citep{FORS2},
VVDS \citep{VVDS},
\citet{Szokoly04},
\citet{Croom01},
and \citet{Strolger04}.
ID numbers below 41000 correspond to B04 \& T04 detections; 
asterisks ({*}) indicate that object definitions have been altered
(\S\ref{sub:detection}).
$z_{b}$ gives the peak of the Bayesian photometric redshift distribution
$P(z)$ along with a 95\% confidence interval, while $\chi_{mod}^{2}$ measures
how poorly the best fitting SED template at $z_{b}$ fits the observed
colors. Magnitudes are ``total'' AB magnitudes with isophotal
colors: NIC3 magnitudes are corrected to the PSF of the ACS images
(\S\ref{sub:photometry}). We have also applied offsets of
($J$: $-0.30\pm0.03$, $H$: $-0.18\pm0.04$) to the NIC3 magnitudes
(\S\ref{sub:NIC3magoffsets}).
And all of our magnitudes have been corrected for galactic extinction
(Table \ref{cap:filters}). Non-detections (listed, for example, as
$>31.05$) quote the 1-$\sigma$ detection limit of the aperture
used on the given object. Magnitudes are left blank where objects
are unobserved (outside the NIC3 FOV) or contain saturated or other
bad pixels. Color images of these objects along with SED fits and
more are available at {\tt http://adcam.pha.jhu.edu/\~{}coe/UDF/zsconf/}.
}
\end{deluxetable}

\clearpage

The UDF ACS \& NIC3 FOVs were oriented to contain Strolger et al.'s
supernova and a spectroscopically confirmed $z=5.8$ object, corresponding
to our \#2225{*}.%
\footnote{Again, the asterisk ({*}) indicates that B04's $i\arcmin$-band
segment for object \#2225 was altered. Here it was replaced with their
$z\arcmin$-band segment (\#31526).}
\#2225{*} was originally detected as an
$i\arcmin$-dropout by \citet{SBM03},
and has since been known as SBM03\#1.
Since then, multiple spectra have been taken of this object
\citetext{\citealt{Dickinson04}, \citealt{Stanway04}, and most recently FORS2}.
The first two papers list SBM03\#1 at $z=5.83$, while FORS2 favors a slightly
lower $z=5.82$. (We find $z_{b}=5.78\pm0.80$.)

Multiple spectra have been obtained for several other objects in the
UDF as well. For all but two of these objects, the different authors
claim nearly identical spectroscopic redshifts ($\Delta z<0.005$).
In both of the discrepant cases, the FORS2 authors reject the earlier ``75\%
confident'' VVDS redshifts in favor of their own {}``solid'' or
{}``likely'' redshifts, citing superior classification of emission
lines. Our {\tt BPZ} values also support the FORS2 values.
These objects are our \#57290 ($z_{b}=0.31_{-.22}^{+.15}$; FORS2
GDS J033242.56-274550.2 $z=0.218$; VVDS \#28150 $z=0.6354$) and
our \#6188{*}  ($z_{b}=1.17\pm0.25$; GDS J033242.38-274707.6 $z=1.314$;
VVDS \#72036 $z=0.6885$).

\subsubsection{UDF NIC3 Recalibration: Empirical Derivation}
\label{sub:NIC3magoffsets}

\begin{figure*}
\plottwo{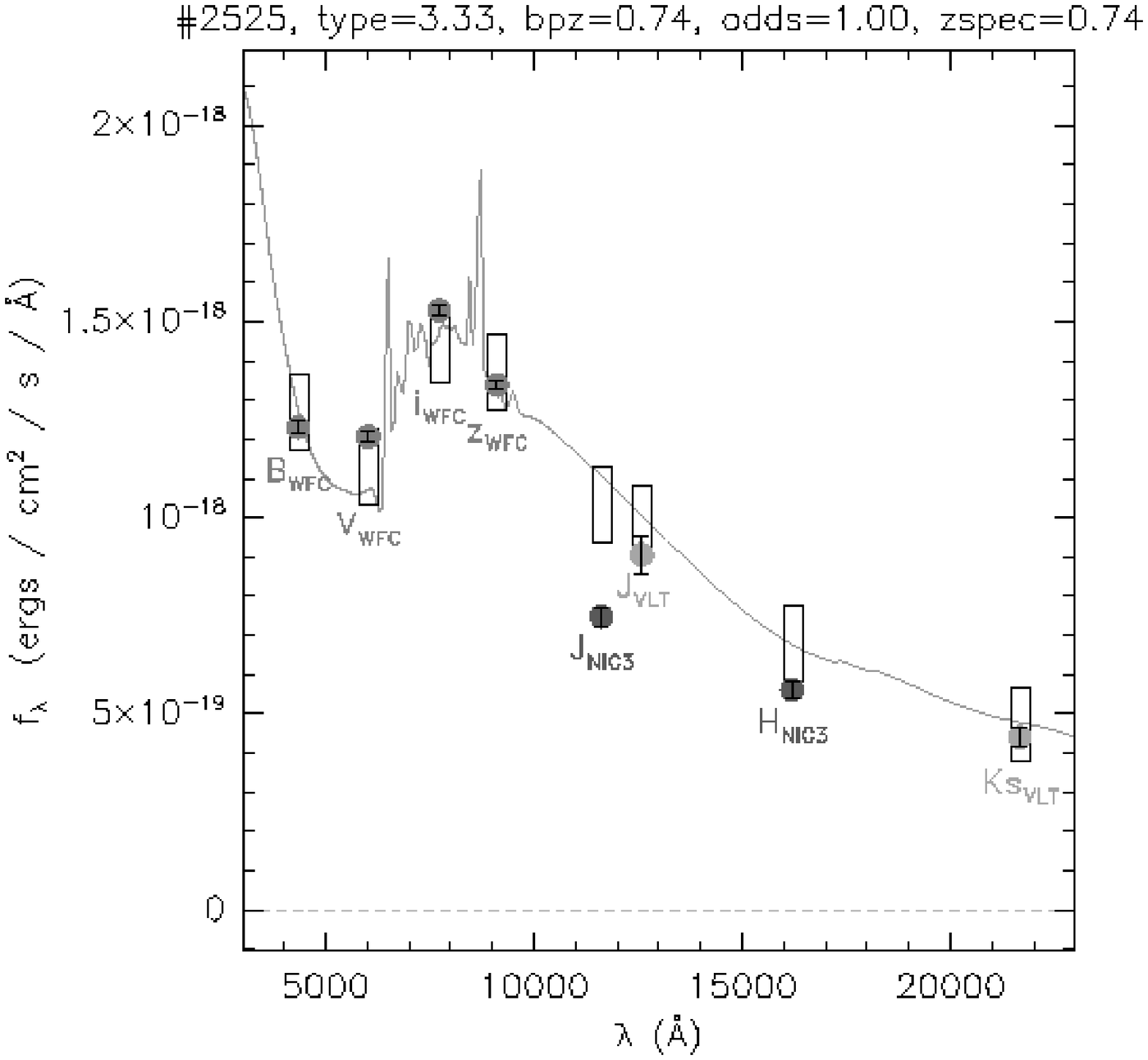}{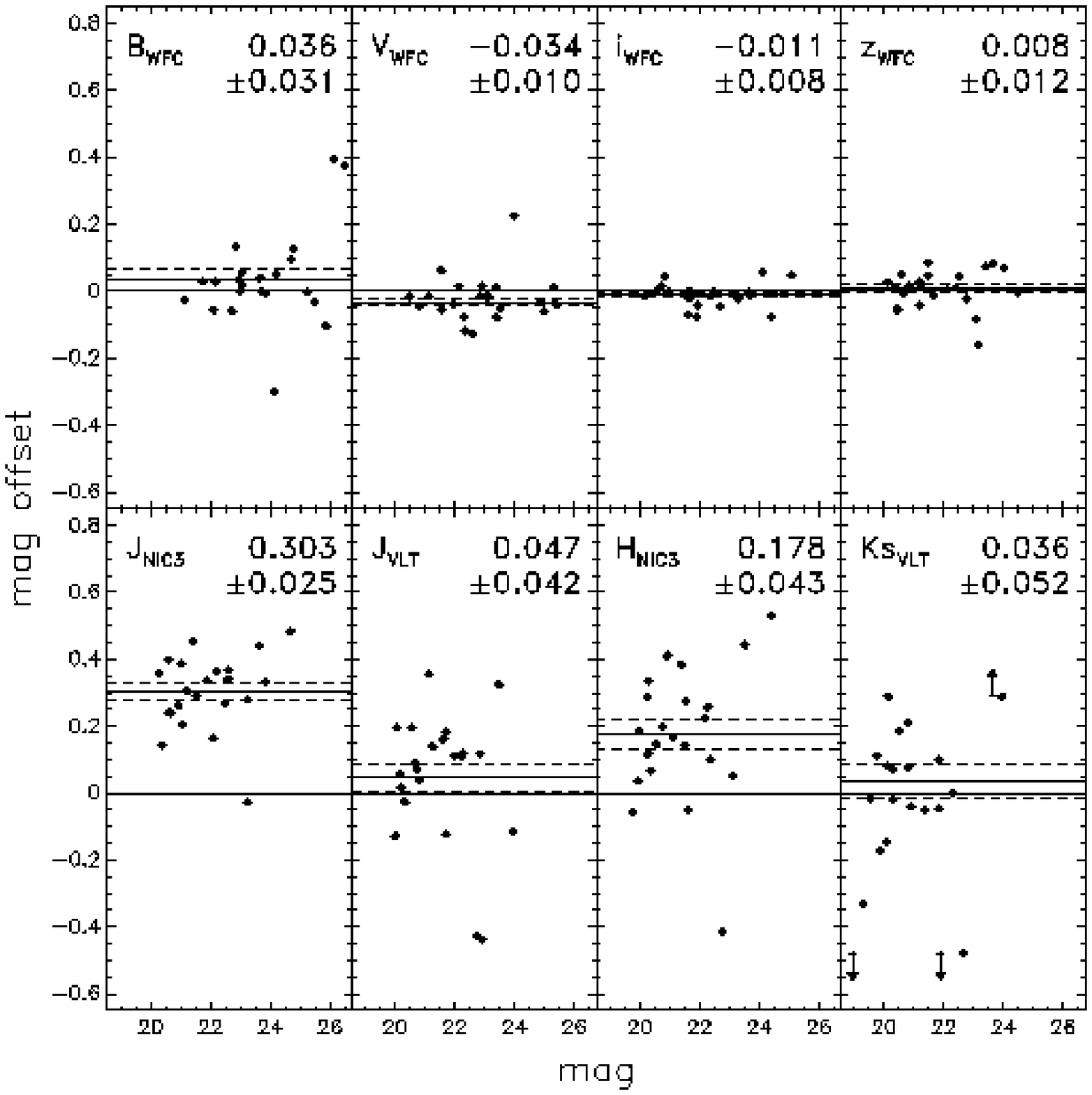}
\caption{\label{cap:SED-offsets}Left: {\tt BPZ} SED fit to an
object with a {}``95\% confident'' spectroscopic redshift in the
VVDS. We supplied {\tt BPZ} with the known redshift, and
{\tt BPZ} chose to fit a hybrid Scd-Im SED
to our observed fluxes (filled circles with filters labeled). The
SED itself is drawn in black. It is integrated over each filter to
yield the model fluxes shown as blue rectangles (given height to represent
uncertainty, \S\ref{sub:bpzhist}). The $J$ \& $H$ NIC3 fluxes
are too low to fit the SED (especially the $J$ flux). Meanwhile,
the $J$ \& $K_{s}$ VLT fluxes fit the SED well. This is typical
of all of the galaxies with {}``confident'' spectroscopic redshifts
available, as we see in the plot on the right, where magnitude offsets
are plotted versus magnitude for each filter. The solid red line indicates
the $\sigma$-clipped mean of the offsets, while the dashed red lines
demark the $\sigma$-clipped 1-$\sigma$ scatter divided by $\sqrt{N}$.
The ACS WFC and VLT magnitudes fit the SEDs well and have no significant
offsets. But the NIC3 magnitudes have weighted average offsets of
$0.30\pm0.03$ for $J$ and $0.18\pm0.04$ for $H$. We correct for
these offsets by adjusting the NIC3 magnitudes.}
\end{figure*}

Here we consider the 23 galaxies within the NIC3 FOV and with confident
spectroscopic redshifts from FORS2 and VVDS. 
When {\tt BPZ} SED templates are fit to the photometry of these galaxies,
we find that the NIC3 fluxes are below those expected given the ACS fluxes
and the known redshifts (see Fig.~\ref{cap:SED-offsets}).
We find weighted average magnitude offsets of $-0.30\pm0.03$ for $J$ 
and $-0.18\pm0.04$ for $H$. This appears to be a normal sample
of relatively bright galaxies ($20.3<z\arcmin<25.2$, with half having $z\arcmin<21.7$).
Thus these magnitude biases cannot be explained by our choice of apertures
or our {}``PSF corrections'', as neither of these significantly
affects the magnitudes of such bright objects (Fig.~\ref{cap:PSFcor}).
The galaxies belong to all of our different SED types, which means that
the problem cannot be traced to a single bad template. 
Few of the galaxies were fit to the new \citetalias{BC03} templates; 
most were instead classified as one of the widely used and 
well-calibrated CWW+SB templates (see \S \ref{sub:faintblue}).

To further test the UDF NIC3 calibration, 
we compared our photometry to photometry we obtained from the VLT $J$ image
(Vandame et al., in prep.)%
\footnote{VLT observations have been carried out at the ESO Paranal Observatory
under Program ID: LP168.A-0485}. 
As the PSF corrections are small for these bright objects, {}``quick
and dirty'' photometry is sufficient here. We simply took our VLT 
{\tt MAG\_AUTO} measurements and added these to our main catalog,
matching objects by position.
(Further tests confirm that these ``quick'' magnitudes are accurate to within $\pm 0.06$ mags.)
A straight comparison between the VLT $J$ and NIC3 $J$-band magnitudes
shows that the NIC3 $J$-band magnitudes are about 0.3 magnitudes too faint.
When we apply corrections to account for the different filter shapes
(VLT $J$ vs.\ NIC3 $J$), this difference is slightly reduced, by 0.05 mags.
(Note that these corrections require assumptions of redshift and SED for each galaxy.)
Given the uncertainties involved, these results are consistent with our above analysis.
We note that nearly identical offsets were independently derived for the UDF
in a similar analysis by \citet[and priv.\ comm.]{Gwyn05}.
Similar deficits in NICMOS fluxes have been observed by \citet{MobasherISR05} in the UDF
and by Adam Riess in observations of supernovae (priv.\ comm.).

\begin{figure*}
\plotone{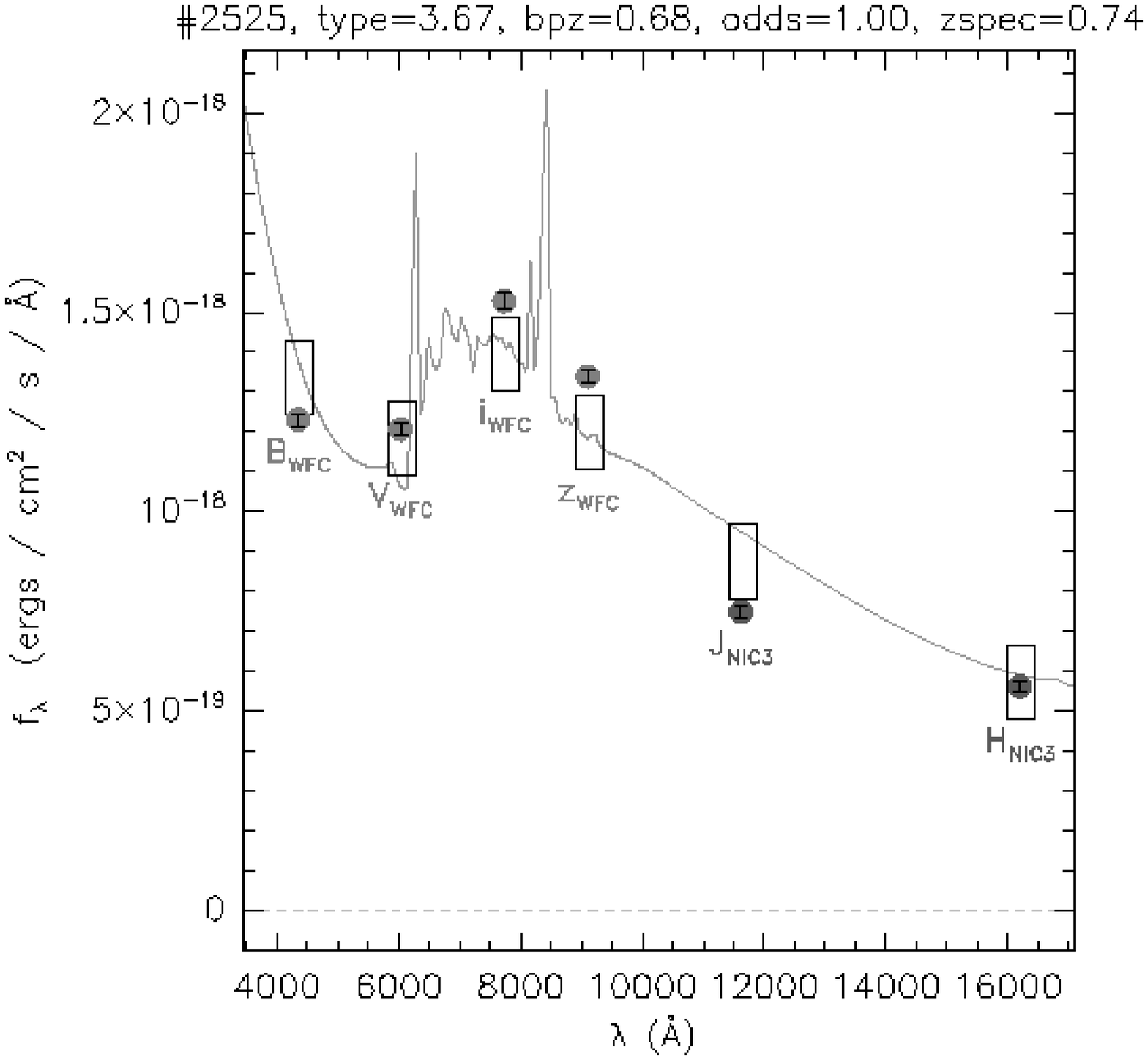}
\caption{\label{cap:splitdiff}{\tt BPZ} SED fit to \#2525 without our NIC3 magnitude offsets applied.
The photometric redshift $z_b = 0.68$ obtained for this galaxy is the same whether our offsets are applied or not.
Here the best fitting SED simply ``splits the difference''
between ACS fluxes that are ``too high'' and NIC3 fluxes that are ``too low''.
Compare to Fig.~\ref{cap:SED-offsets}a
(but keep in mind that the {\tt BPZ} fit in that figure 
was constrained to the correct redshift $z_b = z_{spec} = 0.74$)Accurate photometric redshifts are relatively easy to come by for bright galaxies such as this 
with a spectroscopic redshift available.
Thus this agreement is no guarantee that the underlying photometry is robust.
}
\end{figure*}

Other UDF studies have not questioned the NIC3 calibration;
this can be understood since the uncorrected magnitudes will often produce
photometric redshifts which are roughly correct.
For example, when we revert the NIC3 photometry of galaxy \#2525 to pre-recalibration magnitudes,
the derived redshift remains the same: $z_b = 0.68$
(close to the spectroscopic value of $z_{spec} = 0.74$).
As we see in Fig.~\ref{cap:splitdiff},
the best fit SED simply ``splits the difference''
between NIC3 fluxes that are a bit too low and
ACS fluxes that are a bit too high.
(Compare to Fig.~\ref{cap:SED-offsets}a,
although keep in mind that the {\tt BPZ} fit in that figure 
was constrained to the spectroscopic redshift $z_b = z_{spec} = 0.74$.)
Galaxies such as \#2525 with spectroscopic redshift available are sufficiently bright
that accurate photometric redshifts may often be obtained with less than perfect photometry.
But by looking for and correcting for magnitude offsets in the individual filters (Fig.~\ref{cap:SED-offsets}b),
we help ensure that our photometry is robust for the more challenging fainter galaxies.

Our offsets appear to be supported by a recent recalibration of the UDF NIC3 images
combined with non-linearity measured in NICMOS itself (\S\ref{sub:NIC3recal}).
Thus we are encouraged to proceed with our analysis given our derived offsets:
$-0.30\pm0.03$ in $J$ and $-0.18\pm0.04$ in $H$.

\subsubsection{Spectroscopic Redshift Comparison}
\label{subsub:spec-z_vs}

\begin{figure*}
\plottwo{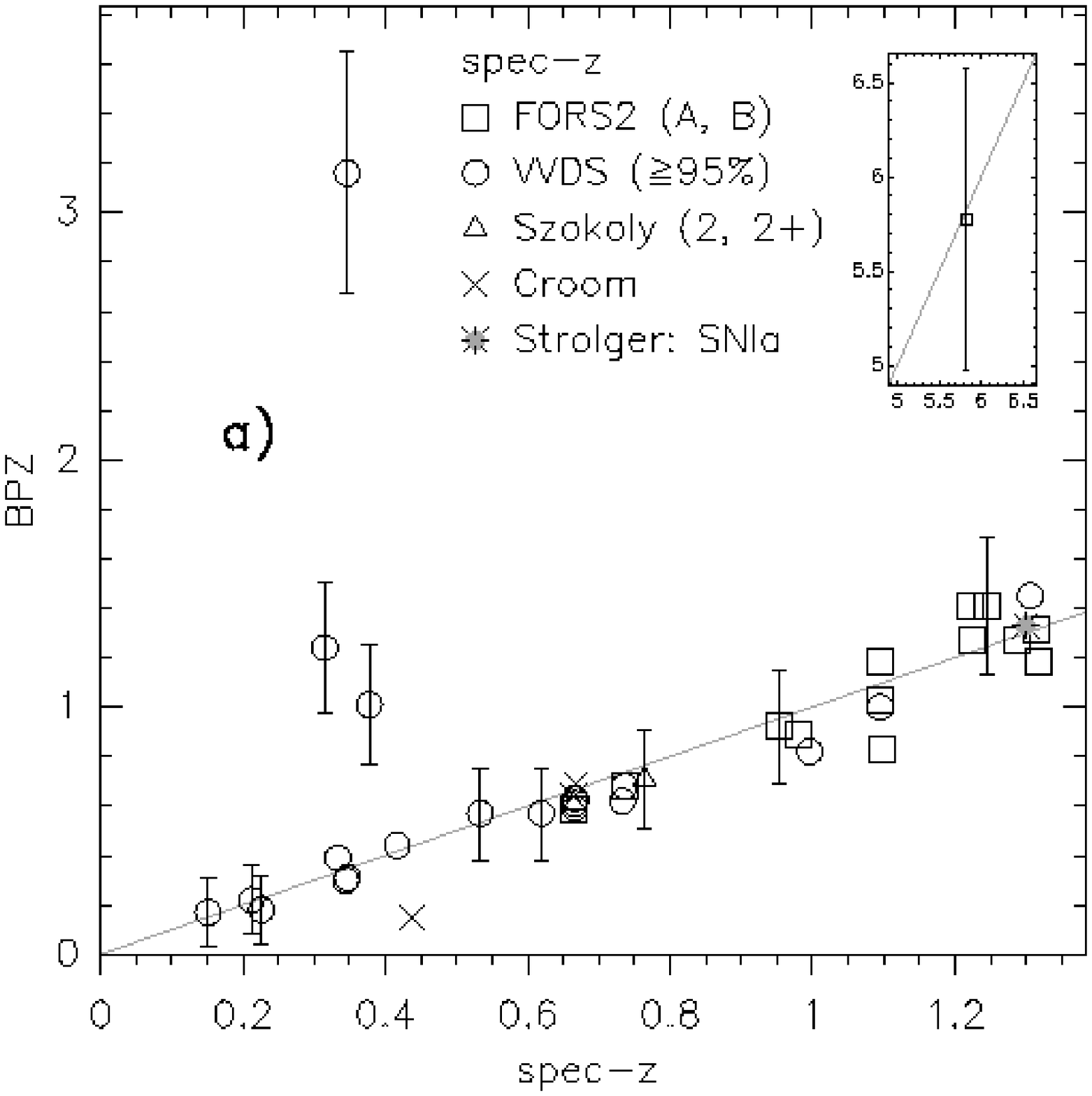}{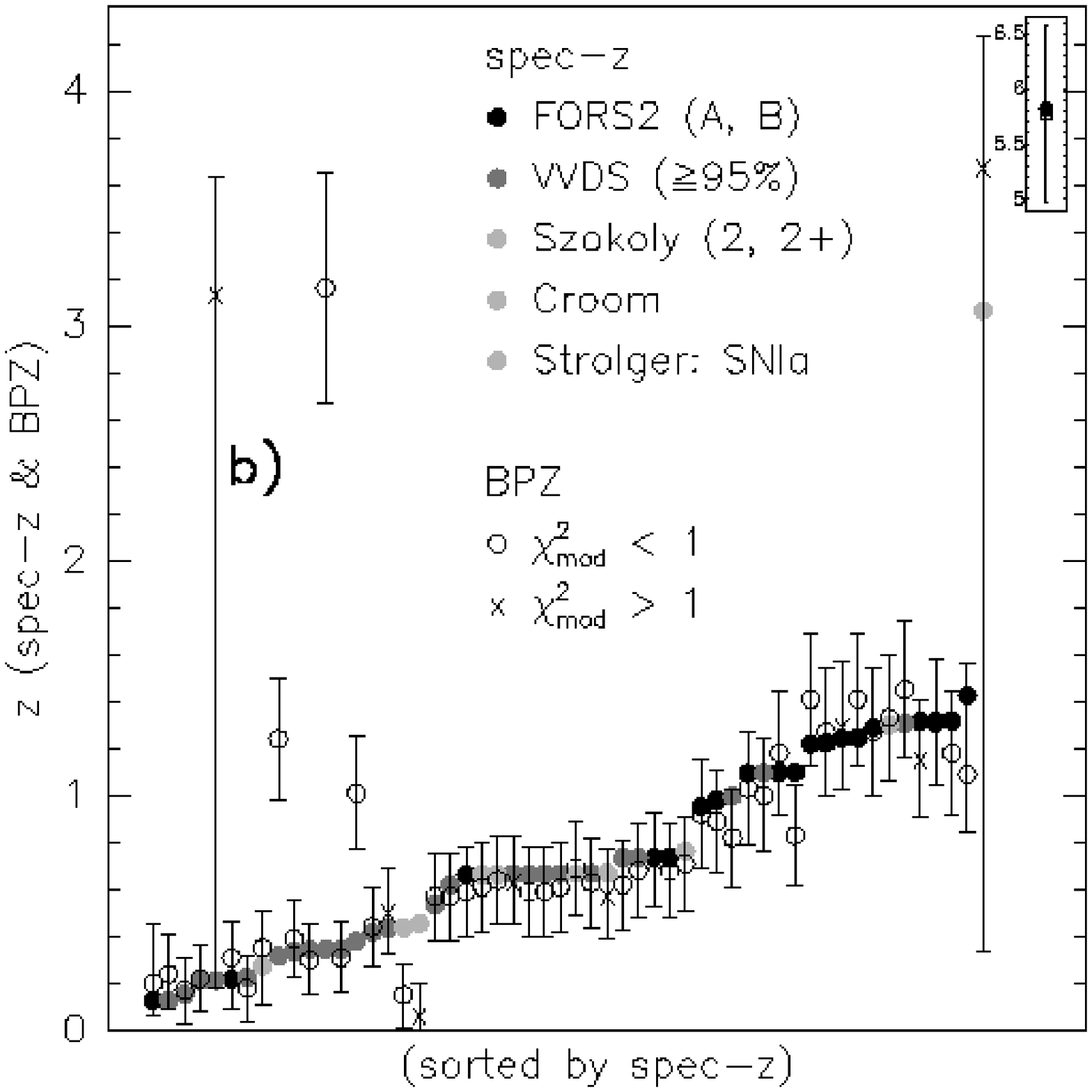}
\caption{\label{cap:ESO}Bayesian photometric redshifts compared to the 55
spectroscopic redshifts described in \S\ref{subsub:spec-z}. On the
left we plot the 41 of those galaxies which have {\tt BPZ}
{\tt ODDS} $\geq0.95$ \& $\chi_{mod}^{2}<1$. The high {\tt ODDS}
values ensure small well-behaved confidence intervals. We plot some
of these 95\% confidence intervals, but we suppress most to avoid
clutter. We zoom in on those galaxies with lower spec-z and reserve
an inset for galaxy \#2225{*} with $z_{spec}=5.82$ ($z_{b}=5.77\pm0.796$).
Colors provide the best (or at least most recent) reference for each
redshift. 
On the right we plot the same data, but all redshifts are plotted
along the $y$-axis. This plot is less cluttered so we are able to
include all 55 galaxies. We are able to plot confidence intervals
for all galaxies, so outliers are clearly identified (and we no longer
need to restrict ${\tt ODDS}\geq0.95$). And we are able to include
galaxies with $\chi_{mod}^{2}>1$ by plotting them with a different
symbol (an {}``x''). To be clear, the $x$-axis is not to scale.
It merely serves to spread galaxies across the plot and sort them
according to spec-z.}
\end{figure*}

After our recalibration of the UDF NIC3 photometry,
our photometric redshifts agree very well 
with the 55 spectroscopic redshifts described in \S\ref{subsub:spec-z}
(see Fig.~\ref{cap:ESO}). Among the 41 galaxies with
{\tt ODDS} $\geq0.95$ \& $\chi_{mod}^{2}<1$, 
we find an RMS of $\Delta z=0.04(1+z_{spec})$, 
but only after we exclude 4 outliers. 
3 of these outliers are from the VVDS,
while one is from \citet{Croom01} (and was not assigned a confidence level).
The outliers are clearly visible in Fig.~\ref{cap:ESO}a.

The $Vi\arcmin z\arcmin$ filter set covers the entire spectral range
covered by these spectroscopic surveys (for example, FORS2: $6000-10800\textrm{\AA}$;
VVDS: $5500-9500\textrm{\AA}$). Inside the NIC3 FOV, our Bayesian photometric
redshifts benefit from three {}``extra'' filters, as the $BVi\arcmin z\arcmin JH$
filter set covers $\sim4000-18000\textrm{\AA}$. And of course the UDF
photometry extends much deeper than the spectroscopic surveys. Thus,
even when discrepancies do occur, it is unclear whether to favor the
photometric or spectroscopic redshifts (especially for those spectroscopic
redshifts that are assigned low confidence). In the case of the HDF,
\citet{FernandezSoto01} showed that most of the discrepancies
were likely the result of incorrect spectroscopic redshifts which were
overruled by more reliable photometric redshifts.

\begin{figure*}
\plottwo{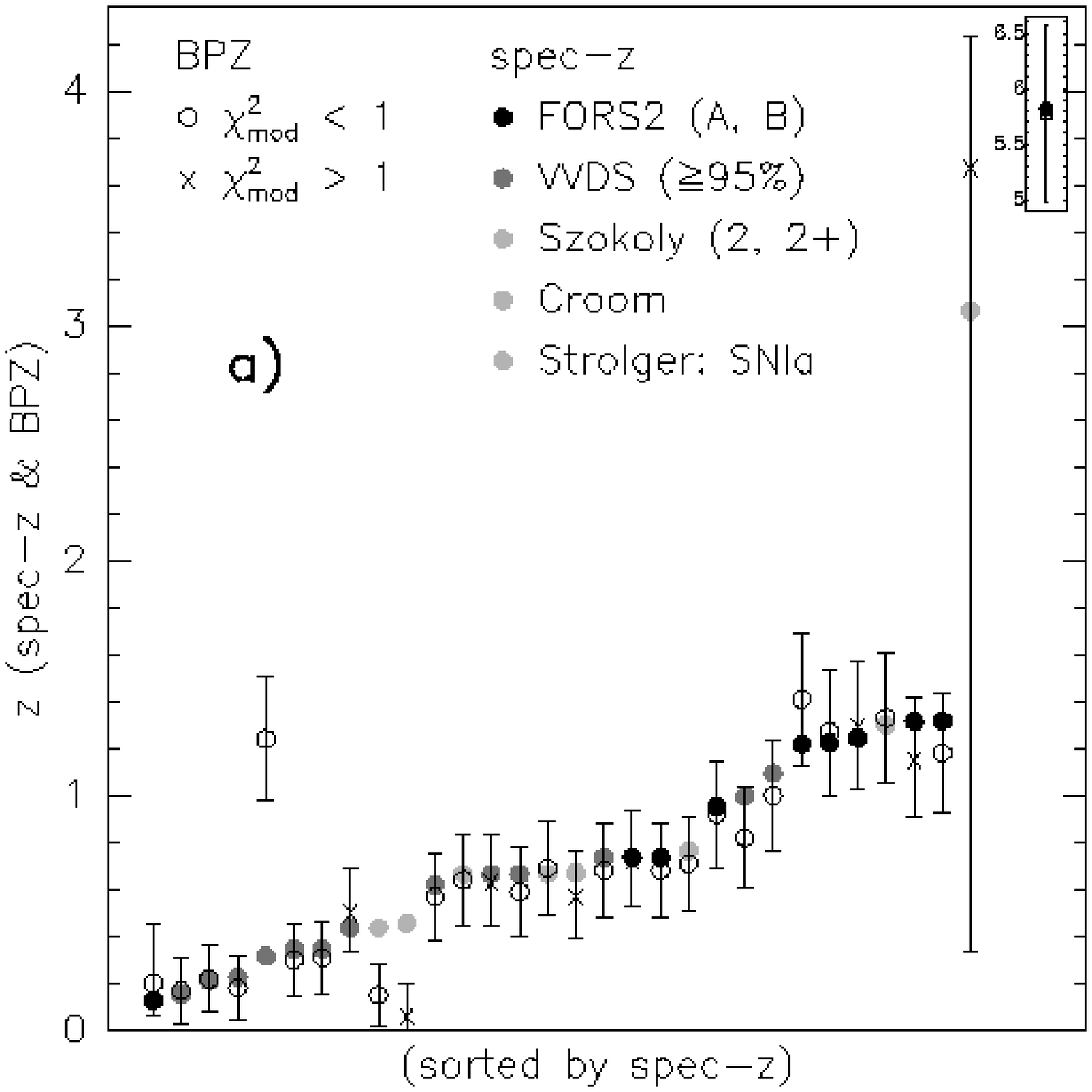}{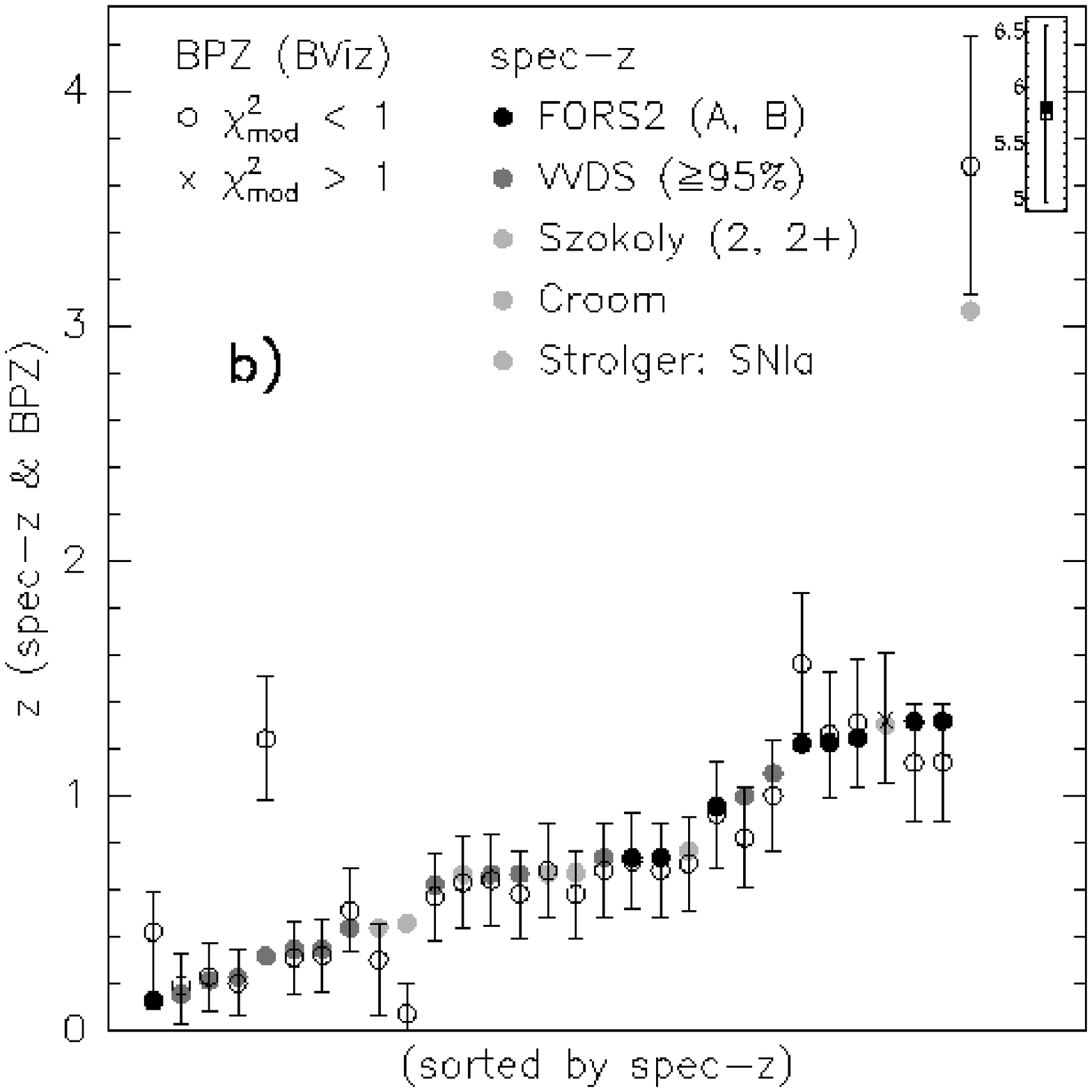}
\caption{\label{cap:zspecNIC3}Left: same as Fig.~\ref{cap:ESO}b, but only
for the 31 galaxies within the NIC3 FOV. Right: {\tt BPZ}
results obtained using $BVi\arcmin z\arcmin$ filters only. The results
degrade very little when the NIC3 filters are omitted, as most of
these galaxies are relatively bright (with high signal-to-noise) in
$BVi\arcmin z\arcmin$.}
\end{figure*}

Outside the NIC3 FOV, we expect the four ACS filters ($BVi\arcmin z\arcmin$)
to continue to deliver high quality photometric redshifts. Fig.~\ref{cap:zspecNIC3}b
presents {\tt BPZ} results obtained using only the ACS
photometry for galaxies with spectroscopic redshifts and within the
NIC3 FOV. The results are on par with those obtained using all 6 filters
(Fig.~\ref{cap:zspecNIC3}a). (Of course galaxies with confident spectroscopic
redshifts are relatively easy tests, as they are usually bright with
dominant spectral features.)

\subsubsection{COMBO-17}

\begin{figure*}
\plottwo{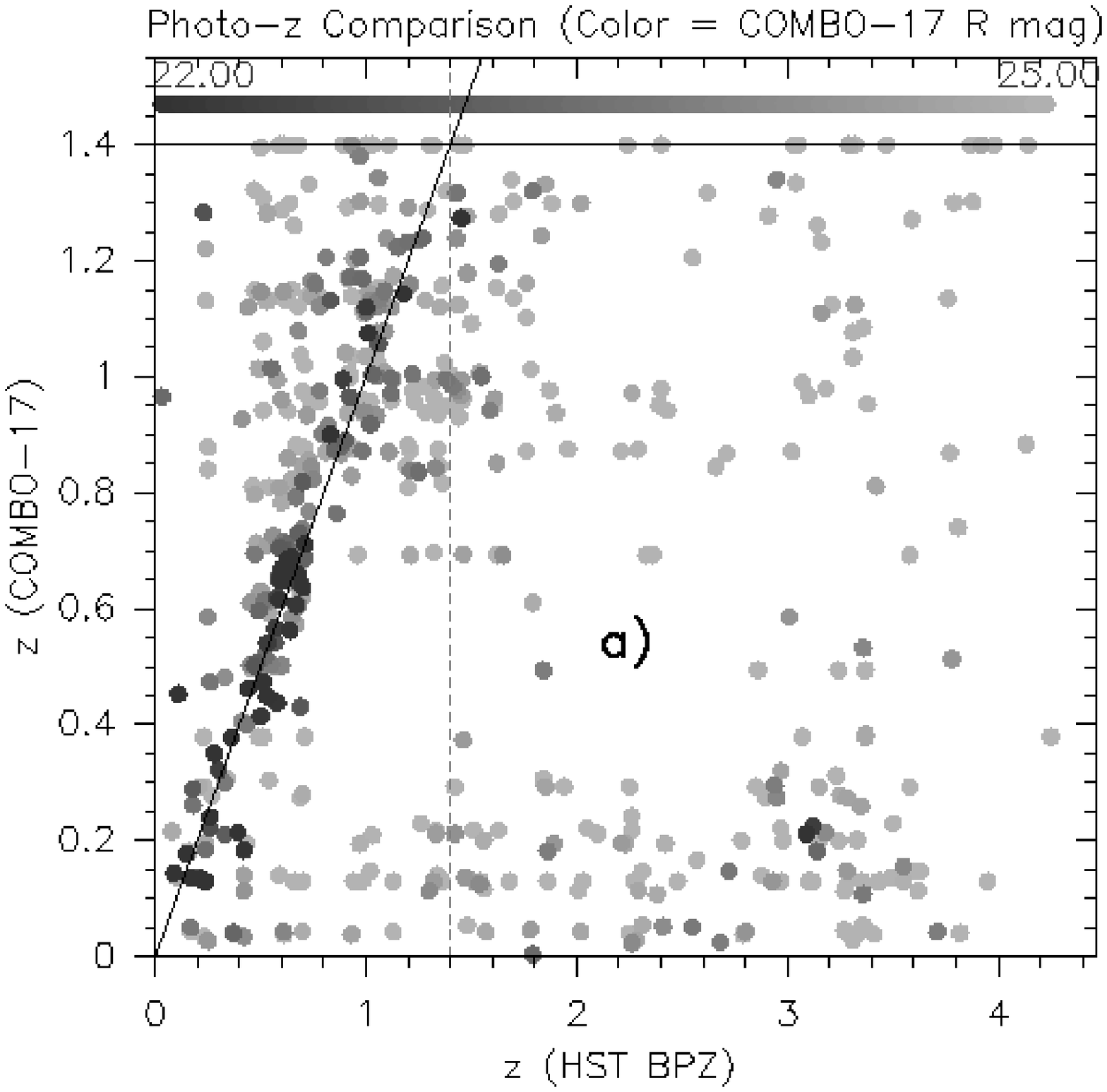}{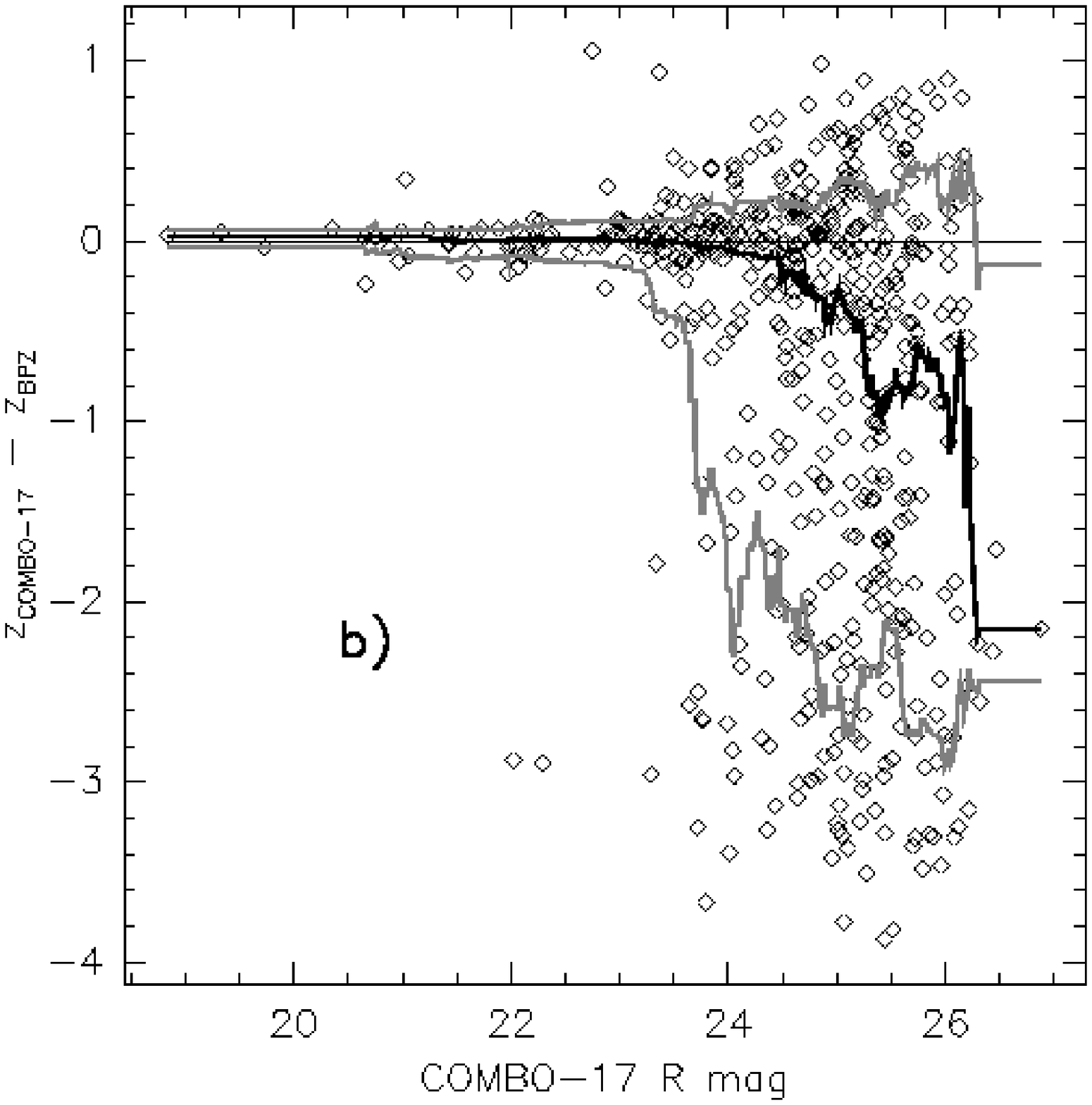}
\caption{\label{cap:COMBO-17}
Left: Photometric redshifts from COMBO-17 plotted
against our Bayesian photometric redshifts. We plot only those galaxies
with {\tt ODDS} $\geq 0.95$ and $\chi_{mod}^{2}<1$. Colors represent
$R$ magnitudes in COMBO-17. COMBO-17 claims to yield reliable redshifts
for $R\lesssim24$, and then only attempts to model redshifts of $z<1.4$.
Note that many of the galaxies with COMBO $z\sim0.13$ have been reassigned
{\tt BPZ} $>1.4$.
All but two of these are faint ($R>23.5$).
Right: The same data points replotted as the difference between our redshifts 
and those obtained by COMBO-17 versus COMBO-17 $R$ magnitude. The inner red line is a
moving average (median) of 100 galaxies (or as few as 10 at the edges),
while the outer magenta lines contain 68\% (1-$\sigma$) of the galaxies.
The relationship is very good for $R\lesssim 23.7$
and especially tight for $R\lesssim 23$ galaxies, with the notable
exception of two outliers with $z_{b}\sim 3$. At fainter magnitudes,
a significant fraction of the COMBO-17 redshifts deviate
far from our values.}
\end{figure*}

We also found good agreement with most of the photometric redshifts
obtained by the COMBO-17 survey \citep{COMBO-17} for bright galaxies
(Fig.~\ref{cap:COMBO-17}).
COMBO-17 covers a wide spectral range with very good resolution for
a photometric redshift survey. But of course, it does not penetrate
nearly as deep as the UDF; COMBO-17 only claims to yield reliable
redshifts up to $R\lesssim24$. In addition, COMBO-17 does not attempt
to model galaxies beyond $z>1.4$. So any galaxies at $z>1.4$ will have
been reassigned $z_{COMBO-17}<1.4$.
Of course as \citet{COMBO-17} point out, 
such bright galaxies are unlikely to be at $z>1.4$.

We find that our best fit redshifts
$z_{b}$ agree well with those of COMBO-17 for $R\lesssim 23.7$ galaxies.
The relationship is especially tight for $R < 23$, with a few notable exceptions. 
Two $R\sim22$, $z_{COMBO-17}\sim0.2$ galaxies (\#5491 \& \#6082) 
are assigned $z_{b} \ga 3$.
(This redshift degeneracy is also documented 
for similar galaxies in Fig.~\ref{cap:P(z)}.)
These two galaxies truly stand out in our catalog.
To be at $z\sim 3$ these $R\sim 22$ galaxies would have to be monsters
($M(1400\AA) \sim -23$; see Paper II).
(Note that {\tt BPZ}'s priors usually help to resolve such redshift degeneracies
in favor of the more reasonable choice, given the galaxy's magnitude.
But in the case of these two galaxies,
the $z\sim 3$ fits were deemed sufficiently superior 
to rule out the more reasonable $z\sim 0.2$ fits.)
Thus for these bright galaxies we are inclined to believe the COMBO-17 results, 
which benefit from observations in many more filters.

\citet{COMBO-17} identify a new galaxy cluster at $z\sim0.15$ 
within the wider CDF-S field.
We are unable to confirm this overdensity in our catalog (\S\ref{sub:Clustering}).
We cannot even confirm the redshifts of the $z_{COMBO-17}\sim0.15$ 
galaxies within the UDF.
All of these galaxies are faint ($R\geq 23.5$),
and {\tt BPZ} reassigns most of them to $z>1.4$,
or outside the redshift range modeled by COMBO-17 (Fig.~\ref{cap:COMBO-17}a).
The {\tt BPZ} results are presumably more reliable than COMBO-17
at these faint magnitudes.
However \citet{COMBO-17} observe the $z_{COMBO-17}\sim0.15$ cluster
even in their brightest $R<21$ galaxy sample,
which we are in no position to question.

While we are unable to confirm any overdensity at $z\sim 0.15$,
we do support COMBO-17's detection of the known overdensity at $z\sim0.67$
(see discussion and references in \S\ref{sub:Clustering})

\subsection{{\tt BPZ} Histogram}
\label{sub:bpzhist}

\begin{figure*}
\begin{center}
\epsscale{0.6}
\plottwo{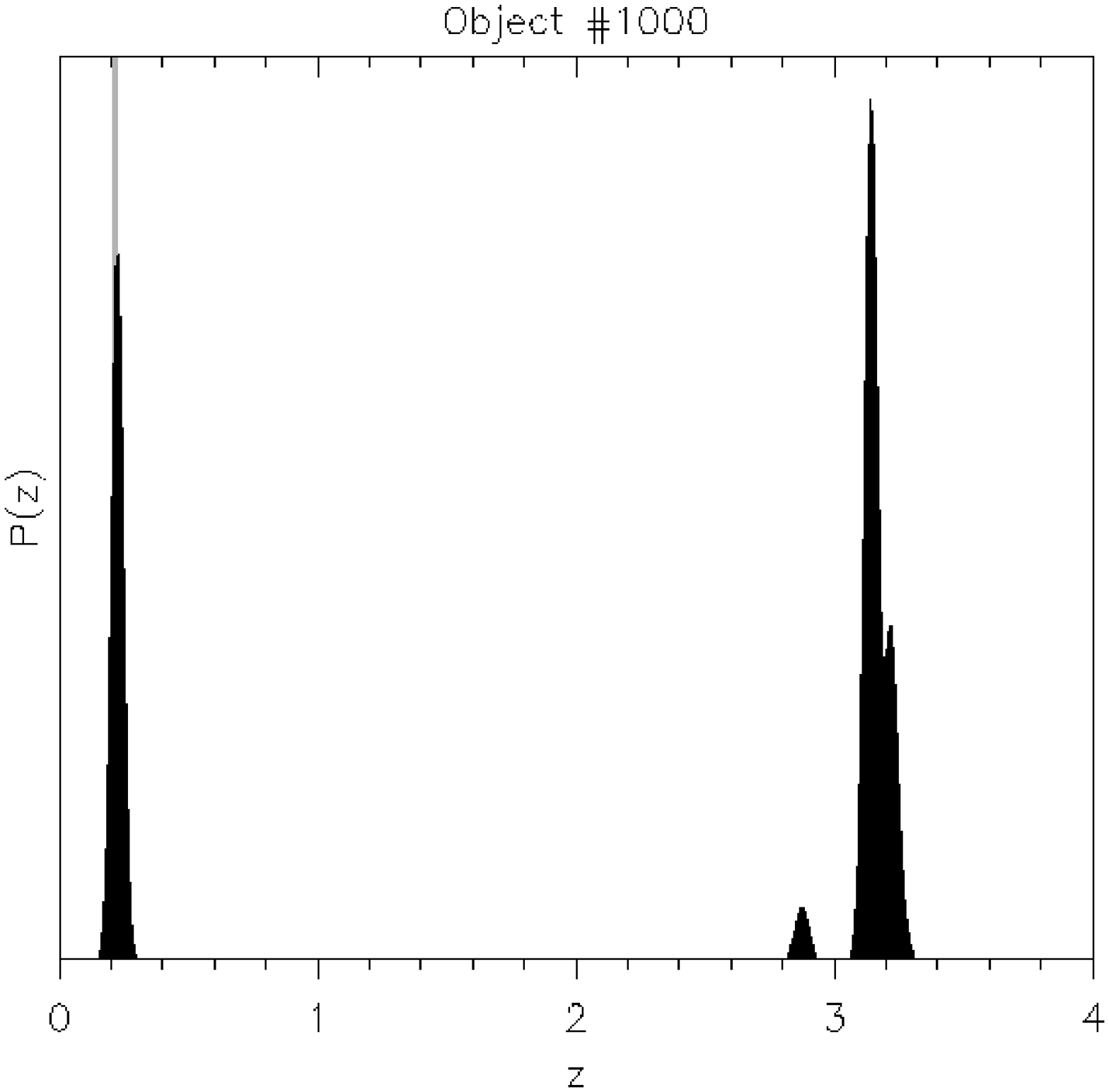}{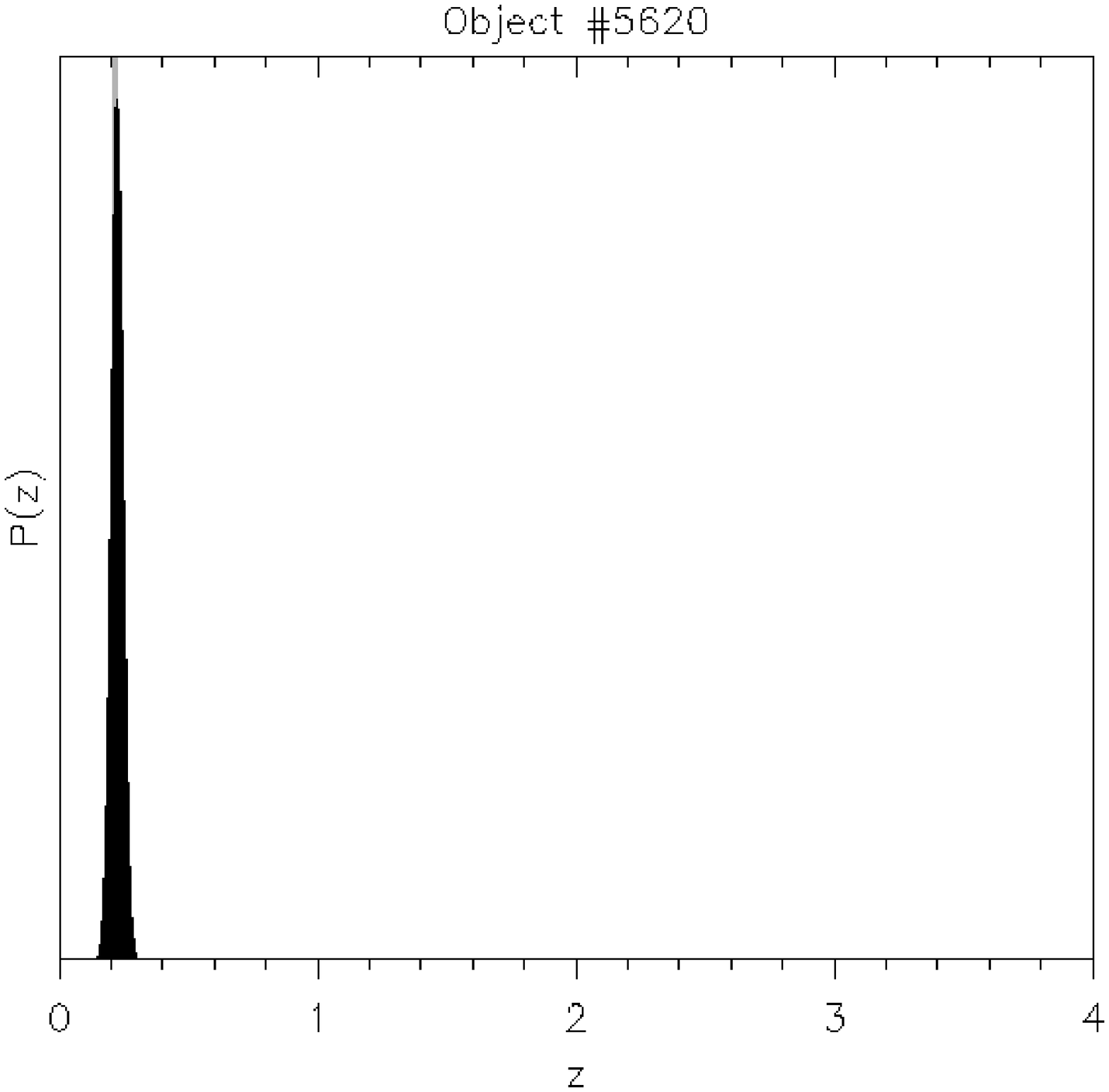}\\
\plottwo{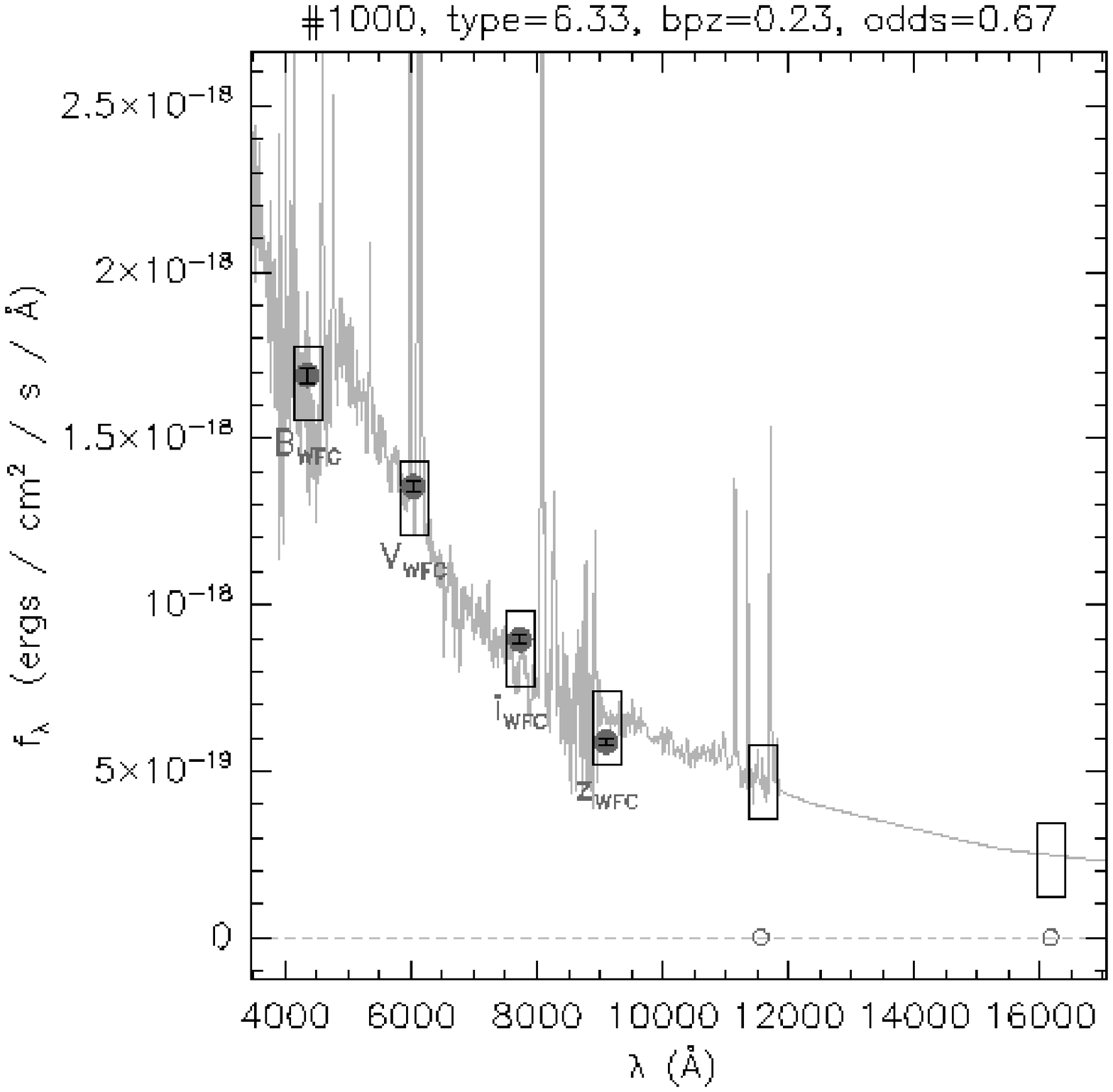}{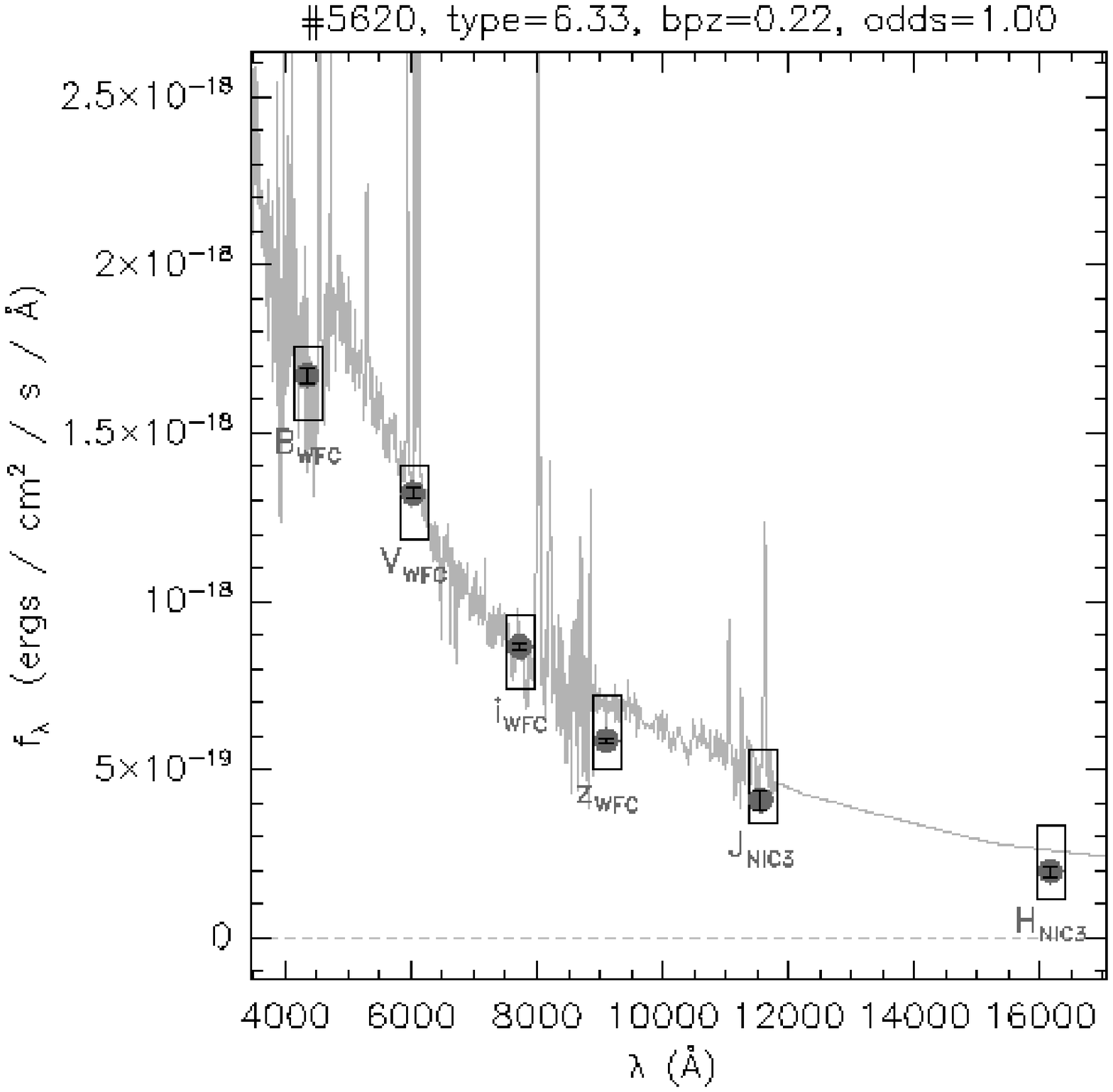}\\
\plottwo{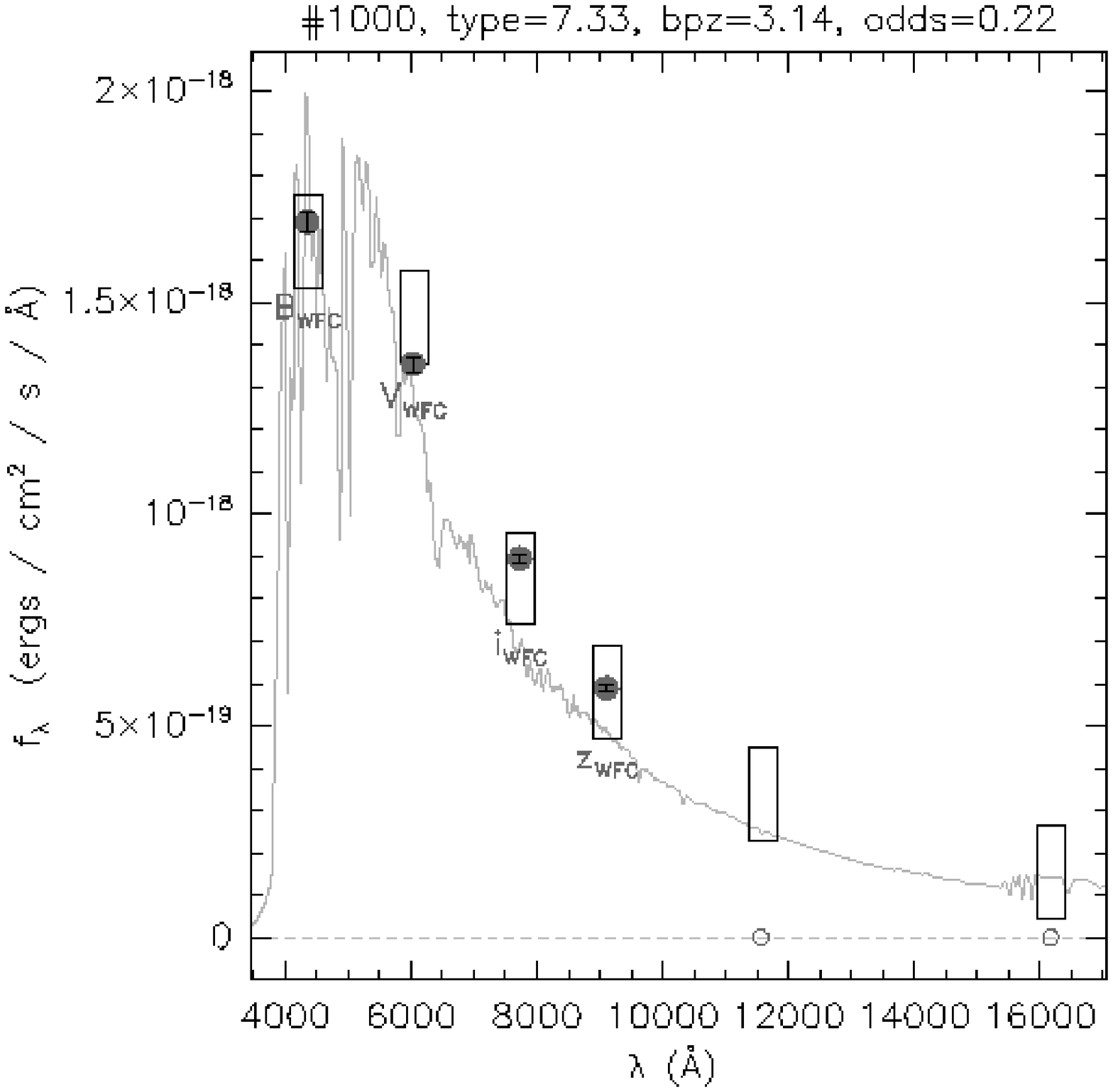}{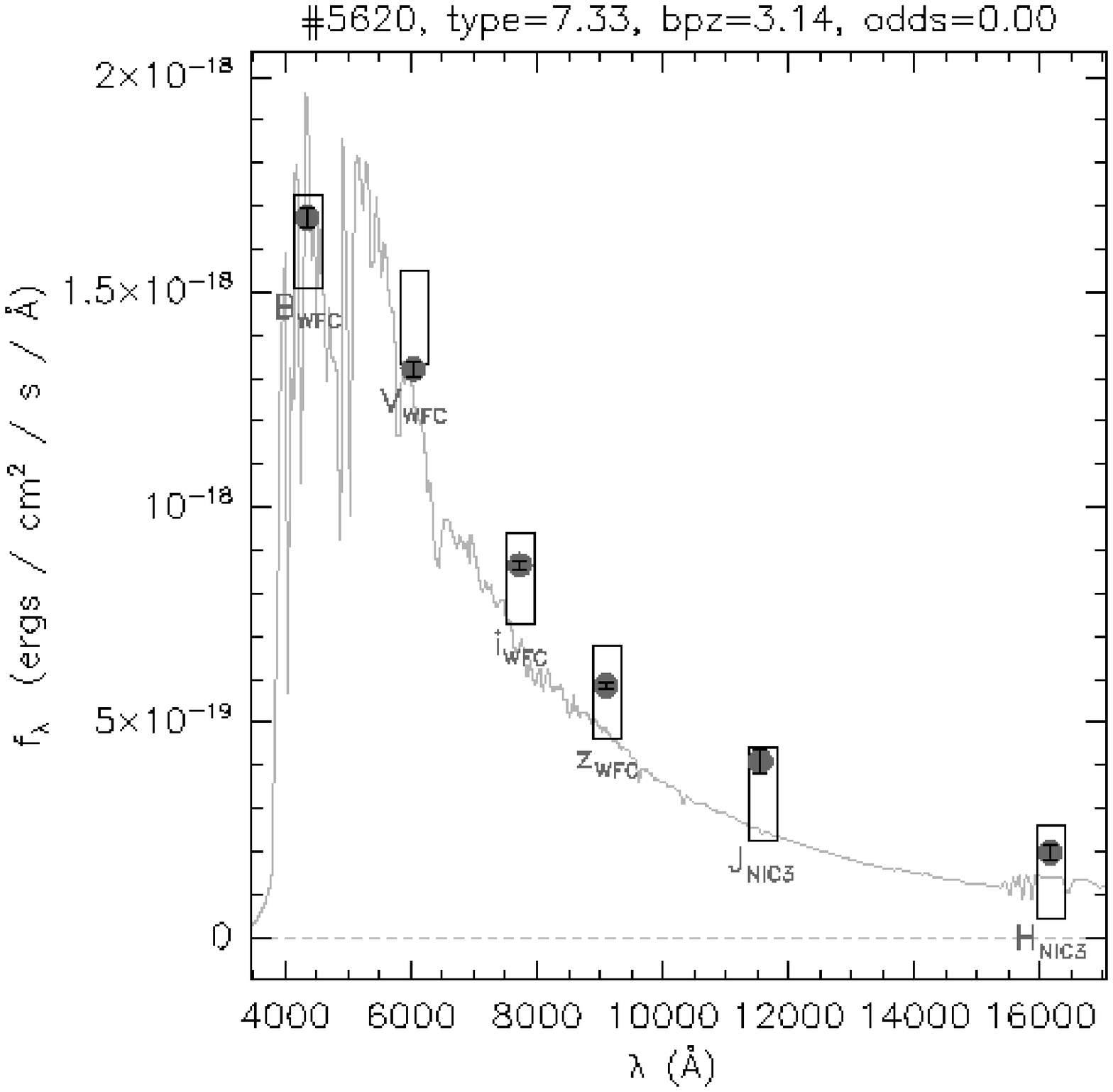}
\epsscale{1.0}
%
%
%
%
%
\end{center}
\caption{{\label{cap:P(z)}Two starburst galaxies (\#1000 \&
\#5620) at a spectroscopic redshift of $z_{spec}=0.21$.} {\tt BPZ}
{returns several possible redshifts for galaxy \#1000.
Four peaks are visible in the probability histogram (top left), each
corresponding to a different SED fit. The highest peak corresponds
to a SB2-25Myr hybrid SED template at $z\sim0.22$ (middle left).} {\tt BPZ}
{is 67\% certain (}{\tt ODDS}{=0.67)
that this is the correct fit. The second highest peak is a 25Myr-5Myr
hybrid at $z\sim3.14$ (bottom left), which also yields a reasonable
fit to the observed $BVi\arcmin z\arcmin$ photometry. But because
this fit is slightly worse, it earns lower} {\tt ODDS}:
a 22\% chance of being the correct fit.
Meanwhile, galaxy \#5620 is within the NIC3 FOV.
{\tt BPZ} assigns \#5620 a single redshift peak at the correct
redshift (top right). The NIC3 photometry helps constrain the SED
fit (middle right). A 25Myr-5Myr SED at $z\sim3.14$ (bottom right)
yields a significantly poorer fit, essentially ruling it out.
}

\end{figure*}

For each galaxy, {\tt BPZ} returns a full probability
distribution $P(z,t)$ (a function of redshift and type). (The new 
version of {\tt BPZ} also returns a catalog
summarizing the redshift, width, and {\tt ODDS} of the three highest
peaks.) Fig.~\ref{cap:P(z)} shows an example of $P(z)$ for two galaxies,
demonstrating how NIC3 photometry helps constrain the fit to a single
redshift.

\begin{figure*}
\includegraphics[width=0.60\linewidth]{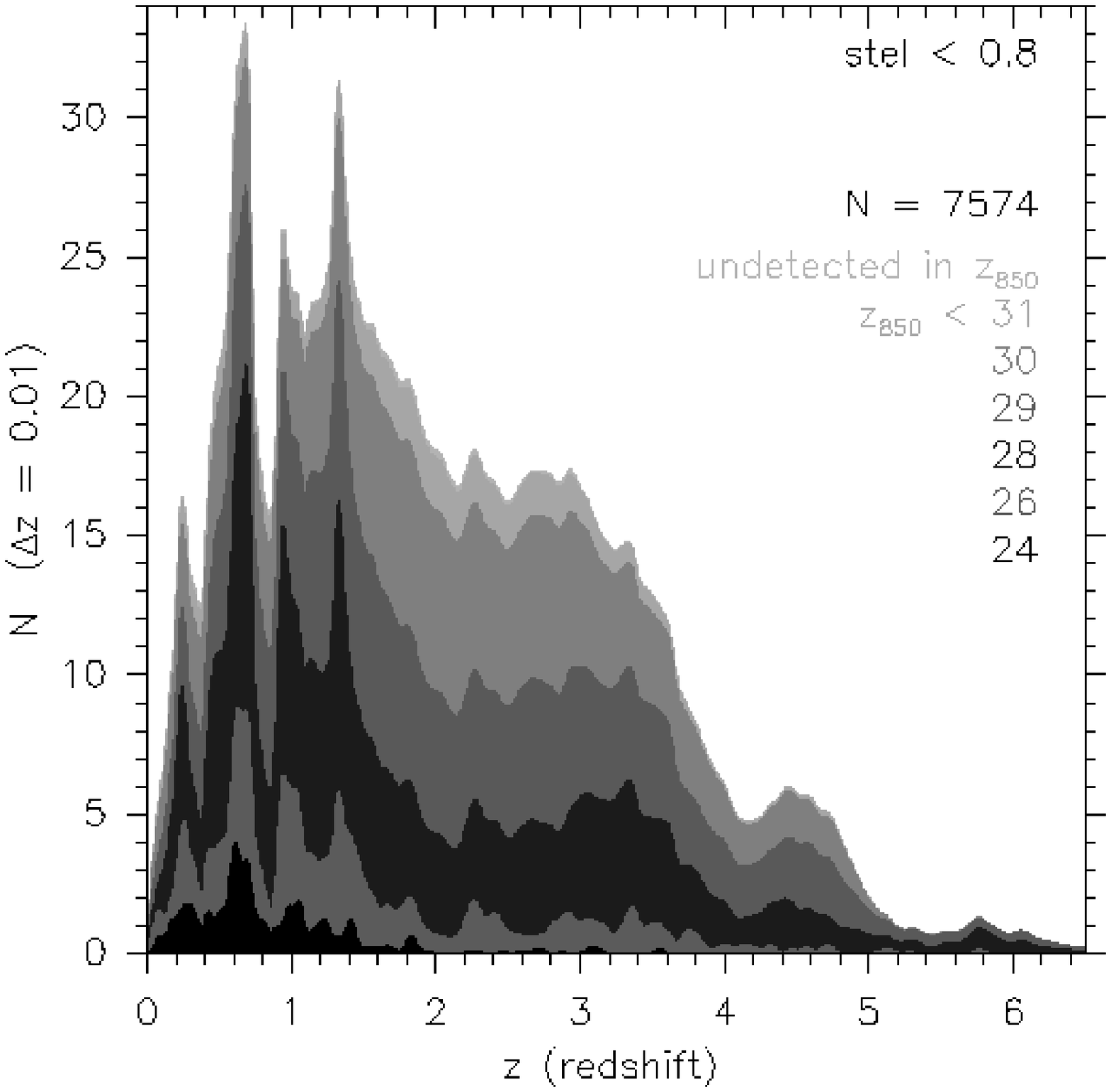} \includegraphics[width=0.39\linewidth]{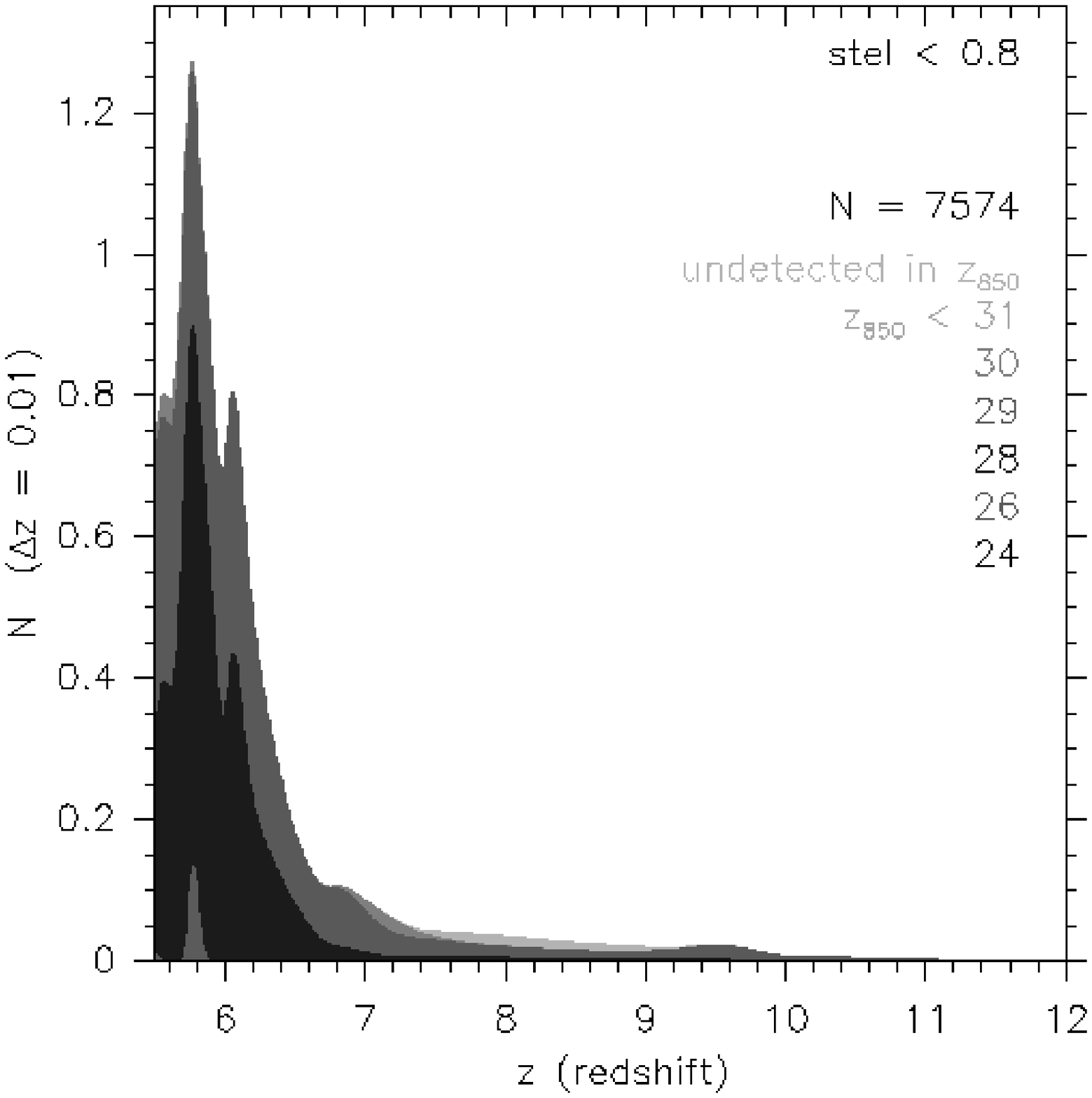}
\caption{\label{cap:bpzphist}{\tt BPZ} probability histogram
obtained by adding the redshift probability distributions ($P(z)$)
for 7,574 galaxies. These galaxies have been detected at the 10-$\sigma$
level in at least one filter or detection image and have ${\tt stellarity}<0.8$
in the $i\arcmin$-band image. We have also excluded galaxies
with particularly egregious SED fits: theoretical fluxes of zero or
infinite. The redshift interval is 0.01. $z\arcmin$-band magnitude
contours are plotted within the histogram at irregular intervals:
undetected in $z\arcmin$, then $z\arcmin<31$, 30, 29, 28, 26, 24.
We plot two different redshift ranges: $0<z<6.5$ and $5.5<z<12$.
A single galaxy is responsible for the peak at $z_{b}=9.58$ (see
\S\ref{sub:high-z}).}
\end{figure*}

By adding the probability histograms $P(z)$ of individual galaxies,
we obtain the redshift probability histogram of the UDF (Fig.~\ref{cap:bpzphist}).
Attempting to construct a histogram by binning $z_{b}$ (each galaxy's
best fit; the peak of $P(z)$) yields a fairly different shape. 
Fig.~\ref{cap:bpzphist} only includes galaxies that are detected at the
10-$\sigma$ level in at least one filter or detection image. And
to exclude stars, we discard all objects with {\tt SExtractor}
${\tt stellarity}\geq0.8$ in the $i\arcmin$-band image. This
eliminates all of the obvious stars (those with diffraction spikes)
without removing any obvious high redshift candidates (which may also
appear to be pointlike). (All of our best $z\geq6$ candidate galaxies
(Table \ref{cap:highz}) have ${\tt stellarity}\leq0.72$.)

We do not need to discard galaxies based on the {\tt BPZ}
output parameter {\tt ODDS}. {\tt ODDS} measures the reliability
of each galaxy's most likely redshift $z_{b}$. A galaxy with high
{\tt ODDS} has $P(z)$ with a narrow single peak. Multiple and/or
broad peaks yield low {\tt ODDS}. But all shapes of $P(z)$ will
be reflected accurately in our histogram which is a sum of galaxies'
$P(z)$'s.

While we don't need to eliminate galaxies based on low {\tt ODDS}
values, we would like to eliminate those galaxies with ill-fitting
SEDs. If {\tt BPZ} was not able to fit a galaxy well to
an SED, then the resulting redshift is probably not accurate. {\tt BPZ}
does return a $\chi^{2}$ goodness-of-fit value for each galaxy. But
low $\chi^{2}$ values (indicating good fits) do not guarantee reliable
redshifts. In fact, the opposite was shown to be true in Fig.~8 in
the original {\tt BPZ} paper \citep{BPZ00}. Galaxies
with high $\chi^{2}$ (poor fits) actually have very reliable redshifts.

\begin{figure*}
\begin{center}
\plottwo{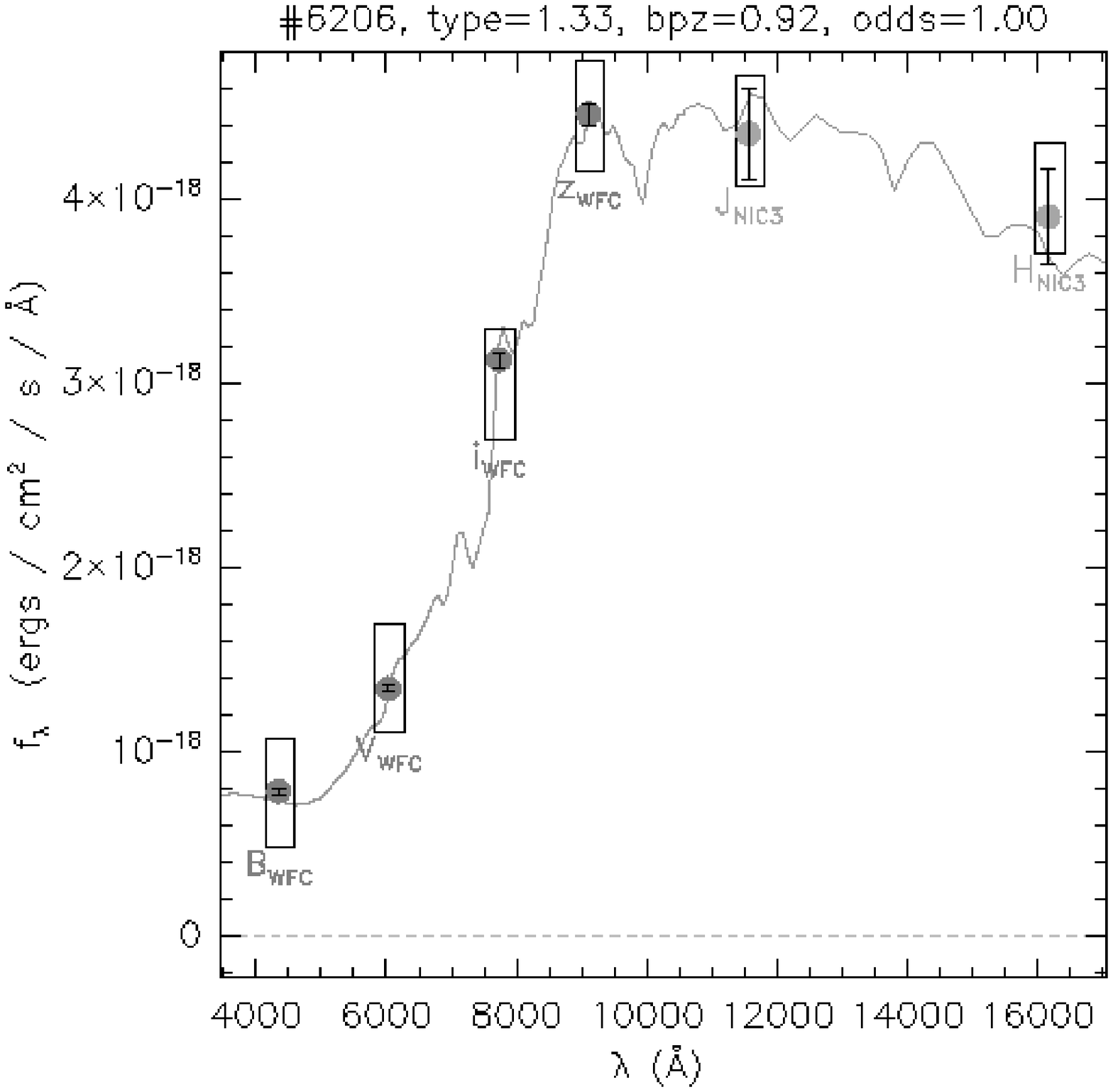}{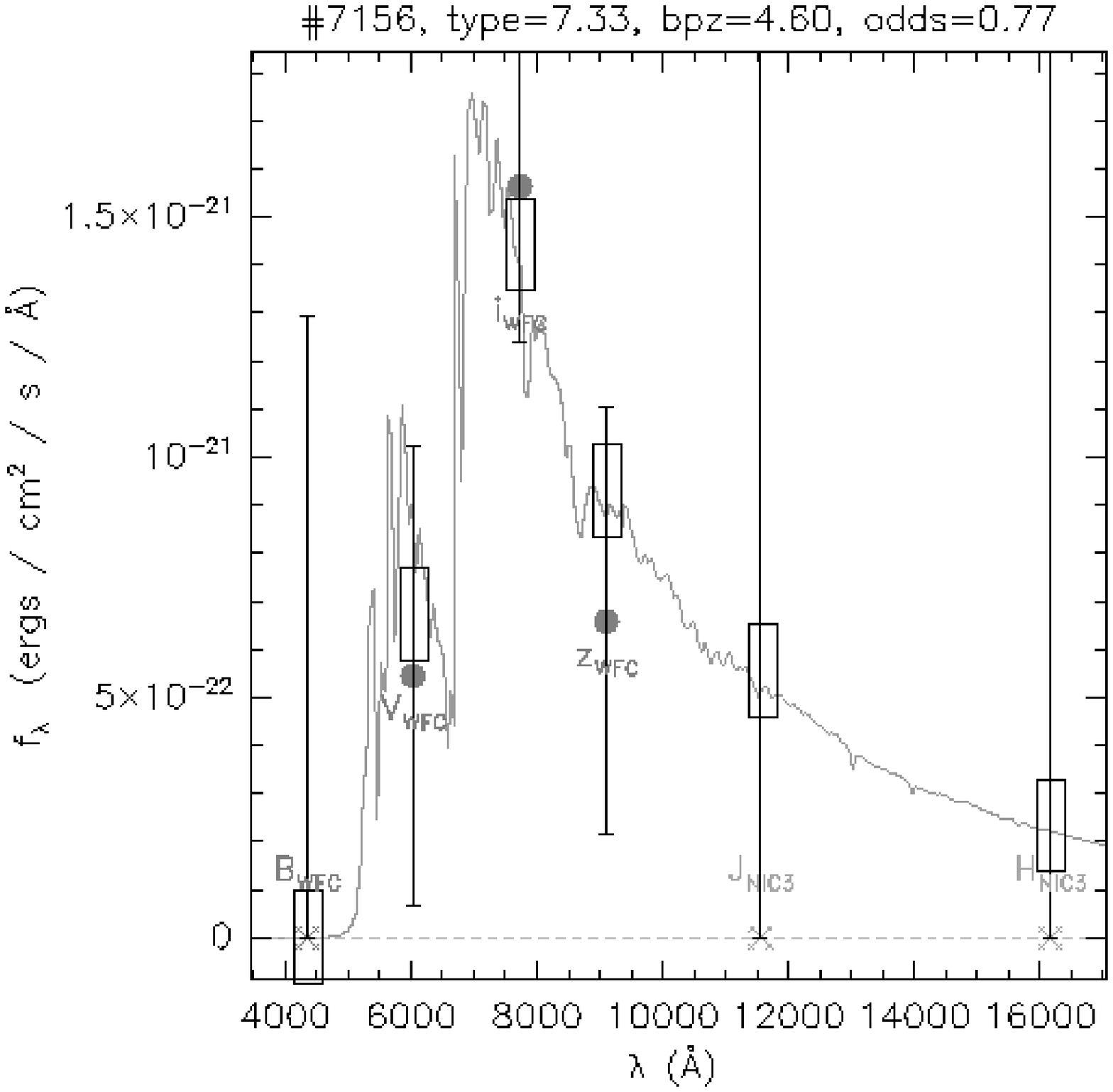}
\end{center}

\caption{\label{cap:thread}SED fits to a bright galaxy (left) and a faint
galaxy (right). Bright galaxies yield more accurate photometric redshifts,
but the small photometric uncertainties almost ensure a high value
of $\chi^{2}$. For this fit, $\chi^{^{2}}=4.27$, even though the
photometric redshift $z_{b}=0.92$ matches the spectroscopic redshift
$z_{spec}=0.95$ very well. Meanwhile, the huge photometric uncertainties
of object \#7156 (only detected at 5-$\sigma$) guarantee a much lower
value of $\chi^{2}$. For this fit, we find $\chi^{^{2}}=0.11$. This
issue is resolved by assigning uncertainty to the model fluxes, as
represented by the heights of the blue rectangles in the plots. The
bright galaxy is now a perfect fit with $\chi_{mod}^{2}=0.03$. The
$\chi^{2}$ value of the faint galaxy rises slightly: $\chi_{mod}^{2}=0.19$.}
\end{figure*}

The reason for this apparent paradox is that all galaxies with high
$\chi^{2}$ are bright galaxies, which have more accurate photometry,
in turn yielding more accurate redshifts. Compare the two SED fits
shown in Fig.~\ref{cap:thread}. The left panel shows the photometry
of a bright galaxy \#6206. The photometric uncertainties are small,
typical of bright galaxies. Object \#6206 has been fit to a hybrid
El-Sbc SED template at a redshift of $z_{b}=0.92$, which agrees well
with the spectroscopic redshift $z_{spec}=0.95$. The SED fits well but was
unable to ``thread the needle'' of small photometric error bars.
Many of the model fluxes are off by several $\sigma$, yielding $\chi^{2}=4.26$.

Faint galaxies on the other hand have much larger error bars, making
high values of $\chi^{2}$ almost impossible to achieve, no matter
how poor the fit. Object \#7156, only detected at 5-$\sigma$, is
an extreme example. The photometric uncertainties are so large, that
no model fluxes can possibly be off by more than 1-$\sigma$, or so.
This guarantees a low value for $\chi^{2}$.

Thus we find more reliable redshifts if we restrict our sample to
those galaxies with high $\chi^{2}$. Bright galaxies are the only
galaxies capable of producing such high values for $\chi^{2}$. But
of course high values of $\chi^{2}$ are supposed to indicate poor
SED fits.

The solution is to assign an uncertainty to the SED itself. This is
represented in Fig.~\ref{cap:thread} by the blue rectangles. These
indicate the model fluxes with uncertainties given by the heights.
Our modified version of $\chi^{2}$ is defined as:

\[
\chi_{mod}^{2}=\sum_{\alpha}\frac{(f_{\alpha}-f_{T\alpha})^{2}}{\sigma_{f_{\alpha}}^{2}+\sigma_{f_{T}}^{2}}\,/\,\textrm{d.o.f.}\]

where $f_{\alpha}$ \& $\sigma_{f_{\alpha}}$ are the observed fluxes
and flux errors and $f_{T\alpha}$ are the model fluxes, normalized
to fit the observed fluxes. $\sigma_{f_{T}}$ serves as our model
flux errors, and we have rather arbitrarily assigned $\sigma_{f_{T}}=\textrm{max}_{\alpha}(f_{T\alpha})/15$.
In other words, $\sigma_{f_{T}}=1/15$ on a scale where the highest
model flux for a given fit is normalized $f_{T\alpha}\equiv1$. This
{}``model thickness'' dominates over the small flux errors of bright
galaxies, yielding a more realistic measure of the goodness of fit.
To obtain a {}``reduced'' $\chi^{2}$, we divide by the number of
degrees of freedom $\textrm{d.o.f.}=\textrm{\# filters observed}-3$.
(3 is the number of fit parameters: $z_{b}$, $t_{b}$, $a$, or redshift,
template, and amplitude. If the object was observed in fewer than
4 filters, we set $\textrm{d.o.f.}=1$.) We now find that object \#6206
is fit almost perfectly by the $z_{b}=0.92$ SED, with $\chi_{mod}^{2}=0.03$.

\begin{figure*}
\includegraphics[width=0.60\linewidth]{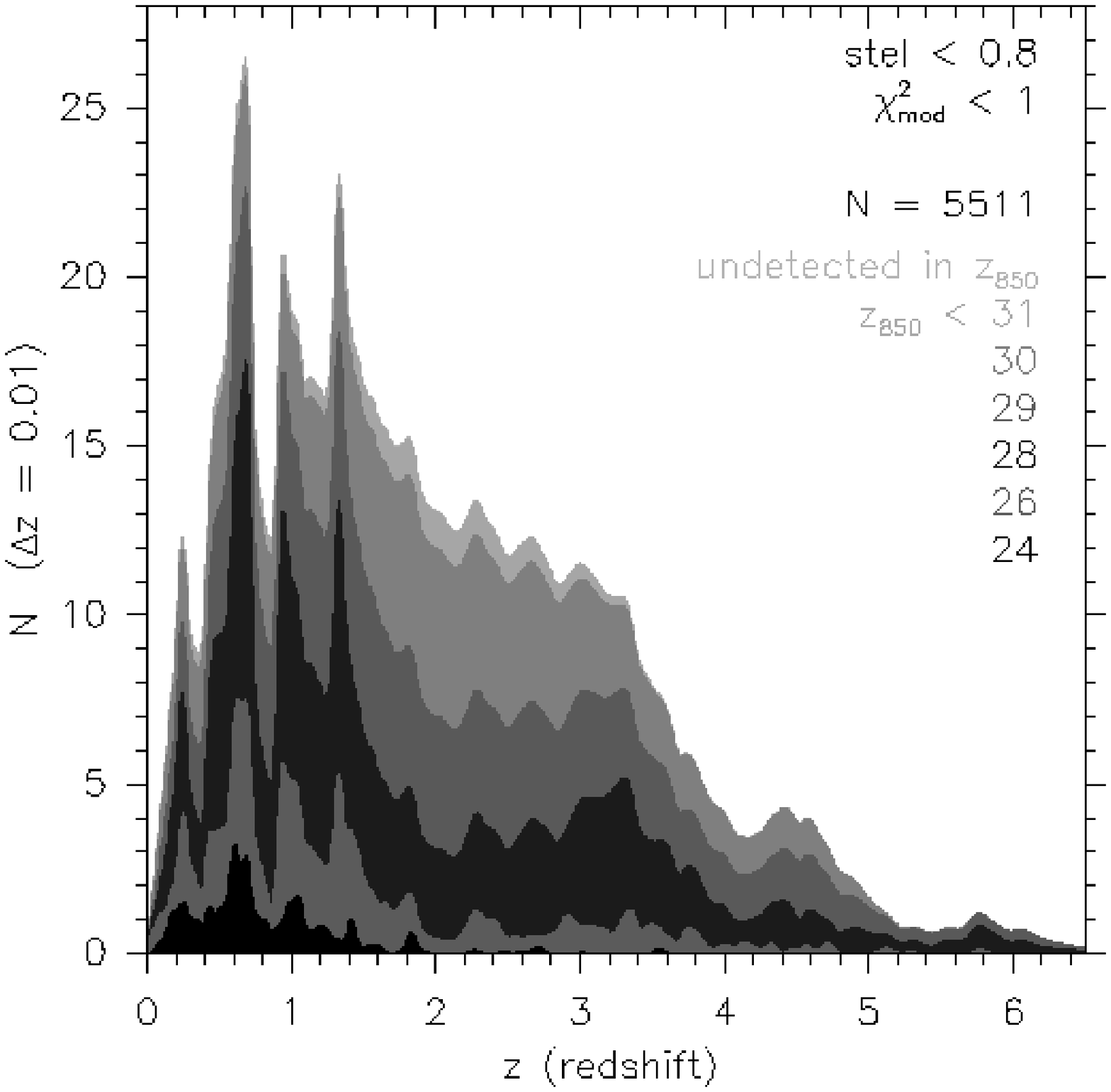} \includegraphics[width=0.39\linewidth]{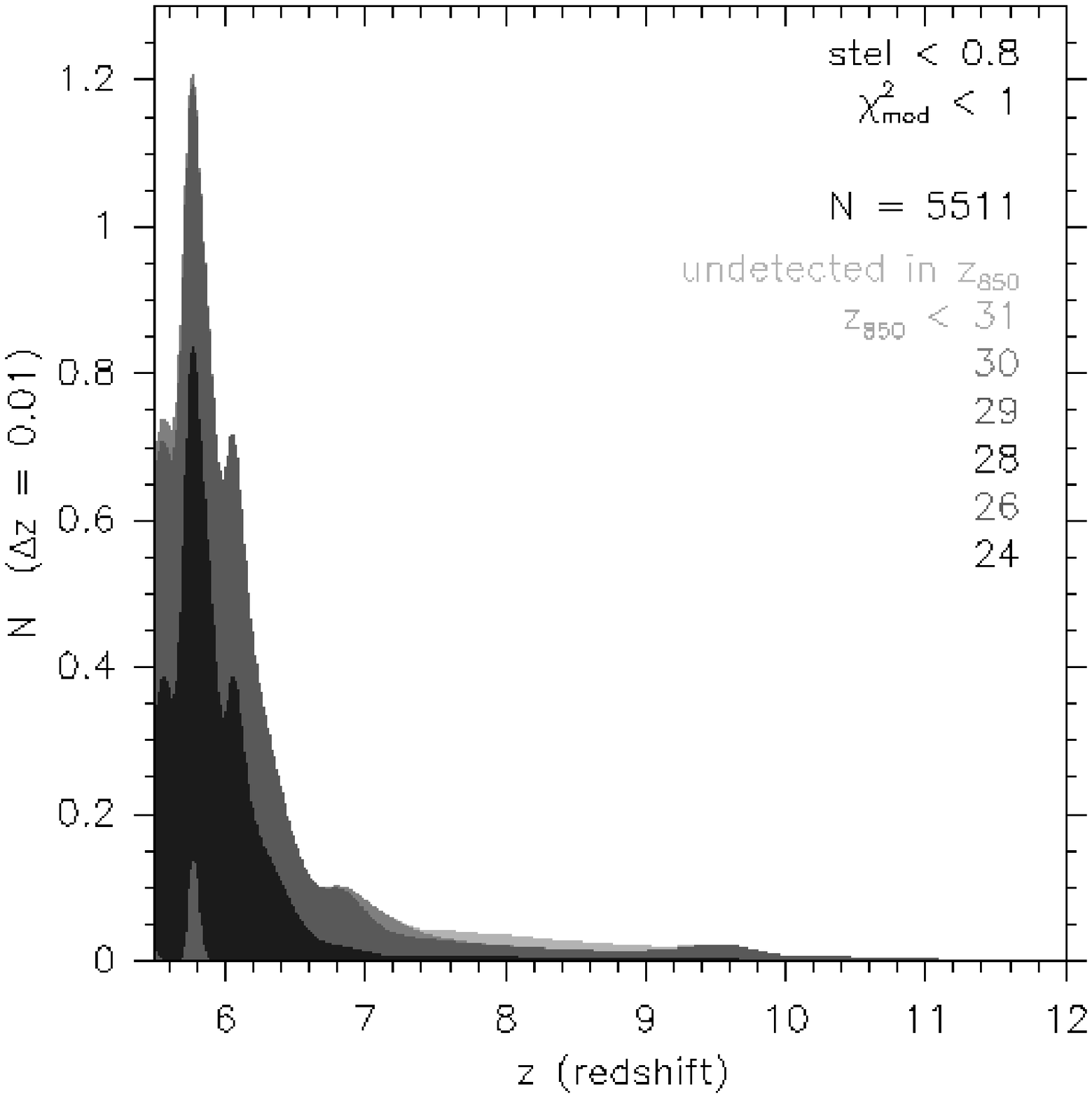}
\caption{\label{cap:bpzphistchisq21}A subset of the galaxies plotted in 
Fig.~\ref{cap:bpzphist}. These 5,511
galaxies all have good SED fits ($\chi_{mod}^{2}<1$).}
\end{figure*}

$\chi_{mod}^{2}<1$ roughly corresponds to all model fluxes fitting
the observed fluxes within their error bars. By eye, we confirm that
galaxies with $\chi_{mod}^{2}>1$ generally have ill-fitting SEDs.
In Fig.~\ref{cap:bpzphistchisq21} we replot our redshift probability
histogram but this time only for those galaxies with $\chi_{mod}^{2}<1$.

\begin{figure}
\plotone{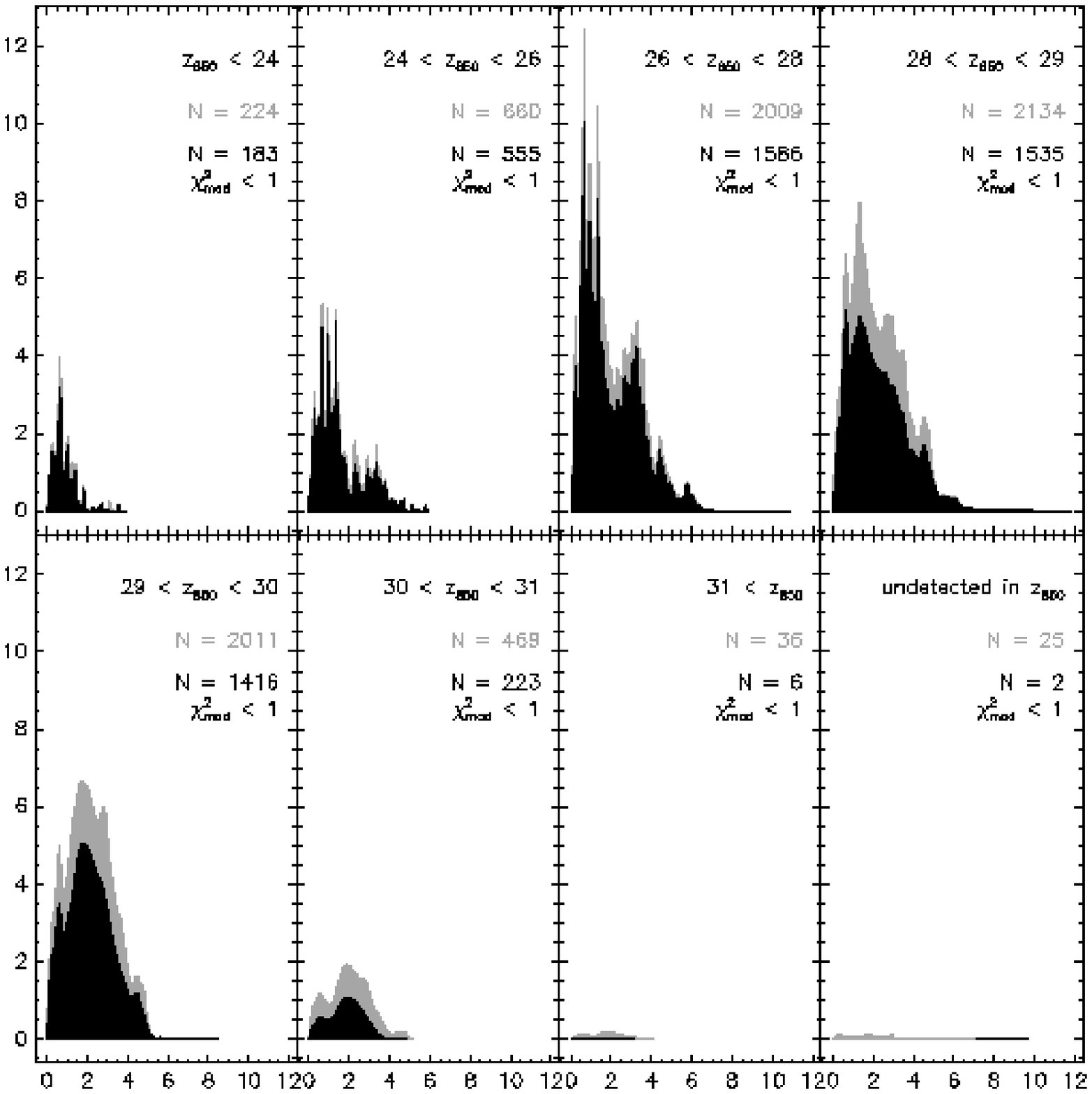}
\caption{\label{cap:bpzphistmag}Same as 
Figs.~\ref{cap:bpzphist} (green) and \ref{cap:bpzphistchisq21} (black),
but broken down into magnitude ranges. The outer green contours plot all
galaxies, while the inner black contours plot only those with good
SED fits ($\chi_{mod}^{2}<1$).}
\end{figure}

\begin{figure}
\plotone{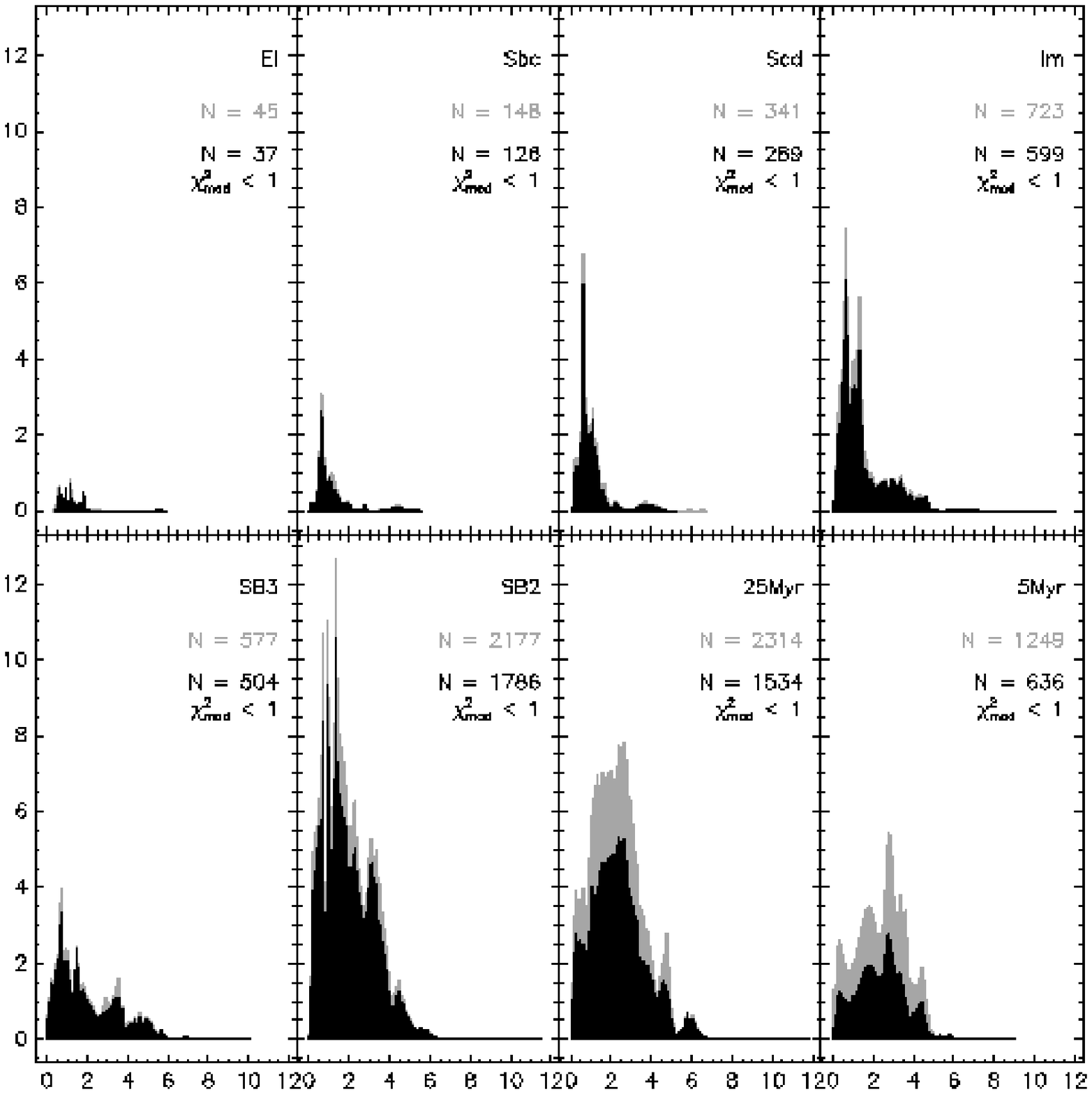}
\caption{\label{cap:bpzphisttbs}Same as 
Figs.~\ref{cap:bpzphist} (green) and \ref{cap:bpzphistchisq21} (black),
but broken down into spectral types, as fit by {\tt BPZ}. Galaxies
fit to {}``interpolated'' spectral types are {}``rounded'' to
the nearest type. The outer green contours plot all galaxies, while
the inner black contours plot only those with good SED fits ($\chi_{mod}^{2}<1$).}
\end{figure}

Fig.~\ref{cap:bpzphistmag} again replots our redshift probability histogram
(Figs.~\ref{cap:bpzphist} and \ref{cap:bpzphistchisq21}),
this time plotting each magnitude range individually. We observe that
the peak of the histogram shifts to higher redshift for fainter galaxies.
And Fig.~\ref{cap:bpzphisttbs} breaks our histogram down by spectral
type.

\begin{figure}
\plotone{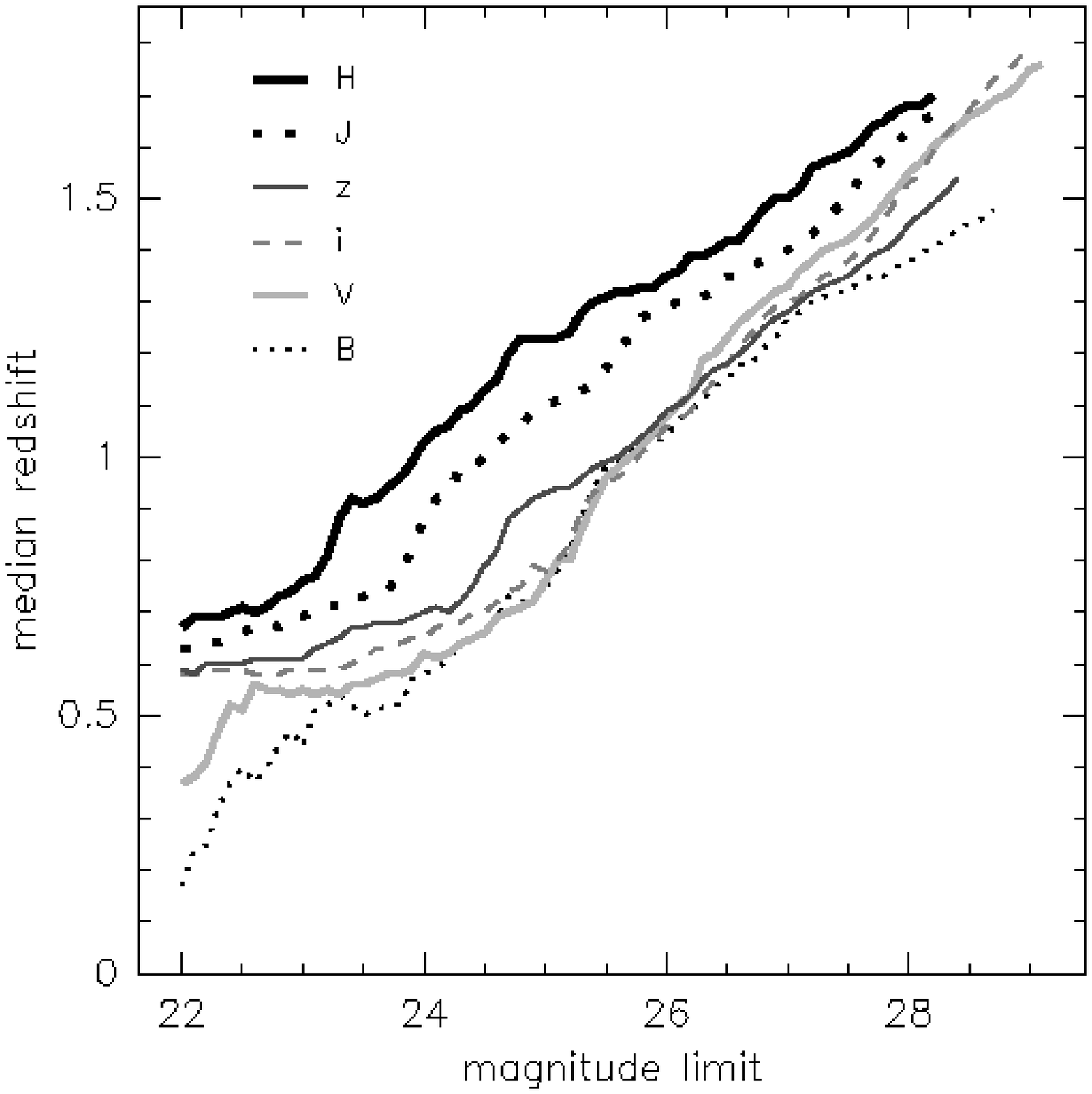}
\caption{\label{cap:zmag}Median redshift as a function of limiting magnitude
(based on magnitude cuts of our redshift probability histogram). For
example, a sample of galaxies complete to $i\arcmin<27$ will have
a median redshift of $z=1.29$. See also Table \ref{cap:zmagtable}.
All curves are truncated at their 10-$\sigma$ completeness limits
(e.g., $i\arcmin=29.01$).
Offsets have been applied to the NIC3 magnitudes (\S\ref{sub:NIC3magoffsets}).}
\end{figure}

Another useful relation is median redshift vs.~limiting magnitude
(Fig.~\ref{cap:zmag} and Table \ref{cap:zmagtable}). For each magnitude
cut ($i\arcmin<27$, for example), we obtain the redshift probability
histogram and find its median. Clustering within the UDF (for example
at $z=0.67$, see \S \ref{sub:Clustering}) may affect these results.
Of course, whenever possible we encourage the use of photometric redshifts
rather than relying on this redshift-magnitude relation.

\clearpage

\begin{deluxetable}{ccccccc}
\tablewidth{0pt}
\tablecaption{\label{cap:zmagtable}Median Redshifts}
\tablehead{
\colhead{magnitude}&
\multicolumn{6}{c}{median($z$)}\\
\colhead{limit (AB)}&
\colhead{$B_{435}$}&
\colhead{$V_{606}$}&
\colhead{$i\arcmin_{775}$}&
\colhead{$z\arcmin_{850}$}&
\colhead{$J_{110}$\tablenotemark{a}}&
\colhead{$H_{160}$\tablenotemark{a}}
}
\startdata
22&
0.17&
0.37&
0.58&
0.59&
0.63&
0.67\\
23&
0.44&
0.55&
0.59&
0.61&
0.69&
0.76\\
24&
0.58&
0.62&
0.65&
0.70&
0.88&
1.03\\
25&
0.75&
0.76&
0.78&
0.93&
1.10&
1.23\\
26&
1.04&
1.08&
1.06&
1.09&
1.29&
1.35\\
27&
1.27&
1.33&
1.30&
1.28&
1.40&
1.50\\
28&
1.38&
1.55&
1.53&
1.45&
1.63&
1.68\\
29&
1.51&
1.75&
1.79&
1.70&
1.74&
1.74
\enddata
\tablecomments{
Example: a sample of galaxies complete down to a limiting magnitude of $i\arcmin<27$
will have a median redshift of $z=1.29$. See also Fig.~\ref{cap:zmag}.}
\tablenotetext{a}{Offsets have been applied to the NIC3 magnitudes (\S\ref{sub:NIC3magoffsets}).}
\end{deluxetable}

\clearpage

\begin{figure}
\plotone{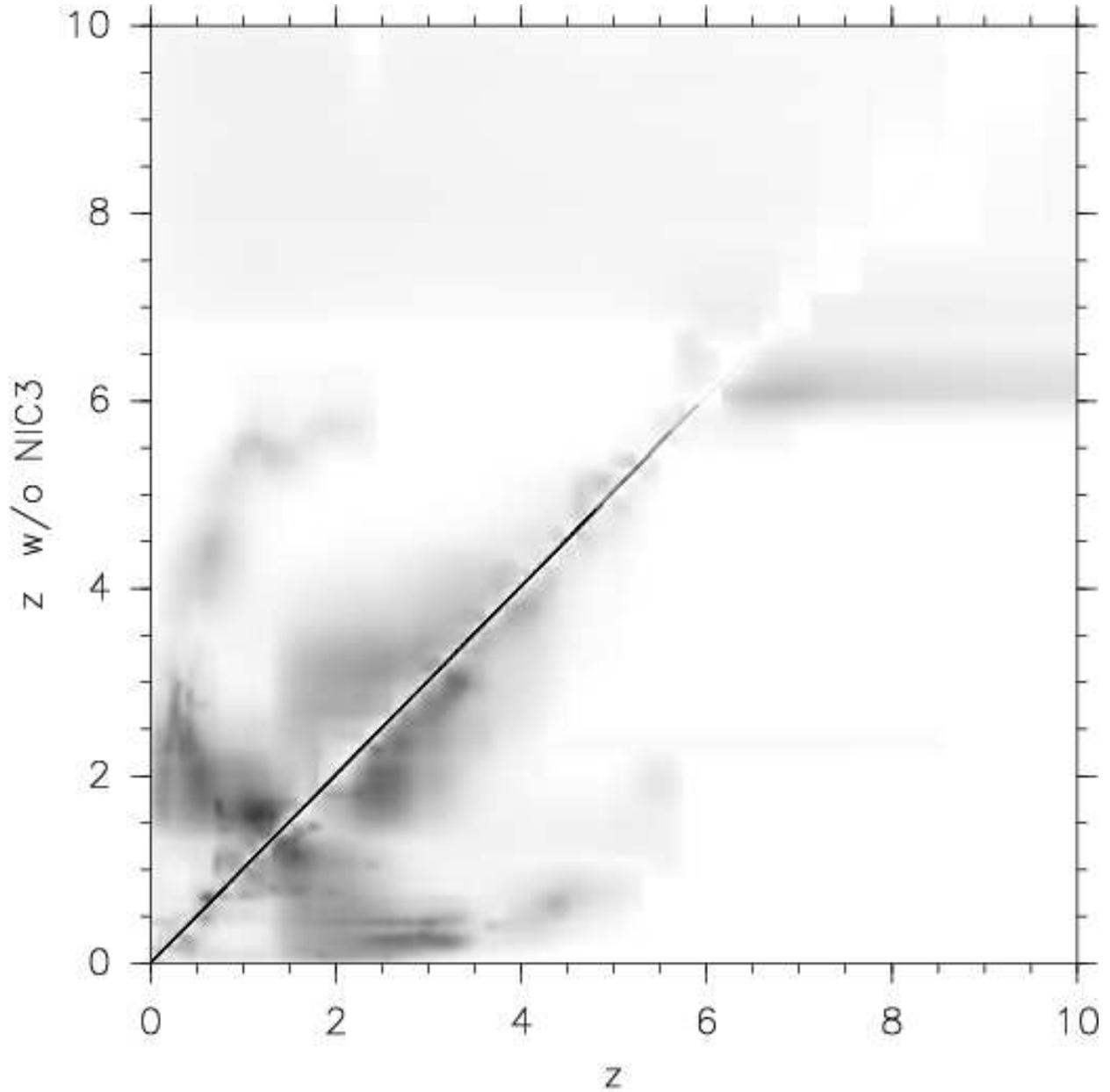}
\caption{\label{cap:BPZwwoNIC3} {\tt BPZ} comparison of the 3,783
galaxies (10-$\sigma$, ${\tt stellarity} < 0.8$) within the NIC3
FOV. {\tt BPZ} was re-run on these galaxies using only
the ACS magnitudes (without NIC3). For each galaxy, we measure the
correlation and migration of $P(z)$ between the $BVi\arcmin z\arcmin$
and $BVi\arcmin z\arcmin JH$ {\tt BPZ} runs. The totals
are plotted here as greyscale: a clipped square root scale is used
to exaggerate the low-level migration.}
\end{figure}

Fig.~\ref{cap:BPZwwoNIC3} examines the 3,783 galaxies
(10-$\sigma$, ${\tt stellarity} < 0.8$)
within the NIC3 FOV and how their {\tt BPZ} results are affected
by the availability of NIC3 photometry. Each galaxy's $P(z)$ is compared
to that obtained when using only the ACS photometry. For most galaxies,
$P(z)$ remains virtually unchanged. This coincident $P(z)$ is plotted
as the dark diagonal line. But we also see migration of $P(z)$, for
example, from $z\sim2.5$ to $z\sim0.3$ and vice versa. This migration
pattern is also plotted in greyscale.%
\footnote{The migration is given by the matrix $A[z,\, z_{ACS}]=\sqrt{dP_{z}(z)\times dP_{zACS}(z_{ACS})}$,
where $dP_{z}(z)$ and $dP_{zACS}(z)$ are the positive and negative
parts respectively of $P(z)-P_{ACS}(z)$.%
} The $BVi\arcmin z\arcmin$ colors of $z\sim2.5$ late type galaxies
are very similar to the $BVi\arcmin z\arcmin$ colors of $z\sim0.3$
earlier type galaxies. Thus without NIC3 photometry, each of these
galaxies is assigned $P(z)$ with two peaks: roughly equal probability
of $z\sim2.5$ and $z\sim0.3$. But the NIC3 photometry is able to
resolve this degeneracy (the nearby earlier type galaxies are brighter
in the near IR), reassigning a single redshift (and uncertainty) to
each galaxy. In Fig.~\ref{cap:BPZwwoNIC3}, this appears as symmetric
migration between $z\sim2.5$ and $z\sim0.3$. We see similar migration
between other pairs of redshifts. And at $z>6$ we see a {}``tail'':
without NIC3 photometry, all $i\arcmin$-dropouts (only detected in $z\arcmin$)
are simply assigned $z\sim6$.

\subsubsection{{\tt BPZ} vs.~Maximum Likelihood}

\begin{figure*}
\plottwo{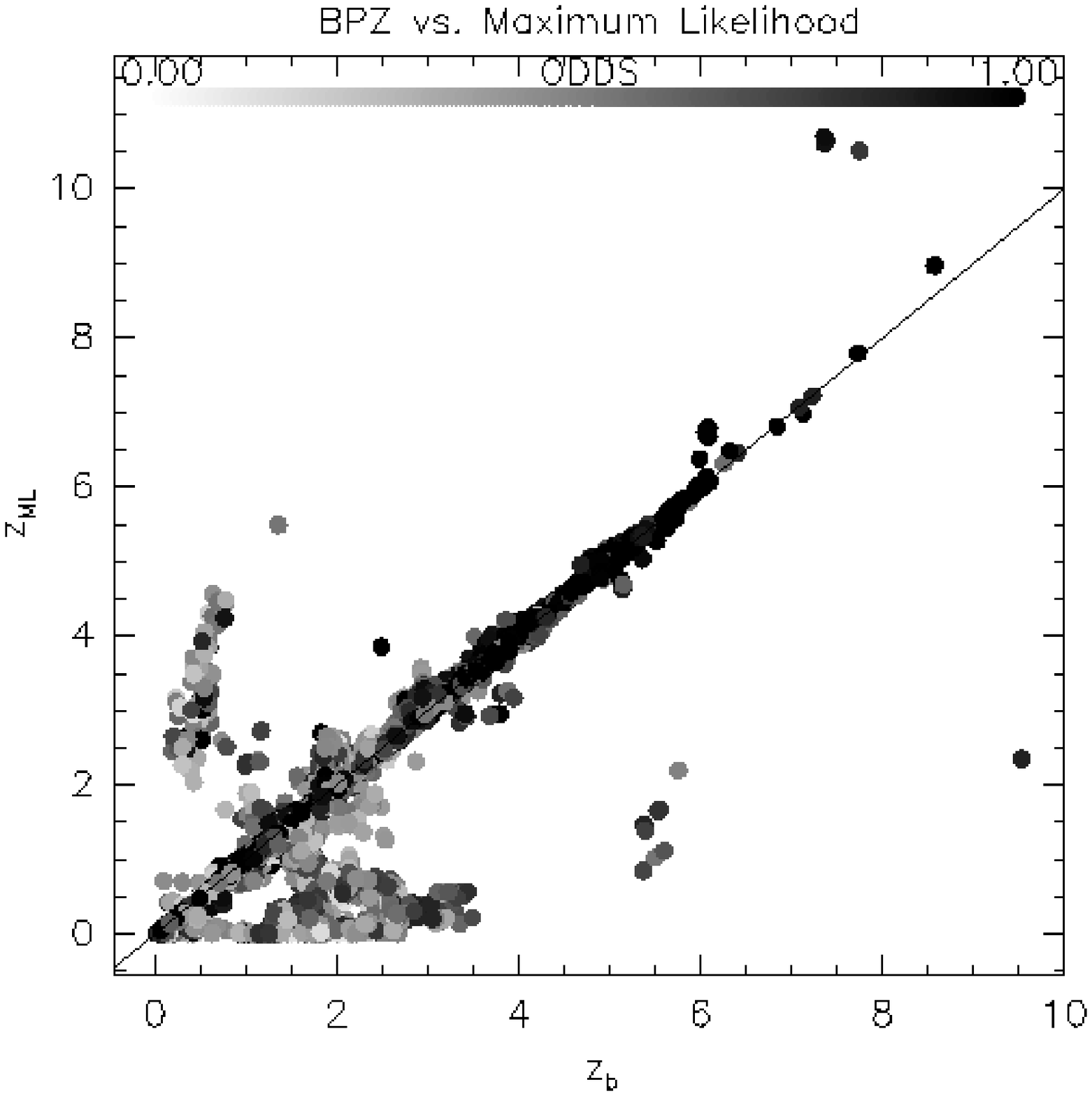}{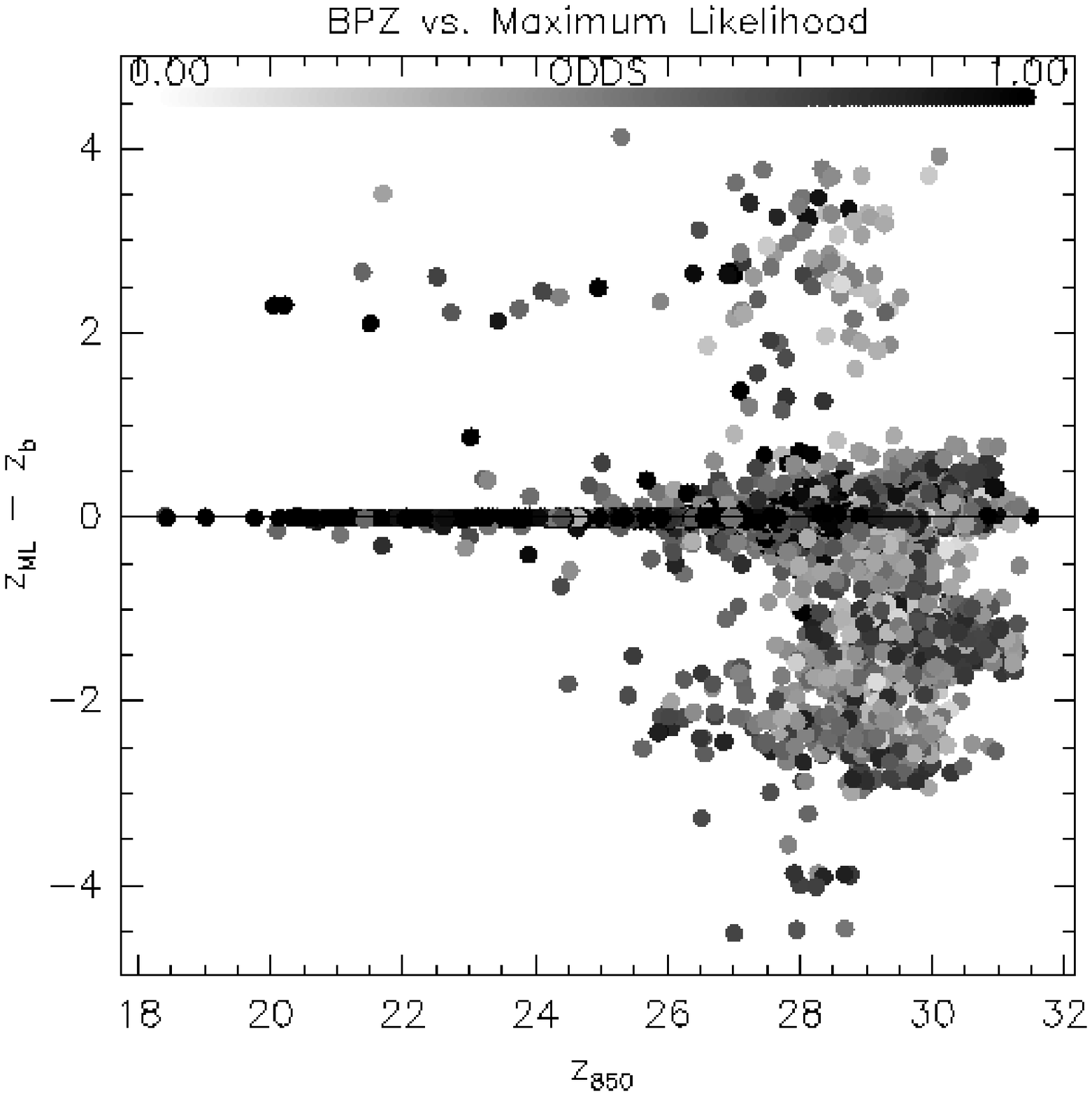}
\caption{\label{cap:BPZvsML} {\tt BPZ} $z_{b}$ vs.~Maximum Likelihood
redshifts for the 10-$\sigma$, ${\tt stellarity} < 0.8$ galaxies.
{\tt BPZ} {\tt ODDS} values are plotted in greyscale. Trimmed
from the plot on the right are galaxies with undetected or {}``unobserved''
(saturated, etc.) $z\arcmin$-band magnitudes and a few extreme outliers
(visible in the plot on the left).}
\end{figure*}

Fig.~\ref{cap:BPZvsML} compares our {\tt BPZ} results
to those obtained from a Maximum Likelihood (ML) approach. Our best
fit $z_{b}$ redshifts generally match very well with the ML redshifts,
especially for bright galaxies. However, at faint magnitudes, there
are a significant number of discrepancies between the two methods.
In most of these cases, {\tt BPZ} realizes its limitations,
assigning low {\tt ODDS} to these galaxies, indicating
a broad or multi-peaked redshift distribution $P(z)$. ML methods
offer no such measure. And of course, ML does not take advantage of
prior knowledge: redshift likelihood as a function of magnitude and
type \citetext{see \citealt{BPZ00} for details}.

\begin{figure*}
\plottwo{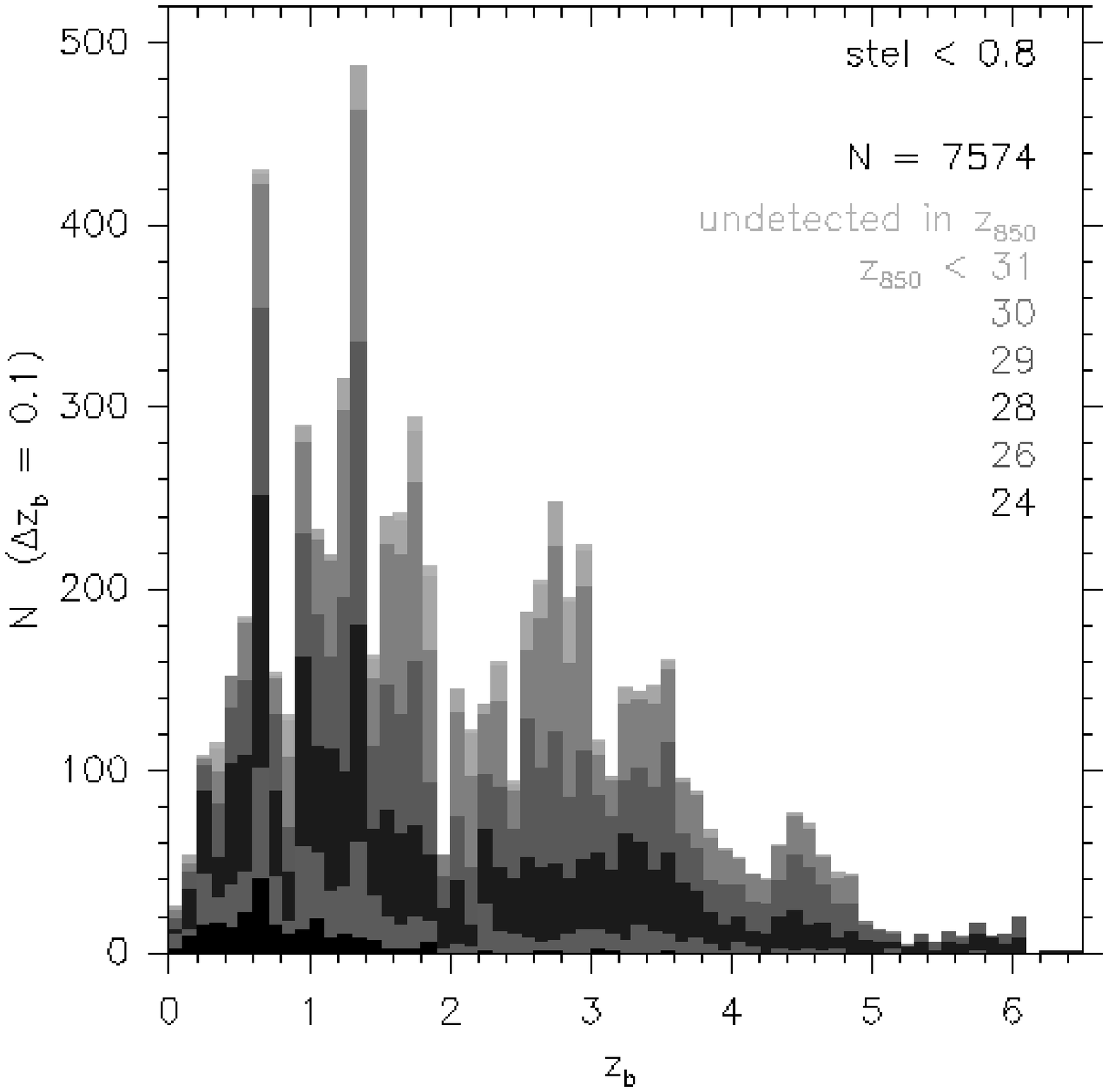}{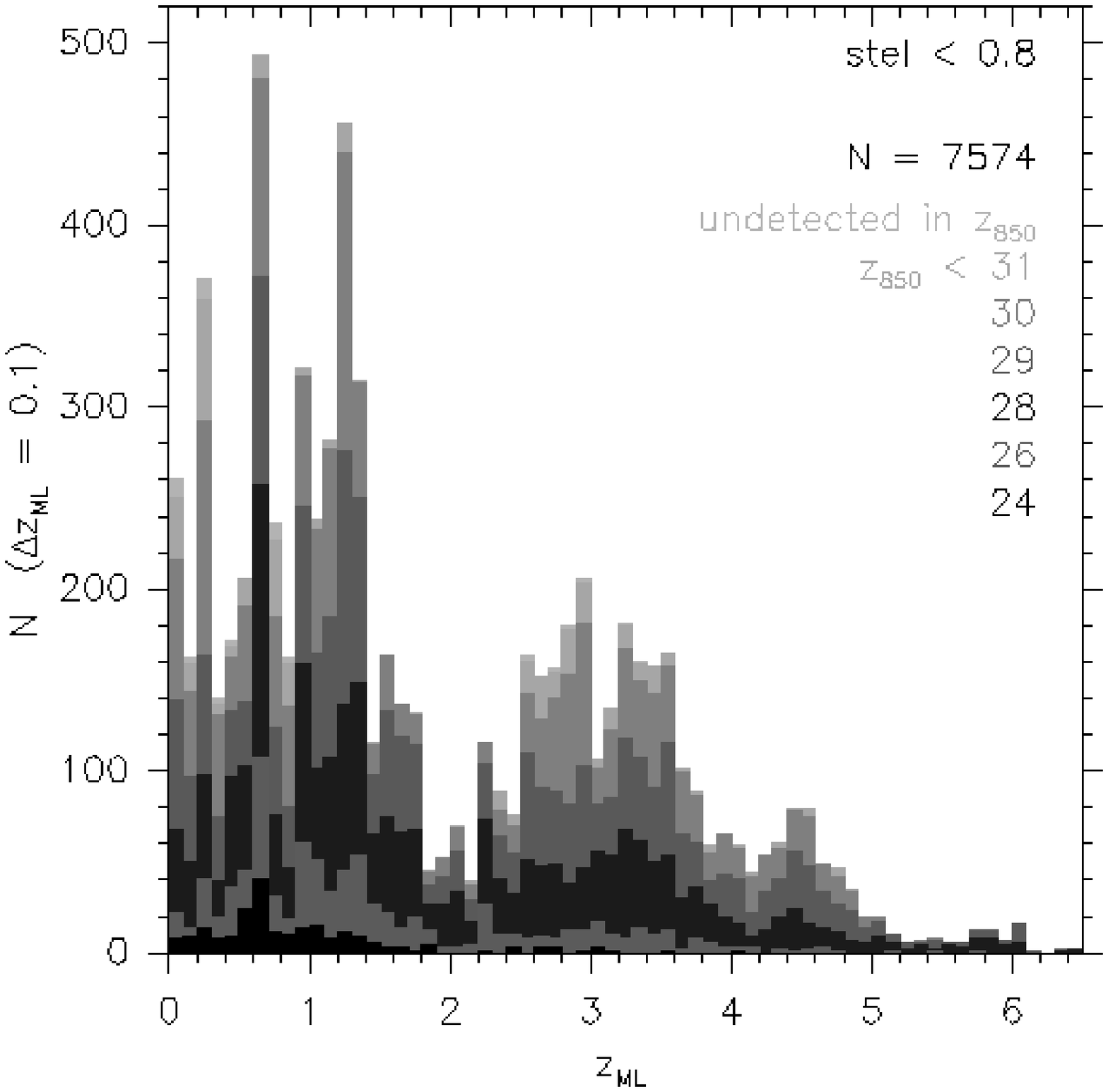}
\caption{\label{cap:zhist}Redshift histograms: same as Fig.~\ref{cap:bpzphist},
except binning single value redshifts rather than probability distributions
$P(z)$. Left: {\tt BPZ} best fit redshifts $z_{b}$.
Right: Maximum Likelihood redshifts $z_{ML}$. The redshift interval
is 0.1.}
\end{figure*}

Fig.~\ref{cap:zhist} compares redshift histograms of the single values
$z_{b}$ and $z_{ML}$ for each galaxy. The $z_{b}$ histogram retains
much of the shape of our full probability distribution histogram 
(Fig.~\ref{cap:bpzphist}), although some differences are apparent. The
$z_{ML}$ histogram is fairly similar for $z\arcmin<28$ galaxies,
but markedly different for fainter galaxies.

\begin{figure}
\plotone{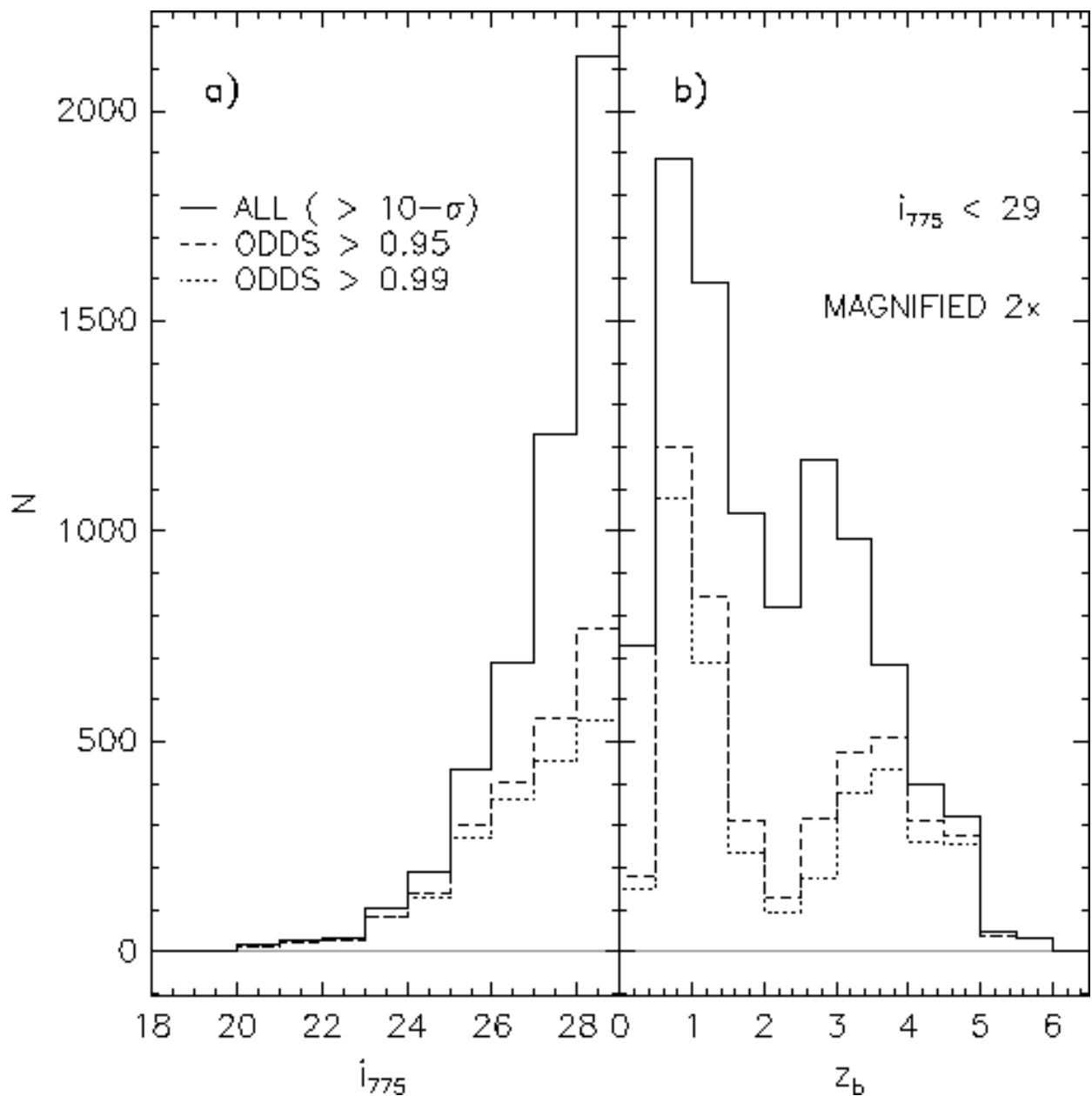}
\caption{\label{cap:oddshist}Left: Histograms of galaxies (10-$\sigma$,
${\tt stellarity} < 0.8$) meeting {\tt ODDS} thresholds of 0.95
and 0.99 as a function of magnitude. Bright galaxies often yield reliable
single value redshifts (high {\tt ODDS}), but fainter galaxies
tend to give a broader probability distribution $P(z)$ (low {\tt ODDS}).
Right: Same as left panel, but as a function of redshift and magnified
$2\times$. A similar figure is plotted for the HDF-N in \citet[Fig.~9]{BPZ00}.}
\end{figure}

Most bright ($i\arcmin<26$) galaxies are well defined by their single
value redshifts $z_{b}$. But fainter galaxies tend to have broader
probability distributions $P(z)$ (i.e., low {\tt ODDS}). 
Fig.~\ref{cap:oddshist}a demonstrates this trend. Fig.~\ref{cap:oddshist}b
replots the same histograms versus redshift $z_{b}$. The low fraction
of high {\tt ODDS} galaxies in the $z_{b}<4$ redshift bins is
simply due to the abundance of faint galaxies with uncertain photometry.
Galaxies are especially hard to pin down to a single redshift between $2<z_{b}<3$. 
But note that higher redshift ($z_{b}>4$) galaxies are typically dropouts, 
leaving little doubt about their redshifts. 

A plot similar to Fig.~\ref{cap:oddshist} can be found in \citet[Fig.~9]{BPZ00}.
The results for the HDF-N were the same: $i\arcmin<26$ galaxies were generally
well defined by their single value redshifts.
$i\arcmin \gtrsim 27$ galaxies were hard to pin down, and unfortunately
they are still hard to pin down in the UDF. There seems to be a strong redshift 
degeneracy affecting such faint galaxies, probably due to the fact that their 
average color becomes very blue. This may represent a barrier to the effectiveness 
of typical photometric redshifts based on a few broad band colors. 

\subsection{Clustering}
\label{sub:Clustering}

\begin{figure*}
\plottwo{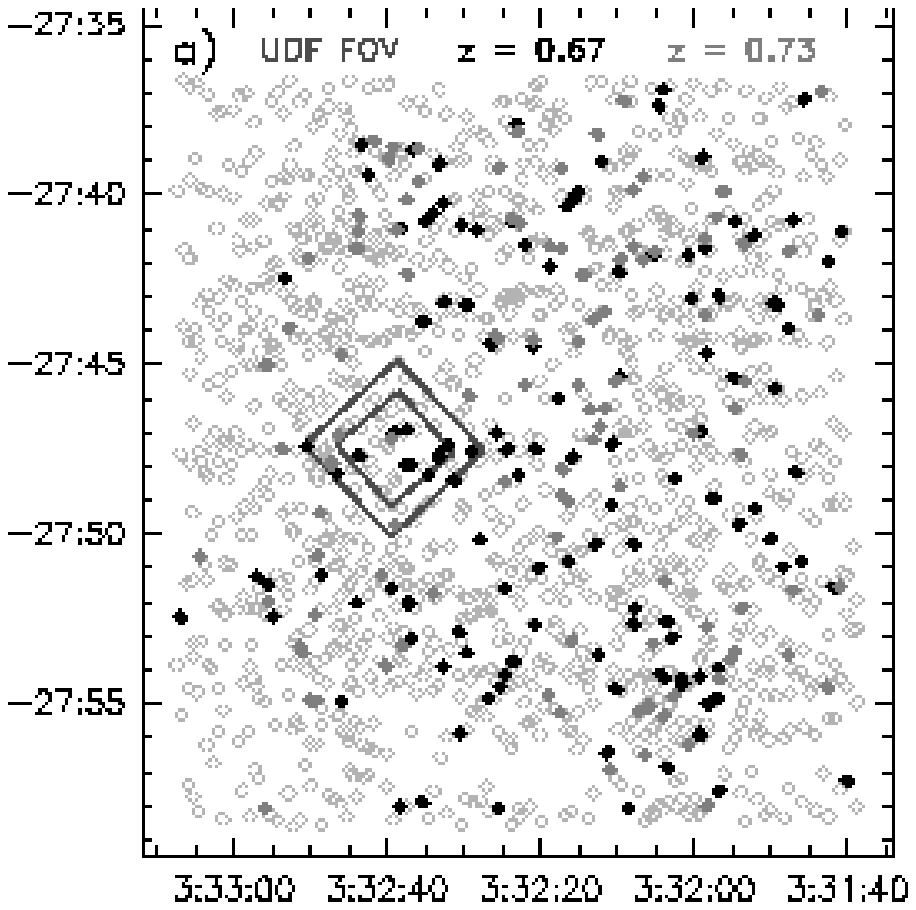}{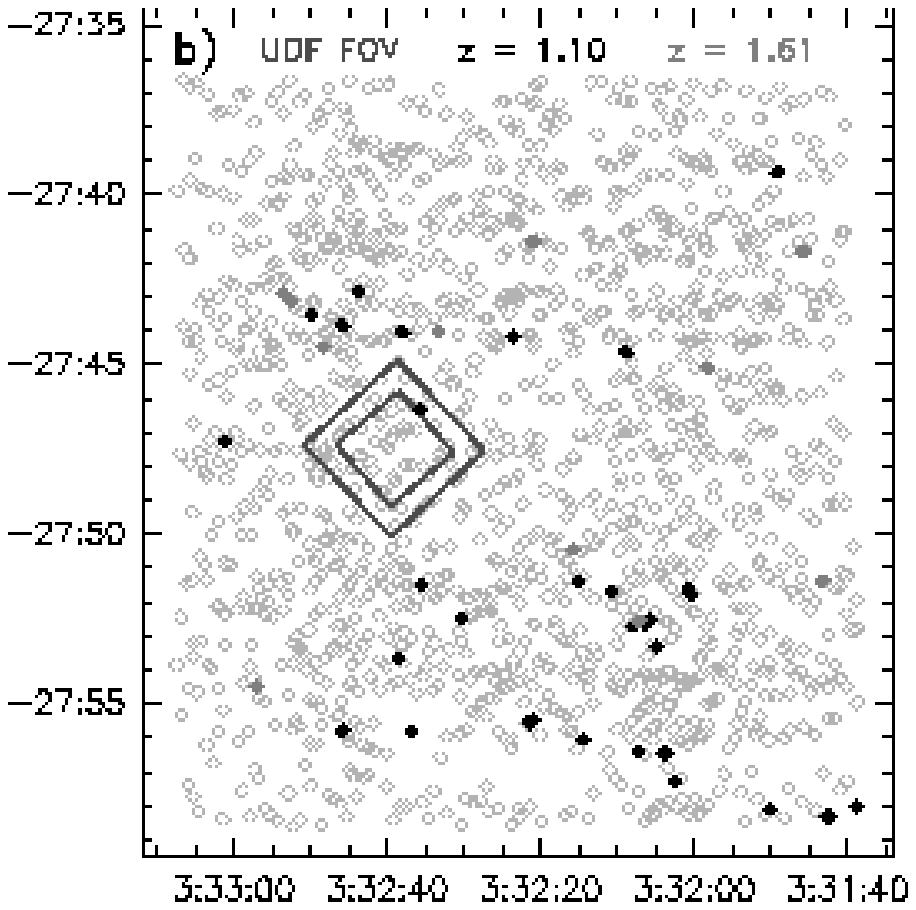}
\caption{\label{cap:clusters}Spectroscopic redshifts obtained by VVDS plotted
on the sky (RA \& Dec J2000). Left: the $z=0.67$ group ($0.66\leq z\leq0.685$)
is plotted in blue. The $z=0.73$ cluster ($0.72\leq z\leq0.74$)
is plotted in red. The WFC \& NIC3 FOVs of the UDF are shown in green.
Note that the UDF appears to be along a filament of the $z=0.67$
group, but perhaps within a void of the $z=0.73$ cluster. Our {\tt BPZ}
results support this. Right: galaxies with $1.09\leq z\leq1.11$ (blue)
and $1.60\leq z\leq1.64$ (red) cluster in wall-like patterns, deftly
avoiding the UDF.}
\end{figure*}

We detect a strong peak of galaxies at $z\sim0.67$ (see Fig.~\ref{cap:bpzphist}).
Previous studies have identified a group of galaxies at $z=0.67$
and a (denser) cluster of galaxies at $z=0.73$ in the wider CDF-S
field \citetext{\citealt{Cimatti02}, \citealt{Gilli03}, \citealt{Croom01}, VVDS, FORS2}.
These galaxies are plotted on the sky in 
Fig.~\ref{cap:clusters}a using the spectroscopic redshifts obtained by
VVDS (which is more densely populated than FORS2). The UDF appears
to be situated along a clump of the $z=0.67$ group but perhaps within
a void of the $z=0.73$ cluster. This claim relies on small number
statistics, but our {\tt BPZ} results do lend some credence
to it. When we zoom in on Fig.~\ref{cap:bpzphist}, we find that our
redshift distribution exhibits a strong peak at $z\sim0.67$, but
then falls off sharply at $z\sim0.73$. However it is questionable
whether our {\tt BPZ} results can be trusted down to this
resolution.

Clusters have also been previously identified within the larger field
at $z=1.04$, 1.10, 1.61, \& 2.57 \citetext{\citealt{Gilli03}, FORS2}. These
clusters also seem to avoid the UDF. The right panel of Fig.~\ref{cap:clusters}
shows two of these overdensities, again using VVDS.
We find no evidence for any of these overdensities within the UDF FOV.

\begin{figure*}
\includegraphics[width=0.32\linewidth]{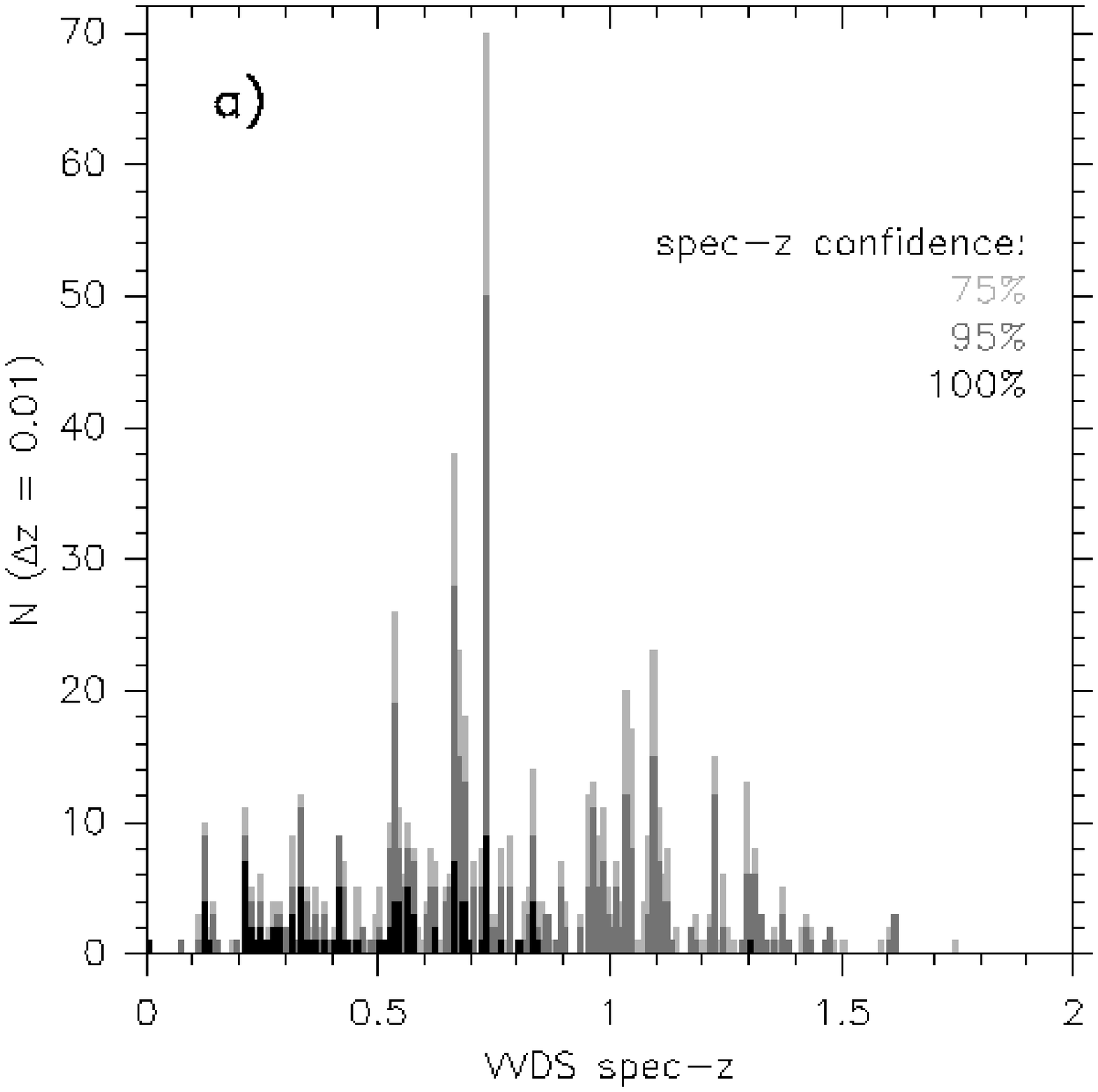}
\includegraphics[width=0.32\linewidth]{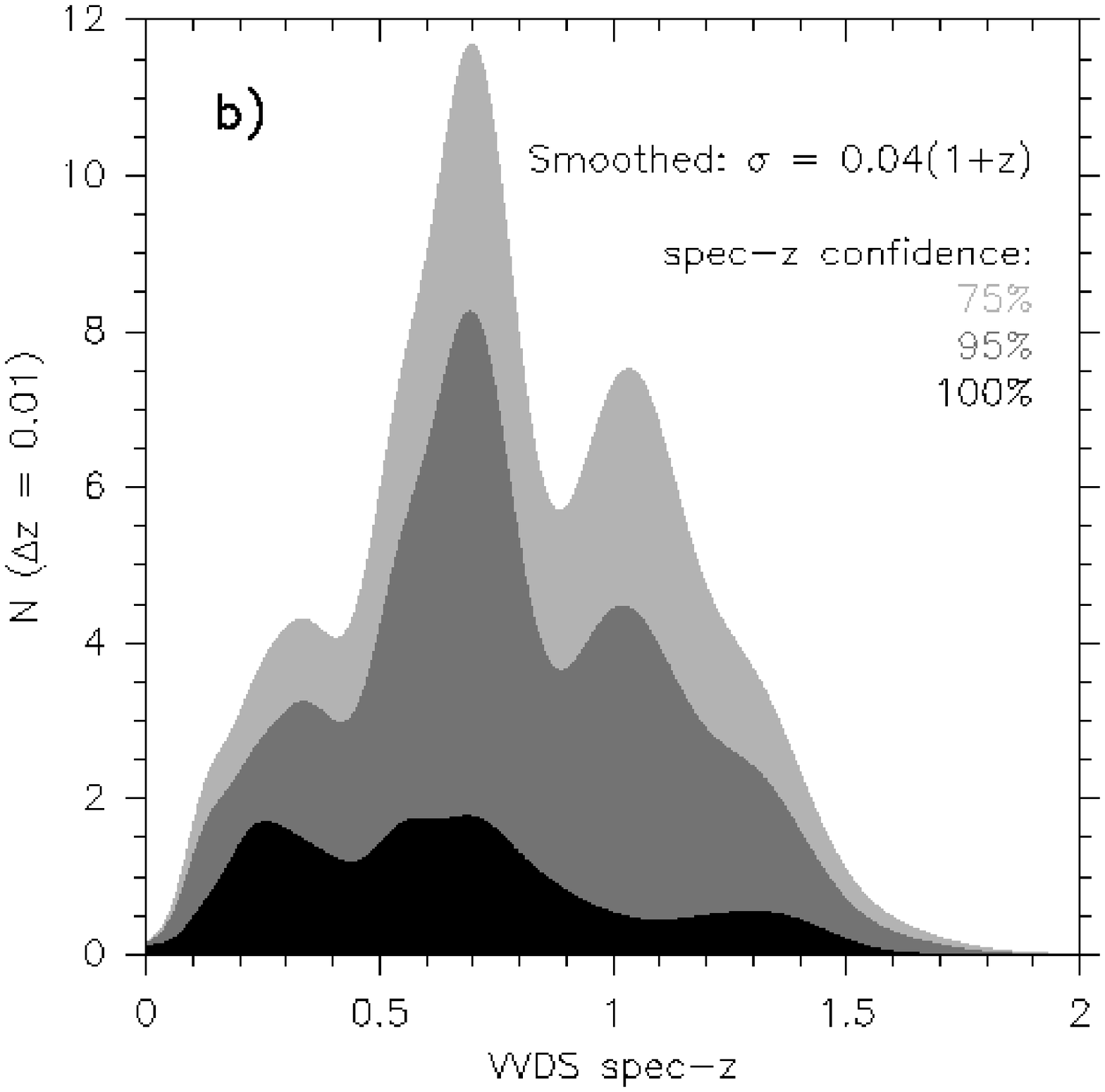}
\includegraphics[width=0.32\linewidth]{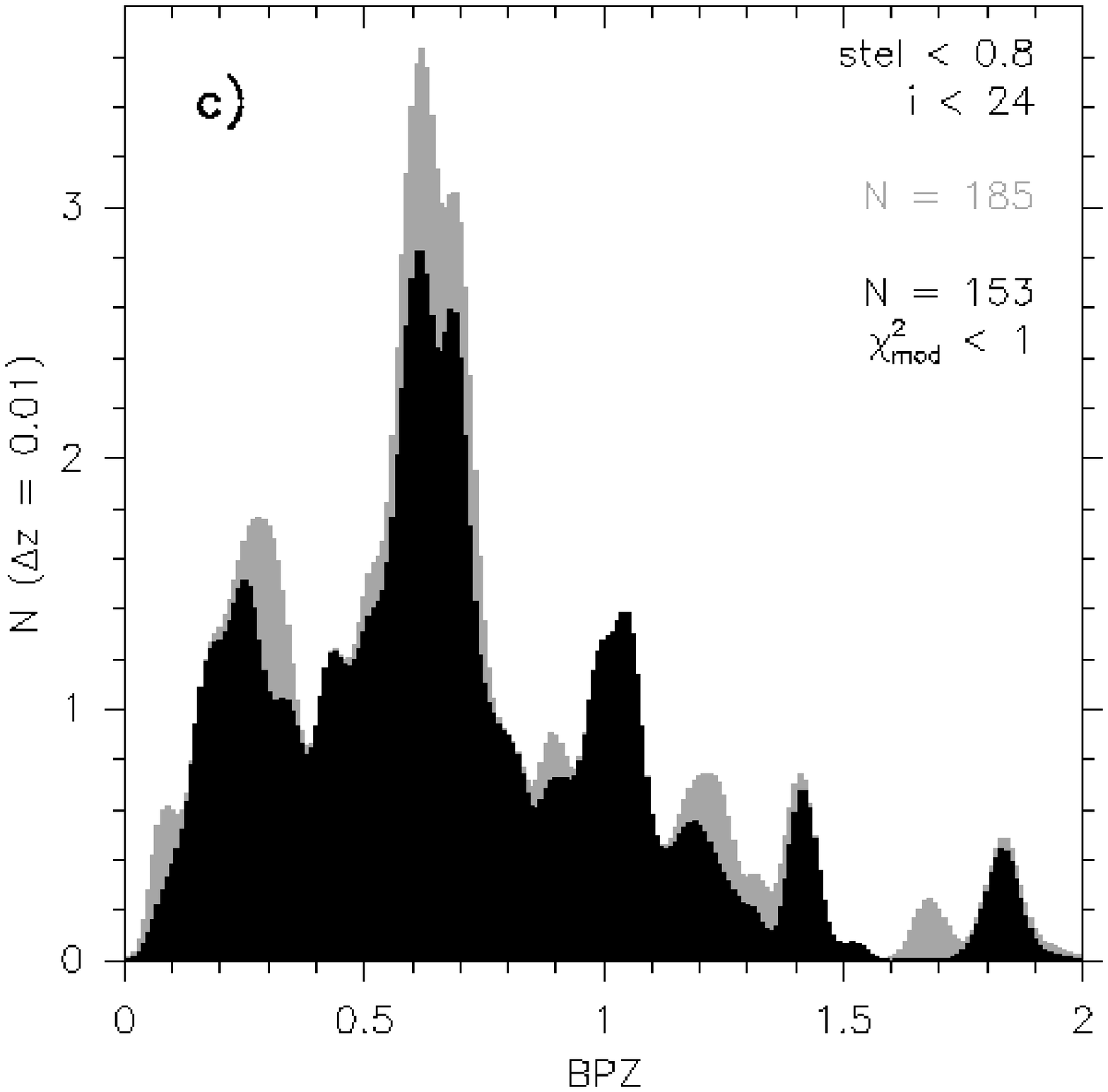}
\caption{\label{cap:VVDSzhist}Left: Histogram of VVDS spectroscopic redshifts for galaxies 
within the larger CDF-S field.  All galaxies are $I<24$ and have spec-z confidence $\geq 75\%$.
Galaxies at $z > 2$ are not plotted.
Center: The same histogram smoothed by $\Delta z=0.04(1+z)$, 
or how it might be observed by {\tt BPZ}.
Right: Our UDF {\tt BPZ} probability histogram (also cut off at $z < 2$) 
for similarly bright galaxies: $i\arcmin < 24$.
Note the general good agreement with the center panel.}
\end{figure*}

But how sensitive is {\tt BPZ} to overdensities?
In Fig.~\ref{cap:VVDSzhist}a we plot a histogram of spectroscopic redshifts from the VVDS.
Fig.~\ref{cap:VVDSzhist}b plots this same histogram as it might be observed by {\tt BPZ},
allowing for redshift uncertainties of $\Delta z=0.04(1+z_{spec})$.
(Note that we should expect {\tt BPZ} to perform this well here,
as the VVDS galaxies are all bright: $I < 24$.)
Note that some of the smaller peaks (e.g., $z = 0.13$) get washed out by the uncertainty.
Redshift peaks only remain at $z \sim 0.7$, 1.05, \& 0.3 (in order of decreasing prominence).
These broad peaks do show up in the UDF FOV, as we see in Fig.~\ref{cap:VVDSzhist}c
which plots our {\tt BPZ} probability histogram for similarly bright galaxies: $i\arcmin < 24$.
But again, some of the narrower peaks are not apparent.

\subsection{High Redshift Candidates}
\label{sub:high-z}

Addressing the issue of reionization, previous authors 
\citep{Bunker04,Stanway05,YW04,Bouwens05}
have identified $i\arcmin$-dropouts in the UDF. Their catalogs
are in good general agreement. When we apply similar criteria to our
catalog, we also find decent agreement. The $i\arcmin$-dropout technique
is robust: {\tt BPZ} is not significantly better at identifying
$z\sim6$ galaxies. (At lower redshifts, photometric redshifts do
yield a more complete catalog than Lyman-break techniques.)

There is little disagreement among the various authors about the number
of $z\sim6$ galaxies, and yet these 50+ objects leave much open to
interpretation. For example, a luminosity function must be assumed
to address the issue of completeness. There is even some debate
surrounding the exact conditions required for reionization \citep{Stiavelli04}.
Sorting out these issues is beyond the scope of this paper
and has been addressed in depth elsewhere \citep[e.g.,][]{Bouwens05}.
For now we simply present our {}``best'' $z\geq6$ candidates within
the NIC3 FOV (Table \ref{cap:highz}).

In Paper II, we will return to all of these issues briefly, 
including detailed comparisons of our $z\sim 6$ results with previous authors.
Yet, there is much to be learned at $z<6$.
The SFR history of the universe remains fairly uncertain between $3 \la z \la 6$.
The UDF provides allows us to probe the $z\sim 4$ luminosity function
with a complete sample of galaxies all the way down to $M_\star + 3$.
Thus Paper II will concentrate on the $z<6$ SFR history in the UDF.

\clearpage

\begin{deluxetable}{rcccccccccccc}
\tabletypesize{\scriptsize}
\rotate
\tablewidth{0pt}
\tablecaption{\label{cap:highz}``Best'' High Redshift ($z_{b}\geq6$) Candidates}
\tablehead{
\colhead{ID}&
\multicolumn{2}{c}{RA \& DEC (J2000)}&
\colhead{$z_b$}&
\colhead{$\chi^2_{mod}$}&
\colhead{$B_{435}$}&
\colhead{$V_{606}$}&
\colhead{$i\arcmin_{775}$}&
\colhead{$z\arcmin_{850}$}&
\colhead{$J_{110}$}&
\colhead{$H_{160}$}
}
\startdata
32521*&
03:32:36.625&
-27:47:50.06&
$ 6.03\pm0.83$&
0.01&
$>30.37$&
$>30.76$&
$29.31\pm0.28$&
$26.83\pm0.06$&
$26.76\pm0.30$&
$26.74\pm0.33$\\
4110*&
03:32:41.573&
-27:47:44.24&
$ 6.03\pm0.83$&
0.21&
$>30.75$&
$>31.17$&
$29.18\pm0.18$&
$26.73\pm0.04$&
$26.84\pm0.19$&
$26.71\pm0.20$\\
32042&
03:32:40.553&
-27:48:02.60&
$ 6.03\pm0.83$&
0.45&
$>31.48$&
$>31.91$&
$31.11\pm0.47$&
$28.43\pm0.09$&
$28.25\pm0.70$&
$>28.64$\\
31496&
03:32:39.127&
-27:48:18.47&
$ 6.04\pm0.83$&
0.13&
$>31.97$&
$>32.41$&
$31.46\pm0.42$&
$28.96\pm0.11$&
$28.42\pm0.67$&
$>28.26$\\
34321&
03:32:44.701&
-27:47:11.57&
$ 6.04\pm0.83$&
0.53&
$30.67\pm0.44$&
$>31.87$&
$30.69\pm0.36$&
$28.06\pm0.07$&
$27.97\pm0.28$&
$28.18\pm0.34$\\
33268&
03:32:34.526&
-27:47:34.84&
$ 6.05\pm0.83$&
0.65&
$>31.75$&
$>32.16$&
$31.24\pm0.43$&
$28.59\pm0.09$&
$28.01\pm0.39$&
$>28.77$\\
34942&
03:32:34.575&
-27:46:58.00&
$ 6.05\pm0.83$&
0.75&
$31.21\pm0.73$&
$31.19\pm0.54$&
$31.56\pm0.75$&
$28.27\pm0.10$&
$27.99\pm0.61$&
$>28.20$\\
8033*&
03:32:36.473&
-27:46:41.45&
$ 6.05\pm0.83$&
0.92&
$>30.63$&
$30.93\pm0.72$&
$28.64\pm0.13$&
$26.05\pm0.02$&
$26.13\pm0.14$&
$25.71\pm0.14$\\
32007&
03:32:42.797&
-27:48:03.24&
$ 6.07\pm0.83$&
0.24&
$>31.59$&
$>32.01$&
$31.16\pm0.46$&
$28.16\pm0.07$&
$27.80\pm0.31$&
$28.22\pm0.49$\\
7730*&
03:32:38.282&
-27:46:17.22&
$ 6.08\pm0.83$&
0.28&
$29.79\pm0.30$&
$30.48\pm0.38$&
$29.84\pm0.25$&
$26.67\pm0.03$&
$26.39\pm0.16$&
$26.25\pm0.17$\\
35616&
03:32:37.690&
-27:46:21.57&
$ 6.10\pm0.84$&
0.21&
$>31.66$&
$>32.05$&
$>31.93$&
$28.63\pm0.11$&
$28.32\pm0.64$&
$28.14\pm0.67$\\
8545*&
03:32:37.465&
-27:46:32.67&
$ 6.26\pm0.85$&
0.39&
$>30.26$&
$>30.66$&
$>30.54$&
$26.66\pm0.06$&
$26.19\pm0.12$&
$26.22\pm0.13$\\
33003*&
03:32:35.053&
-27:47:40.18&
$ 6.32\pm0.86$&
0.06&
$>31.20$&
$>31.60$&
$31.20\pm0.63$&
$27.81\pm0.07$&
$27.11\pm0.17$&
$26.96\pm0.17$\\
4050*&
03:32:33.427&
-27:47:44.88&
$ 6.40\pm0.87$&
0.17&
$30.35\pm0.45$&
$31.30\pm0.67$&
$29.63\pm0.20$&
$27.29\pm0.05$&
$26.60\pm0.16$&
$26.68\pm0.17$\\
34987*&
03:32:42.560&
-27:46:56.62&
$ 6.84\pm0.92$&
0.02&
$>30.90$&
$>31.31$&
$>31.18$&
$28.04\pm0.12$&
$26.54\pm0.16$&
$26.02\pm0.14$\\
41066&
03:32:42.558&
-27:47:31.39&
$ 7.13\pm0.96$&
0.04&
$>31.20$&
$>31.62$&
$>31.50$&
$29.53\pm0.32$&
$27.51\pm0.23$&
$26.76\pm0.19$\\
41092&
03:32:38.798&
-27:47:07.11&
$ 7.73\pm1.03$&
0.30&
$>31.06$&
$>31.48$&
$>31.37$&
$>30.67$&
$26.81\pm0.19$&
$26.66\pm0.19$\\
41107&
03:32:40.937&
-27:47:41.83&
$ 8.57\pm1.12$&
0.00&
$>30.82$&
$31.09\pm0.67$&
$>31.18$&
$>30.35$&
$24.86\pm0.16$&
$23.71\pm0.18$
\enddata
\tablecomments{
We select here only galaxies detected in
multiple filters including at least one NIC3 filter, with {\tt ODDS}
$\geq0.95$, and with $\chi_{mod}^{2}<1$. 
ID numbers below 41000 correspond to B04 \& T04 detections; 
asterisks ({*}) indicate that object definitions have been altered (\S\ref{sub:detection}).
Magnitudes are ``total'' AB magnitudes with isophotal colors: 
NIC3 magnitudes are corrected to the PSF of the ACS images (\S\ref{sub:photometry}). 
We have also applied offsets of ($J$: $-0.30\pm0.03$, $H$: $-0.18\pm0.04$) 
to the NIC3 magnitudes (\S\ref{sub:NIC3magoffsets}). And all of our
magnitudes have been corrected for galactic extinction (Table \ref{cap:filters}).
Non-detections (listed, for example, as $>31.05$) quote the
1-$\sigma$ detection limit of the aperture used on the given object.
$z_{b}$ gives the peak of the Bayesian photometric redshift distribution
$P(z)$, while $\chi_{mod}^{2}$ measures how poorly the best fitting
SED template at $z_{b}$ fits the observed colors. Color images of
these objects along with SED fits and more are available at {\tt http://adcam.pha.jhu.edu/\~{}coe/UDF/z6g/}.}
\end{deluxetable}

\clearpage

\section{Summary}
\label{sec:Summary}

We have presented a catalog of photometry, Bayesian photometric redshifts,
and morphological parameters for galaxies in the UDF
(8,042 of which are detected at 10-$\sigma$).
Our comprehensive catalog combines $i\arcmin$, $z\arcmin$, $J$+$H$,
and $B$+$V$+$i\arcmin$+$z\arcmin$ detections. To facilitate comparison
with catalogs released by B04 and T04, most
of our object definitions are taken directly from their segmentation maps. 
Our robust photometric method corrects the $z\arcmin$, $J$, \& $H$ magnitudes
for the wider PSFs observed in those bands.
NIC3 magnitudes proved too faint relative to ACS magnitudes of galaxies with known spectroscopic redshift.  
To correct for this, magnitude offsets ($J$: $-0.30\pm0.03$, $H$: $-0.18\pm0.04$) were applied to our catalog.
Part of these offsets ($J$: 0.08, $H$: 0.09) have since been attributed to
a slight miscalibration of the filter response curves used to produce the NICMOS Treasury catalog \citep{Thompsoncal}.
While the rest of the offsets appear to stem from 
a count-rate dependent non-linearity in NICMOS \citep{NICMOScal}.

The UDF reveals a large population of faint blue galaxies (presumably
young starbursts), bluer than those observed in the original Hubble
Deep Fields (HDF). 
We present a redshift histogram derived from full {\tt BPZ} probability distributions.
A strong peak is observed at $z\sim 0.67$, 
corresponding to a known group of galaxies in the wider CDF-S.
Our results and software packages are available at {\tt http://adcam.pha.jhu.edu/\~{}coe/UDF/}.

\acknowledgements{We are indebted to Steven Beckwith, Rodger Thompson, 
their teams, and to everyone who has worked to 
provide the astronomical community with this tremendous dataset.
We would also like to thank Roelof de Jong, Rodger Thompson, Bahram Mobasher, Louis Eddie Bergeron, and Adam Riess
for numerous valuable conversations regarding the UDF NIC3 calibration.
And we thank Stephen Gwyn for discussing his independent derivation of identical empirical offsets.
We especially thank the referee for helpful comments that have improved the paper.
ACS was developed under NASA contract NAS 5-32865, 
and this research is supported by NASA grant NAG5-7697. 
We are grateful for an equipment grant from the Sun Microsystems, Inc.
This work has also been supported by the European Commission Marie Curie International
Reintegration Grant 017288-BPZ and the PNAYA grant AYA2005-09413-C02.}

\clearpage

\appendix

\section{PSF Matching}

We used the {\tt DAOPHOT II/ALLSTAR} software package
\citep{daophot} to determine the $i\arcmin$, $J$, \& $H$ PSFs in
the reduced images provided by STScI (that is after drizzling but
before any re-mapping). The brightest non-saturated stars in the $i\arcmin$
image were sufficient for {\tt DAOPHOT} to compute an
average PSF. The $i\arcmin$ PSF was accepted as the PSF for all WFC
images, including the $B$+$V$+$i\arcmin$+$z\arcmin$ detection
image $d$. (We verified that the PSFs for all five images are nearly
identical. The PSF of the $z\arcmin$ image is slightly worse, and
we deal with it separately in \S\ref{sub:zapcor}.)

Unfortunately these stars alone could not be used to determine the
NIC3 PSFs. The NIC3 images of these stars are highly asymmetric (perhaps
due to the image reduction). So we allowed {\tt DAOPHOT}
to average many much fainter {}``stars'' to determine the NIC3 PSFs.
Many of these objects are not stars at all; they are resolved in the
WFC images as extended objects. But these objects are still narrow
enough to appear as point sources to NIC3. Thus they are suitable
for use in the NIC3 PSF determination. 

Once the $i\arcmin$, $J$, \& $H$ PSFs are determined, the $J$
\& $H$ PSFs are re-mapped to the WFC frame
(using {\tt IRAF}'s {\tt wregister}, {\tt interp=spline3}).
(The re-mapping process is controlled to ensure that each re-mapped 
NIC3 PSF is centered on a pixel in the WFC frame.)
Then {\tt IRAF}'s {\tt psfmatch} is used to determine
the kernels necessary to degrade the $i\arcmin$-band PSF to match
the re-mapped NIC3 PSFs. We use these psfmatching kernels to degrade
$d$ (which has a PSF nearly identical to $i\arcmin$) to the NIC3
PSFs, yielding images $d^{J}$ \& $d^{H}$. Meanwhile, the NIC3 images
are re-mapped to the WFC frame, yielding images $J^{A}$ \& $H^{A}$
(see Fig.~\ref{cap:apertures}). Finally, by training identical apertures
on a given galaxy in $d^{J}$ \& $J^{A}$, we obtain a robust measure
of the $d-J$ color for that galaxy. Similarly, we obtain the $d-H$
color. The $d-B$, $d-V$, $d-i\arcmin$, \& $d-z\arcmin$ colors
are all obtained without the need for such PSF matching gymnastics,
and we arrive at 6 colors referenced against the same image, $d$.
(Again, the $z\arcmin$ PSF correction is handled separately in \S\ref{sub:zapcor}.)

A much simpler PSF matching approach proves unreliable: instead of determining
the PSFs explicitly, one could simply use {\tt IRAF}'s
{\tt psfmatch} to degrade $d$ directly to the PSF of the remapped NIC3 images
$J^A$ \& $H^A$. But {\tt psfmatch} has
a difficult time determining the $J$ \& $H$ PSFs from the re-mapped
NIC3 images. Each star in the NIC3 images is pixelized and therefore
slightly asymmetric. These asymmetries are greatly magnified by the
re-mapping. {\tt psfmatch} returns highly distorted kernels, with
very significant effects on the magnitudes of faint objects. The asymmetries
can be averaged out much more effectively \emph{before} re-mapping.

\section{Morphology Simulations}
\label{sub:morphsims}

To determine the limits of reliability of the extracted galaxy profile
parameters we have performed a set of simulations. Simulated galaxies
were created using \galfit\ assuming a S\'ersic profile and fixing
all the model parameters to a set of input values. Particular care
was taken to resample the noise histogram and noise pattern seen in
the UDF ACS dithered images, as described in \citet{Sanchez04}.
Our first simulation reproduces the UDF, with all galaxy positions
and morphological parameters as derived from the UDF $i\arcmin$-band
image. Our second simulation retains the galaxy positions (preserving
the observed clumpiness) but shuffles the galaxies among them. Also,
random scatter is added to the observed morphological parameters.
Each galaxy's effective radius $R_{e}$, semiaxis ratio $a/b$, and S\'ersic
index $n$ were changed randomly within 20\% of their original values.
Magnitude $i\arcmin$ was altered within $\pm0.25$ mags, and
the position angle $\theta$ was randomly rotated. Our third and final
simulation bears little resemblance to the UDF. The observed catalog
is scrapped in favor of new galaxies which are distributed homogeneously throughout the
image. Galaxy parameters are chosen randomly with flat distributions:
$21<i\arcmin<31$,~ $0.2<b/a<0.8$,~ $0.2<n<8$,~ $0<\theta<2\pi$.
The effective radii follow a relation $\log_{10}(R_{e})\sim5.62-0.18\times i\arcmin\pm0.5$,
where $R_{e}$ is measured in pixels, and a random scatter of 0.5
has been added to the logarithm. Due to the flat magnitude distribution,
this last simulation yields many more large and bright galaxies. To
avoid overcrowding the field, we create only 5,000 galaxies.

\begin{figure*}
\epsscale{0.9}
\plotone{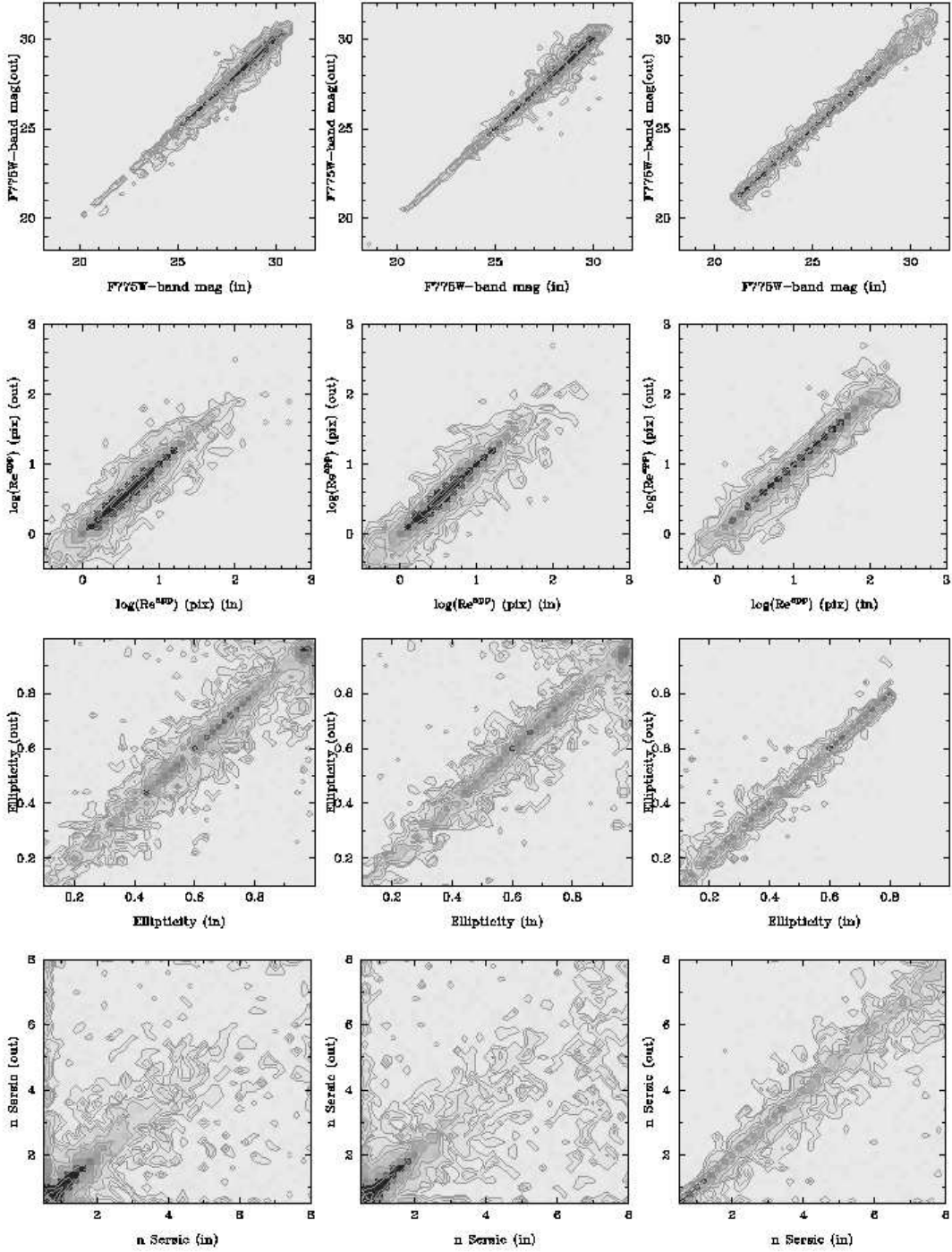}
\epsscale{1.0}
\caption{\label{cap:morphsims}Distribution of the output parameters as a
function of the input parameters for the three simulations. The panels
include from top to bottom the distributions of the $i\arcmin$-band
magnitudes, the effective radii $R_{e}$, the ellipticities and the
S\'ersic indices $n$, and from left to right, the results from
the first, second and third simulations.}
\end{figure*}

The simulated images are analyzed by running \sex\  and \galfit\ 
as was done for the UDF $i\arcmin$-band image. Output galaxies are
matched to input galaxies by object position (centroids within $R_{e}/4$
and within 10 pixels). We find agreement between input
and output parameters, with no appreciable biases and the following scatters:
$\Delta i\arcmin=0.35$,~ $\Delta R_{e}/R_{e}\sim0.2$,~ $\Delta(b/a)=0.17$,~ $\Delta n=1.6$
(see Fig.~\ref{cap:morphsims}).

\begin{figure*}
\plotone{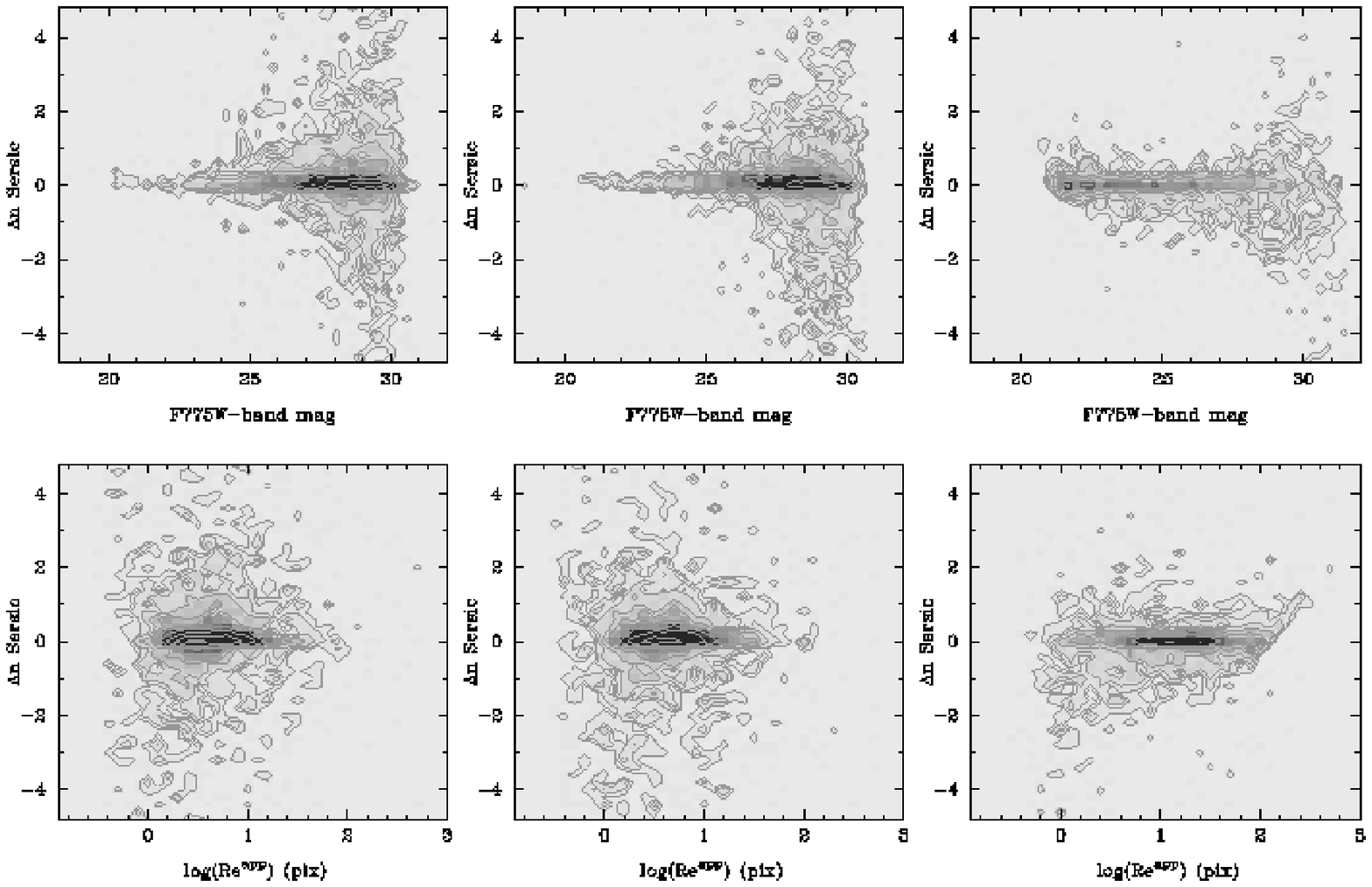}
\caption{\label{cap:deltan}
S\'ersic index errors  $\Delta n = n_{out} - n_{in}$
as a function of the $i\arcmin$-band magnitudes (top row) 
and as a function of the effective radii $R_{e}$ (bottom row)
for the three simulations (left to right).}
\end{figure*}

\begin{figure*}
\plotone{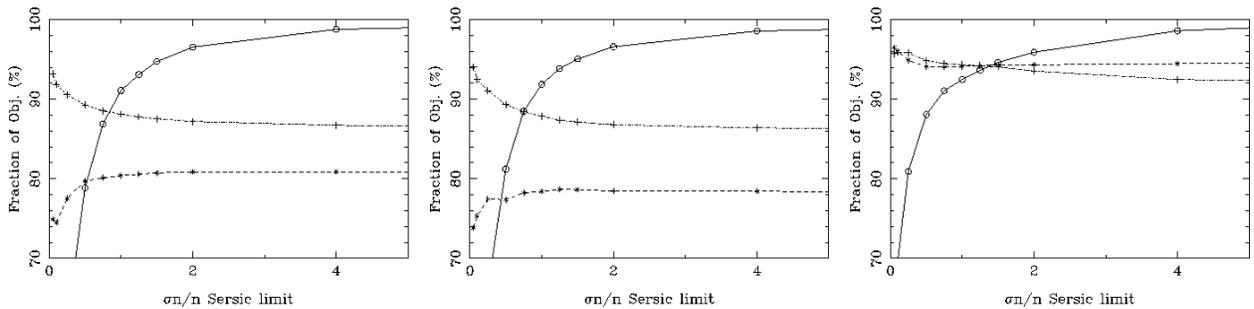}
\caption{\label{cap:latevsearly} 
Effects of pruning the catalogs resulting from the three simulations (left to right)
based on S\'ersic index uncertainty (e.g., $\sigma n/n<1$).
The solid black lines (with circles) show the total fraction of galaxies spared by the cuts.
The dot-dashed green (stars) and dashed red (crosses) lines show
the fraction of well classified late and early type galaxies in the pruned catalogs
(assuming that $n=2.5$ is a perfect discriminator).}
\end{figure*}

As mentioned in \S\ref{sub:morph}, we adopt $n=2.5$ as the dividing
line between late ($n<2.5$) and early ($n>2.5$) type galaxies. But
our ability to classify galaxies as late or early type depends on
the \galfit\  uncertainty $\sigma n/n$. (Note that \galfit's $\sigma n$
is not defined as a typical RMS uncertainty.)
From our simulations, we find we retrieve $n$ to within $\Delta n=1$
for bright ($i\arcmin<26$) and large ($R_{e}>10$ pixels) galaxies
and to within $\Delta n=2$ for faint ($i\arcmin>28$) and small ($R_{e}<3$
pixels) galaxies (Fig.~\ref{cap:deltan}). If we assume that $n=2.5$
is a perfect discriminator between late and early type galaxies, then
we can make some simple predictions about our ability to classify
galaxies. For example, if we select only those galaxies with $\sigma n/n<1$,
then 80-95\% of them will be correctly classified as late ($n<2.5$)
or early ($n>2.5$) (Figs.~\ref{cap:latevsearly}a \& \ref{cap:latevsearly}b).
(Fig.~\ref{cap:latevsearly}c is based on simulation \#3 which contains
too many bright, and thus easily classifiable, galaxies.) This $\sigma n/n<1$
cut only discards $\sim8$\% of the catalog. Thus we recommend it
as a good compromise between selecting the maximum number of objects
and selecting a reliable sample.

\begin{figure}
\includegraphics[width=0.5\linewidth, angle=270]{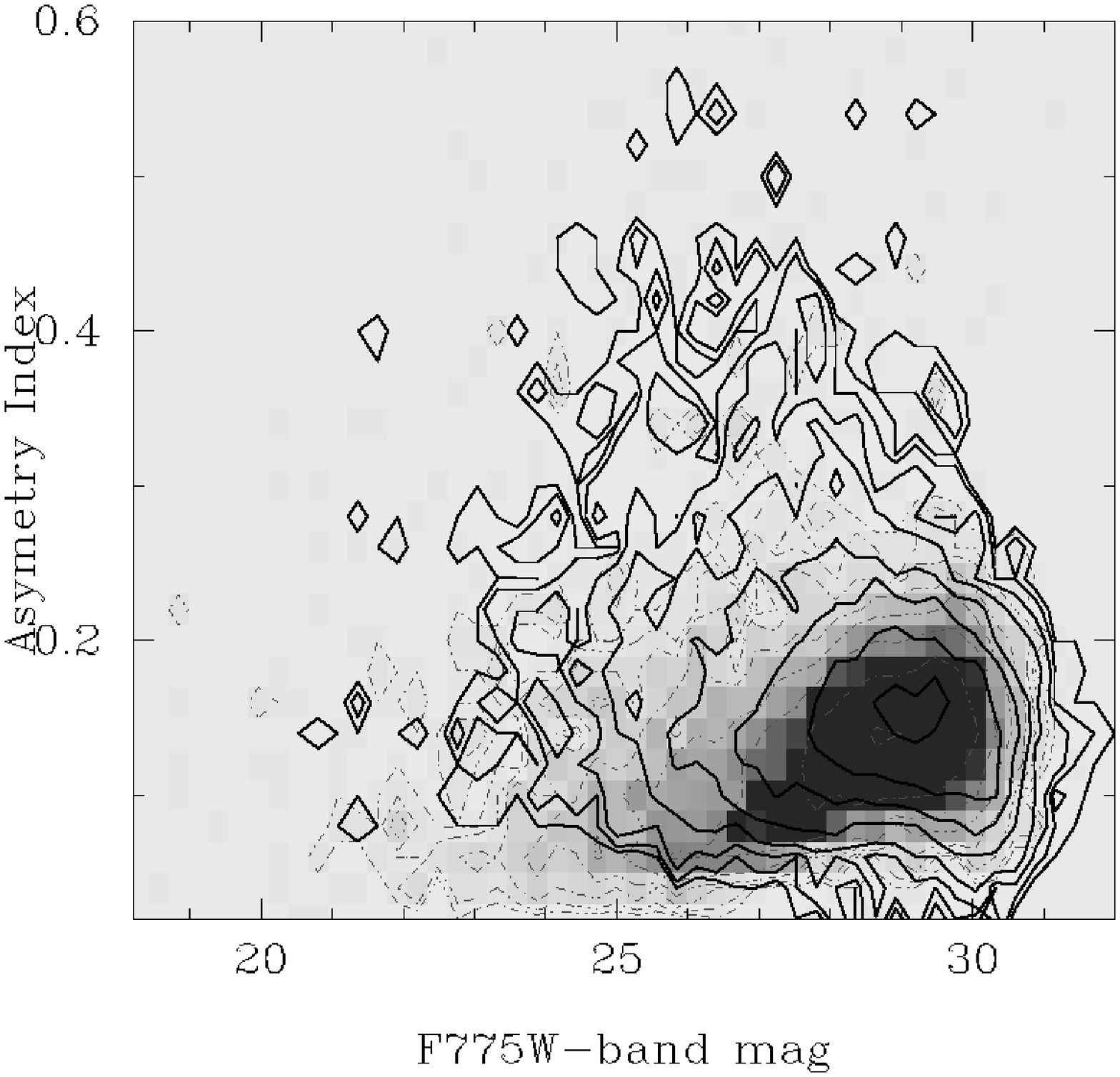}
\caption{\label{asym} Distribution of the asymmetry indices as a function
of $i\arcmin$-band magnitude for the UDF galaxies (continuous contours)
overlaid on top of the distribution for galaxies in simulation \#1
(greyscale and discontinuous contours).}
\end{figure}

Figure \ref{asym} compares the UDF galaxy asymmetries with those
from simulation \#1. The simulated galaxies are intrinsically symmetric,
so any observed asymmetry is due to noise. Asymmetry measurements
of real galaxies that fall below those typical of simulated galaxies
of similar magnitude should not be considered reliable.

\clearpage

\begin{thebibliography}{}
\bibitem[Abraham et al.(1996)]{Abraham96} Abraham, R.~G., van den 
Bergh, S., Glazebrook, K., Ellis, R.~S., Santiago, B.~X., Surma, P., \& 
Griffiths, R.~E.\ 1996, \apjs, 107, 1 
\bibitem[Andredakis et al.(1995)]{Andredakis95} Andredakis, Y.~C., 
Peletier, R.~F., \& Balcells, M.\ 1995, \mnras, 275, 874 
\bibitem[Beckwith et al.\ (2003)]{UDFACS}Beckwith, S., Somerville, R., Stiavelli, M. 2003, STScI Newsletter vol 20 issue 04
\bibitem[Ben{\'{\i}}tez et al.\ (1999)]{Benitez99} Ben{\'{\i}}tez, N., Broadhurst, T., 
Bouwens, R., Silk, J. \& Rosati, P. 1999, \apjl, 515, L65
\bibitem[Ben{\'{\i}}tez(2000)]{BPZ00} Ben{\'{\i}}tez, N.\ 
2000, \apj, 536, 571 
\bibitem[Ben{\'{\i}}tez et al.(2004)]{Benitez04} Ben{\'{\i}}tez, 
N., et al.\ 2004, \apjs, 150, 1 
\bibitem[Bertin \& Arnouts(1996)]{SExref} Bertin, E., \& 
Arnouts, S.\ 1996, \aaps, 117, 393
\bibitem[Bohlin et al.(2005)]{Bohlin05} Bohlin, R.~C., Lindler, D.~J., \& Riess, A.\ 2005 NICMOS Instrument Science Report 2005-002 (Baltimore: STScI)
\bibitem[Bohlin et al.(2006)]{Bohlin06} Bohlin, R.~C., Riess, A., \& de Jong, R.\ 2006 NICMOS Instrument Science Report 2006-002 (Baltimore: STScI)
\bibitem[Bouwens et al.(2005)]{Bouwens05} Bouwens, R.~J., 
Illingworth, G.~D., Blakeslee, J.~P., \& Franx, M.\ 2005, ArXiv 
Astrophysics e-prints, arXiv:astro-ph/0509641 
\bibitem[Bruzual \& Charlot(2003)]{BC03} Bruzual, G., \& 
Charlot, S.\ 2003, \mnras, 344, 1000 
\bibitem[Bunker et al.(2004)]{Bunker04} Bunker, A.~J., Stanway, 
E.~R., Ellis, R.~S., \& McMahon, R.~G.\ 2004, \mnras, 355, 374 
\bibitem[Casertano et al.(2000)]{drizzleRMS} Casertano, S., et 
al.\ 2000, \aj, 120, 2747 
\bibitem[Cimatti et al.(2002)]{Cimatti02} Cimatti, A., et al.\ 
2002, \aap, 392, 395 
\bibitem[Coleman et al.(1980)]{CWW80} Coleman, G.~D., Wu, 
C.-C., \& Weedman, D.~W.\ 1980, \apjs, 43, 393
\bibitem[Conselice et al.(2000)]{Conselice00} Conselice, C.~J., 
Bershady, M.~A., \& Jangren, A.\ 2000, \apj, 529, 886 
\bibitem[Conselice et al.(2003)]{Conselice03} Conselice, C.~J., 
Bershady, M.~A., Dickinson, M., \& Papovich, C.\ 2003, \aj, 126, 1183
\bibitem[Cowie et al.(2004)]{Cowie04} Cowie, L.~L., Barger, 
A.~J., Hu, E.~M., Capak, P., \& Songaila, A.\ 2004, \aj, 127, 3137 
\bibitem[Croom et al.(2001)]{Croom01} Croom, S.~M., Warren, 
S.~J., \& Glazebrook, K.\ 2001, \mnras, 328, 150 
\bibitem[Dickinson et al.(2004)]{Dickinson04} Dickinson, M., et 
al.\ 2004, \apjl, 600, L99 
 \bibitem[Fern{\'a}ndez-Soto et al.(1999)]{FLY99} 
Fern{\'a}ndez-Soto, A., Lanzetta, K.~M., \& Yahil, A.\ 1999, \apj, 513, 34 (FLY99)
\bibitem[Fern{\'a}ndez-Soto et al.(2001)]{FernandezSoto01} 
Fern{\'a}ndez-Soto, A., Lanzetta, K.~M., Chen, H.-W., Pascarelle, S.~M., \& 
Yahata, N.\ 2001, \apjs, 135, 41 
\bibitem[Ford et al.(2002)]{ACS} Ford, H.~C., et al.\ 2002, 
Bulletin of the American Astronomical Society, 34, 675 
\bibitem[Fruchter \& Hook(2002)]{drizzle} Fruchter, A.~S., \& 
Hook, R.~N.\ 2002, \pasp, 114, 144  
\bibitem[Giavalisco et al.(2004)]{GOODS} Giavalisco, M., et 
al.\ 2004, \apjl, 600, L93 
 \bibitem[Gilli et al.(2003)]{Gilli03} Gilli, R., et al.\ 2003, 
\apj, 592, 721 
\bibitem[Gwyn \& Hartwick(1996)]{Gwyn96} Gwyn, S.~D.~J., \& 
Hartwick, F.~D.~A.\ 1996, \apjl, 468, L77 
\bibitem[Gwyn \& Hartwick(2005)]{Gwyn05} Gwyn, S.~D.~J., \& 
Hartwick, F.~D.~A.\ 2005, \aj, 130, 1337 
\bibitem[de Jong et al.(2006a)]{lamptests} de Jong, R.~S., et al.\ 2006, NICMOS Instrument Science Report 2006-001 (Baltimore: STScI)
\bibitem[de Jong et al.(2006b)]{NICMOScal} de Jong, R.~S., et al.\ 2006, 2005 HST Calibration Workshop (Baltimore: STScI)
\bibitem[de Jong(2006)]{NICMOScor} de Jong, R.~S.\ 2006, NICMOS Instrument Science Report 2006-003 (Baltimore: STScI)
\bibitem[Kinney et al.(1996)]{Kinney96} Kinney, A.~L., Calzetti, 
D., Bohlin, R.~C., McQuade, K., Storchi-Bergmann, T., \& Schmitt, H.~R.\ 
1996, \apj, 467, 38 
\bibitem[Koo(1985)]{Koo85} Koo, D.~C.\ 1985, \aj, 90, 418 
\bibitem[Kauffmann et al.(2004)]{Kauffmann04} Kauffmann, G., White, 
S.~D.~M., Heckman, T.~M., M{\'e}nard, B., Brinchmann, J., Charlot, S., 
Tremonti, C., \& Brinkmann, J.\ 2004, \mnras, 353, 713 
\bibitem[Le F{\`e}vre et al.(2004)]{VVDS} Le F{\`e}vre, O., 
et al.\ 2004, \aap, 428, 1043
\bibitem[Mobasher \& Riess(2005)]{MobasherISR05} Mobasher, B., \& Riess, A.\ 2005 NICMOS Instrument Science Report 2005-003 (Baltimore: STScI)
\bibitem[Mobasher et al.(2005)]{Mobasher05} Mobasher, B., et al.\ 2005, \apj, 635, 832 
\bibitem[Peng et al.(2002)]{galfit} Peng, C.~Y., Ho, L.~C., 
Impey, C.~D., \& Rix, H.-W.\ 2002, \aj, 124, 266 
\bibitem[Rix et al.(2004)]{GEMS} Rix, H.-W., et al.\ 2004, 
\apjs, 152, 163 
\bibitem[S{\'a}nchez et al.(2004)]{Sanchez04} S{\'a}nchez, S.~F., 
et al.\ 2004, \apj, 614, 586 
\bibitem[Schade et al.(1995)]{Schade95} Schade, D., Lilly, 
S.~J., Crampton, D., Hammer, F., Le Fevre, O., \& Tresse, L.\ 1995, \apjl, 
451, L1  
\bibitem[Schlegel et al.(1998)]{dustmaps} Schlegel, D.~J., 
Finkbeiner, D.~P., \& Davis, M.\ 1998, \apj, 500, 525 
\bibitem[Sersic(1968)]{Sersic}{S{\'e}rsic, J.~L.\ 1968, Atlas de Galaxias Australes (C{\'o}rdoba, Argentina: Press of the Universidad Nacional)}
\bibitem[Shen et al.(2003)]{Shen03} Shen, S., Mo, H.~J., 
White, S.~D.~M., Blanton, M.~R., Kauffmann, G., Voges, W., Brinkmann, J., 
\& Csabai, I.\ 2003, \mnras, 343, 978 
\bibitem[Sirianni et al.(2003)]{Sirianni03} Sirianni, M., et al.\ 
2003, American Astronomical Society Meeting Abstracts, 202,  
\bibitem[Stanway et al.(2003)]{SBM03} Stanway, E.~R., Bunker, 
A.~J., \& McMahon, R.~G.\ 2003, \mnras, 342, 439 
\bibitem[Stanway et al.(2004)]{Stanway04} Stanway, E.~R., Bunker, 
A.~J., McMahon, R.~G., Ellis, R.~S., Treu, T., \& McCarthy, P.~J.\ 2004, 
\apj, 607, 704 
\bibitem[Stanway et al.(2005)]{Stanway05} Stanway, E.~R., 
McMahon, R.~G., \& Bunker, A.~J.\ 2005, \mnras, 359, 1184 
\bibitem[Steidel \& Hamilton(1992)]{Steidel92} Steidel, C.~C., \& 
Hamilton, D.\ 1992, \aj, 104, 941 
\bibitem[Stetson(1994)]{daophot} Stetson, P.~B.\ 1994, \pasp, 
106, 250 
\bibitem[Stiavelli et al.(2004)]{Stiavelli04} Stiavelli, M., Fall, 
S.~M., \& Panagia, N.\ 2004, \apjl, 610, L1 
\bibitem[Strolger et al.(2004)]{Strolger04} Strolger, L.-G., et 
al.\ 2004, \apj, 613, 200 
\bibitem[Szokoly et al.(2004)]{Szokoly04} Szokoly, G.~P., et al.\ 
2004, \apjs, 155, 271 
\bibitem[Thompson et al.(2005)]{UDFNIC3} Thompson, R.~I., et 
al.\ 2005, ArXiv Astrophysics e-prints, arXiv:astro-ph/0503504 
\bibitem[Thompson et al.(2006)]{Thompsoncal} Thompson, R.~I., et 
al.\ 2006, \apj, submitted
\bibitem[Vanzella et al.(2001)]{VanzellaHDF-S} Vanzella, E., et al.\ 
2001, \aj, 122, 2190 
\bibitem[Vanzella et al.(2005)]{FORS2} Vanzella, E., et al.\ 
2005, \aap, 434, 53
\bibitem[Williams et al.(1996)]{HDF-N} Williams, R.~E., et 
al.\ 1996, \aj, 112, 1335
\bibitem[Williams et al.(1998)]{HDF-S} Williams, R., et al.\ 
1998, Bulletin of the American Astronomical Society, 30, 1366 
\bibitem[Wirth et al.(2004)]{Wirth04} Wirth, G.~D., et al.\ 
2004, \aj, 127, 3121 
\bibitem[Wolf et al.(2004)]{COMBO-17} Wolf, C., et al.\ 2004, 
\aap, 421, 913 
\bibitem[Wolf et al.(2005)]{Wolf05} Wolf, C., et al.\ 2005, 
\apj, 630, 771 
\bibitem[Yan \& Windhorst(2004)]{YW04} Yan, H., \& 
Windhorst, R.~A.\ 2004, \apjl, 612, L93 
\end{thebibliography}
\end{document}